\documentclass[a4paper,12pt]{article}
 
\PassOptionsToPackage{original}{pict2e}
\usepackage{amsmath}
\usepackage{dsfont}
\usepackage{amsfonts}
\usepackage{amssymb}
\usepackage{color}
\usepackage{mathrsfs}
\usepackage{calc}
\usepackage{mathbbol}
\usepackage{colonequals}
\usepackage{pstricks}
\usepackage{lmodern}
\usepackage[english]{babel}
\usepackage{xcolor}
\usepackage{pst-plot}
\usepackage{upgreek}
\usepackage{babel,blindtext,scrpage2,eso-pic}
\usepackage{graphicx,pspicture}
\usepackage{colortbl}
\usepackage{mathtools}
\usepackage{array}
\usepackage{wrapfig}
\usepackage{subfigure}
\usepackage[makeroom]{cancel}
\usepackage{parcolumns}
\usepackage{framed}
\usepackage{booktabs}
\usepackage{psfrag}
\usepackage[margin=10pt,font=small,labelfont=bf,
labelsep=period]{caption}
\usepackage[bottom]{footmisc} 
\usepackage{indentfirst} 
\usepackage{upref} 
\usepackage{cite} 
\usepackage[rightcaption]{sidecap}
\usepackage{psfrag}

\definecolor{rhsFormColor}{rgb}{0.05,0.05,0.35}
\definecolor{Formel}{rgb}{0.05,0.05,0.35}
\definecolor{Formel2}{rgb}{0.40,0.40,0.55}
\definecolor{SideNote}{rgb}{0.35,0.35,0.4}
\definecolor{fSubDeriv}{rgb}{0.1,0.1,0.1}
\definecolor{fSubDerivA}{rgb}{0.35,0.35,0.50}
\definecolor{Text}{rgb}{0,0,0}
\definecolor{can1}{rgb}{0.7,0.1,0.1}
\definecolor{can2}{rgb}{0.45,0.45,0.15}
\definecolor{can3}{rgb}{0.05,0.3,0.05}
\definecolor{can4}{rgb}{0.5,0.1,0.1}
\definecolor{can5}{rgb}{0.1,0.5,0.1}
\definecolor{can6}{rgb}{0.15,0.45,0.45}
\definecolor{can7}{rgb}{0.3,0.05,0.05}
\definecolor{can8}{rgb}{0.35,0.1,0.35}
\definecolor{can9}{rgb}{0.5,0.3,0.4}
\definecolor{SumWeylCol}{rgb}{0.05,0.05,0.35}
\definecolor{SumGeoCol}{rgb}{0.05,0.05,0.35}
\colorlet{TableR1}{Formel}
\colorlet{TableR2}{Formel!75}
\colorlet{TableC1}{SideNote}
\definecolor{TableC2}{rgb}{0.8,0.8,0.8}
\colorlet{TableC3}{SideNote!10}
\definecolor{TableNot}{rgb}{0.8,0.8,0.8}
\definecolor{TableYes}{rgb}{1,1,1}
\definecolor{pageNumberC}{rgb}{0,0,0}

\numberwithin{equation}{section}

\renewcommand{\baselinestretch}{1.258}
\newcommand{\RegA}{R}
\newcommand{\arbit}{\text{arbitrary}}

\newcommand{\CutO}{{\cal R}}

\makeatletter

\newcommand{\Rmnum}[1]{\expandafter\@slowromancap\romannumeral #1@}
\makeatother

\newcommand{\selfcon}{\text{selfcon}}

\newcommand{\noR}{\mathds{R}}

\newcommand{\On}[1]{O(#1)}

\renewcommand{\Re}{\operatorname{Re}}

\newcommand{\MaF}{\mathcal{M}} 

\newcommand{\MaFs}{\mathcal{M}}  

\newcommand{\Ric}{R}     
\newcommand{\SR}{R}     
\newcommand{\EinT}{G}    
\newcommand{\Ricb}{\bar{\Ric}}     
\newcommand{\SRb}{\bar{\SR}}     
\newcommand{\Z}{D}      

\newcommand{\bg}{\bar{g}}   
\newcommand{\bZ}{\bar{\Z}}   

\newcommand{\iM}{H}  
\newcommand{\EC}{K}  

\newcommand{\cUV}{\mathscr{S}_{\text{UV}}}
 
\newcommand{\id}{\delta} 
\newcommand{\idm}{\Eins} 
\newcommand{\var}{\delta} 
\newcommand{\varFunc}{\delta}     
\newcommand{\fD}[2]{\frac{\varFunc #1}{\varFunc #2}} 
\newcommand{\tr}{\text{tr}}   
\newcommand{\Tr}{\text{Tr}}   
\newcommand{\Order}[1]{\mathcal{O}(#1)}  
\newcommand{\OrderX}[2]{\mathcal{O}^{#2}(#1)}
\newcommand{\GF}[1]{\Gamma\left(#1\right)}   
 
\newcommand{\md}{\text{d}}
 
\newcommand{\NMW}{1cm} 
\newcommand{\FEW}{0.3cm} 
 
\newcommand{\Ii}{i}      
\newcommand{\LHS}{\text{LHS}}     
\newcommand{\RHS}{\text{RHS}}     

\newcommand{\const}{\text{const}}    
 
\definecolor{shadecolor}{rgb}{0.98,.98,1}
\definecolor{framecolor}{rgb}{0.5,0.5,1}
\definecolor{shadecolorII}{rgb}{0.98,0.98,0.98}
\definecolor{framecolorII}{rgb}{0.6,0.6,0.6}

\newenvironment{bshaded}{%
 \MakeFramed {\FrameRestore}}%
 {\endMakeFramed}

 {\endMakeFramed}

\newcommand{\SW}{S}  

\newcommand{\EAA}{\Gamma}  
\newcommand{\CGF}{W}   
 \newcommand{\ParFD}{\mathds{Z}}  
 \newcommand{\frEny}{F}				
 
\newcommand{\Ghx}{\xi}   
\newcommand{\GhAx}{\bar{\xi}}  

\newcommand{\CosmConst}{\lambda}
\newcommand{\Kk}{\CosmConst}  
\newcommand{\Kkbar}{\bar{\CosmConst}}  
\newcommand{\KkEC}{\CosmConst^{\partial}}  
\newcommand{\KkbarEC}{\bar{\CosmConst}^{\partial}}  

\newcommand{\Nk}{G}  
\newcommand{\NkEC}{G^{\partial}}  
 
\newcommand{\ThrfA}[3]{\Phi^{#1}_{#2}\left({ #3}\right)} 
\newcommand{\ThrfB}[3]{\widetilde{\Phi}^{#1}_{#2}\left(#3\right)}

\newcommand{\Mdim}[1]{\left[#1\right]}  
 
\newcommand{\background}{\text{B}}

\newcommand{\lin}{\text{lin}}
\newcommand{\biM}{\bar{\iM}}
\newcommand{\bEC}{\bar{\EC}}
\newcommand{\RegGr}{R^{\text{grav}}} 
\newcommand{\RegGh}{R^{\text{gh}}} 
\newcommand{\RegX}[1]{R^{(0)}\left(#1\right)} 
\newcommand{\RegXd}[1]{R^{(0)\prime}\left(#1\right)} 

\newcommand{\aDzG}{\eta}
 
\newcommand{\aDz}{\eta_{\text{N}}}  

\newcommand{\aDzEC}{\eta^{\partial} _{\text{N}}}  

\newcommand{\PotAkA}{V^{(0)}_k(A)}
\newcommand{\PotAkB}{V^{(1)}_k(A)}
\newcommand{\PotAd}[1]{V^{#1}_k(A)}

\newcommand{\tg}{g}
\newcommand{\nkA}{G_k^{(0)}}
\newcommand{\nkAB}{G_k^{(0,\partial)}}
\newcommand{\nkB}{G_k^{(1)}}
\newcommand{\nkBB}{G_k^{(1,\partial)}}
\newcommand{\kkA}{\Lambda_k^{(0)}}
\newcommand{\kkAB}{\Lambda_k^{(0,\partial)}}
\newcommand{\kkB}{\Lambda_k^{(1)}}
\newcommand{\xiA}{\xia_k}
\newcommand{\xiaS}{\xi^{(0,\partial)}}
\newcommand{\xiAS}{\xiaS_k}
\newcommand{\xibS}{\xi^{(1,\partial)}}
\newcommand{\xiBS}{\xibS_k}
\newcommand{\xiBg}{\xibg_k}
\newcommand{\xiBR}{\xibR_k}
\newcommand{\xia}{\xi^{(0)}}
\newcommand{\xibg}{\xi^{(1,\text{\Rmnum{1}})}}
\newcommand{\xibR}{\xi^{(1,\text{\Rmnum{2}})}}

\newcommand{\eka}{E_k}   

\newcommand{\TOp}[2]{W_{#1}\left(#2;A\right)}
\newcommand{\ns}{n_{\text{s}}}
\newcommand{\bmk}{\bar{m}_k}
\newcommand{\mk}{m_k}

\newcommand{\mka}{m^{(0)}}
\newcommand{\mkB}{m^{(0)\,2}_k}
\newcommand{\mkC}{m^{(1)\,2}_k}

\newcommand{\omgA}{u_k}
\newcommand{\bomg}{\bar{u}_k}

\newcommand{\AS}{A^{\partial}} 

\newcommand{\EAAbulk}{\EAA^{\text{bulk}}}
\newcommand{\EAAS}{\EAA^{\partial}}
\newcommand{\flcb}{\bar{h}}
\newcommand{\FPop}{\mathcal{M}}
\newcommand{\ud}[2]{^{#1}_{\phantom{#1}#2}}

\newcommand{\Id}{I}
\newcommand{\calE}{{\cal E}}

\begin{document}

\addtolength{\topmargin}{-10mm} 
\begin{titlepage}
\enlargethispage{1cm}
\renewcommand{\baselinestretch}{1.1} 
\title{\begin{flushright}
\normalsize{MZ-TH/12-14}
\vspace{1cm}
\end{flushright}
Running boundary actions, Asymptotic Safety, and black hole thermodynamics}
\date{}
\author{Daniel Becker and Martin Reuter\\
{\small Institute of Physics, University of Mainz}\\[-0.2cm]
{\small Staudingerweg 7, D-55099 Mainz, Germany}}
\maketitle\thispagestyle{empty}

\begin{abstract} 
Previous explorations of the Asymptotic Safety scenario in Quantum Einstein Gravity (QEG) by means of the effective average action and its associated functional renormalization group (RG) equation assumed spacetime manifolds which have no boundaries. Here we take a first step towards a generalization for non-trivial boundaries, restricting ourselves to action functionals which are at most of second order in the derivatives acting on the metric. We analyze two examples of truncated actions with running boundary terms: full fledged QEG within the single-metric Einstein-Hilbert truncation, augmented by a scale dependent Gibbons-Hawking surface term, and a bi-metric truncation for gravity coupled to scalar matter fields. The latter contains 17 running couplings, related to both bulk and boundary terms, whose beta-functions are computed in the induced gravity approximation (large $N$ limit). We find that the bulk and the boundary Newton constant, pertaining to the Einstein-Hilbert and Gibbons-Hawking term, respectively, show opposite RG running; proposing a scale dependent variant of the ADM mass we argue that the running of both couplings is consistent with gravitational anti-screening. We discuss the status of the `bulk-boundary matching' usually considered necessary for a well defined variational principle within the functional RG framework, and we explain a number of conceptual issues related to the `zoo' of (Newton-type, for instance) coupling constants, for the bulk and the boundary, which result from the bi-metric character of the gravitational average action. In particular we describe a simple device for counting the number of field modes integrated out between the infrared cutoff scale and the ultraviolet. This method makes it manifest that, in an asymptotically safe theory, there are effectively no field modes integrated out while the RG trajectory stays in the scaling regime of the underlying fixed point.
As an application, we investigate how the semiclassical theory of Black Hole Thermodynamics gets modified by quantum gravity effects and compare the new picture to older work on `RG-improved black holes' which incorporated the running of the bulk Newton constant only. We find, for instance, that the black hole's entropy vanishes and its specific heat capacity turns positive at Planckian scales.

\end{abstract}

\end{titlepage}
\addtolength{\topmargin}{10mm} 

\section{Introduction}

It is well known that Einstein's field equation of classical General Relativity can be obtained by requiring the Einstein-Hilbert action functional 
\begin{align}
S_{\text{EH}}\left[g_{\mu\nu}\right]&=-\frac{1}{16\pi G} \int_{\MaFs}\md^d x \sqrt{g} \, \SR \label{eqn:01_01}
\end{align}
to become stationary, {\it provided} the spacetime manifold, $\MaFs$, has no boundary.
Trying to generalize the variational principle to spacetimes with a non-empty boundary $\partial\MaFs$ one faces the difficulty that $S_{\text{EH}}$ responds to a change $\var g_{\mu\nu}$ of the metric, vanishing on $\partial\MaFs$, by producing a certain surface term, over and above the desired `bulk' term containing the Einstein tensor $\EinT^{\mu\nu}$:
\begin{align}
\var S_{\text{EH}} &=\frac{1}{16\pi G}\left(\int_{\MaFs}\md^d x \sqrt{g} \,\, \EinT^{\mu\nu}\,\var g_{\mu\nu} + \int_{\partial\MaFs} \md^{d-1}x \sqrt{\iM}\,\, \iM^{\alpha\beta} n^{\mu} \partial_{\mu} \,\var g_{\alpha\beta}   \right) \label{eqn:01_02}
\end{align}
Here $\iM_{\alpha\beta}$ denotes the metric induced on the boundary, and $n^{\mu}$ is the corresponding normal vector field. When we vary the metric we keep its boundary values fixed ; hence, by assumption, $\var g_{\mu\nu}$ vanishes on $\partial\MaFs$. Then the derivatives of $\var g_{\mu\nu}$ in directions tangential to $\partial\MaFs$ will vanish as well, but not necessarily its normal derivative: $\left.n^{\mu}\partial_{\mu}\, \var g_{\alpha\beta}\right|_{\partial\MaFs}\neq 0$. For this reason the surface term in \eqref{eqn:01_02} is non-zero in general, and the condition of stationarity $\var S_{\text{EH}}=0$ is not equivalent to Einstein's equation $\EinT^{\mu\nu}=0$.

Difficulties of this kind are not uncommon in Lagrangian or Hamiltonian systems. What they call for is an additional surface contribution to the action whose variation cancels the unwanted surface term originating from its volume part. 

In General Relativity, the most popular proposal\footnote{For recent proposals of different surface corrections see \cite{AA-surface}; for a general discussion see \cite{pad-book,pad-review}.} for a surface correction with this property is the Gibbons-Hawking term \cite{BH-Therm},
\begin{align}
 S_{\text{GH}}&=\frac{1}{16\pi G}\int_{\partial \MaFs}\md^{d-1}x \sqrt{\iM}\, \left(-2 \EC\right) \label{eqn:01_03}
\end{align}
where $\EC$ denotes the trace of the extrinsic curvature of the boundary.%
\footnote{In the literature, $S_{\text{GH}}$ is usually normalized by replacing $\EC\rightarrow \EC-\EC_0$ in \eqref{eqn:01_03}, with $\EC_0$ the trace of the extrinsic curvature tensor appropriate for an embedding of the boundary manifold in flat space. While subtracting $\EC_0$ does not change $\var S_{\text{GH}}$, we shall refrain from it here for reasons to be discussed below.}  
The variation of $S_{\text{GH}}$ cancels indeed the second term on the right hand side ($\RHS$) of \eqref{eqn:01_02} so that the stationary points of the total action are precisely those metrics satisfying Einstein's equation:
$\var\left(S_{\text{EH}}+S_{\text{GH}}\right)=0 \, \Leftrightarrow \, \EinT^{\mu\nu}=0$.

The Gibbons-Hawking term has played an important r\^{o}le in the Euclidean functional integral approach to black hole thermodynamics \cite{BH-Therm, Eucl-Book}.
In the leading order of the semiclassical expansion the black hole's free energy is given by the `on-shell' value of the classical action functional. Since, for vacuum solutions, $\Ric_{\mu\nu}=0$ implies a vanishing contribution from the bulk term, the free energy, and hence all derived thermodynamical quantities such as the entropy, for instance,  stem entirely from the surface term. It is important to ask how this picture presents itself in full fledged quantum gravity \cite{Kiefer}. There exist already detailed investigations within Loop Quantum Gravity \cite{A,R,T}, for instance \cite{LQG-BH}.

The present paper is devoted to an analysis of surface terms in Quantum Einstein Gravity (QEG) within a different approach to quantum gravity, Asymptotic Safety, \cite{wein, NJP, livrev}. More generally, we shall be interested in the renormalization group (RG) evolution of scale dependent gravitational actions which include surface terms. Concretely we shall use the effective average action to formulate a diffeomorphism invariant and, most importantly, background independent coarse graining flow on the `theory space' of action functionals for the metric \cite{mr}. While in the past this approach was limited to the quantization of gravity on spacetimes $\MaFs$ without a boundary\footnote{See however \cite{AS-surf-eps} for an early perturbative calculation of the induced surface term in $2+\epsilon$ dimensional gravity.}, we shall now allow for a non-trivial boundary, $\partial\MaFs\neq \emptyset$.

A pivotal building block of the approach in\cite{mr} is the `paradoxical' implementation of background independence by first introducing a background metric $\bg_{\mu\nu}$, decomposing the bare, dynamical metric $\gamma_{\mu\nu}$ as $\gamma_{\mu\nu}=\bg_{\mu\nu}+h_{\mu\nu}$, quantizing then the fluctuations $h_{\mu\nu}$ similar to a matter field in the classical background spacetime equipped with $\bg_{\mu\nu}$, and, ideally, verifying at the very end that the observable predictions of the resulting quantum field theory are independent of the metric $\bg_{\mu\nu}$ chosen. This theory is fully described by an effective action $\EAA[g_{\mu\nu},\bg_{\mu\nu}]$ where $\bg_{\mu\nu}$ does not refer to any concrete geometry, but rather is a second free argument, almost on a par with the dynamical metric, i.e. the expectation value $g_{\mu\nu}=\langle \gamma_{\mu\nu} \rangle = \bg_{\mu\nu} + \flcb_{\mu\nu}$ with $\flcb\equiv\langle h_{\mu\nu}\rangle$. (In this section we suppress matter fields and Faddeev-Popov ghosts in the list of arguments.) Often it is more natural to consider $\flcb_{\mu\nu}$ rather than $g_{\mu\nu}$ the dynamical field argument of the action, and one sets $\EAA[\flcb_{\mu\nu}; \bg_{\mu\nu}]\equiv \EAA[\bg_{\mu\nu}+\flcb_{\mu\nu}, \bg_{\mu\nu}]$.

The advantage of the background field technique \cite{DeWitt-books} is that it sidesteps many of the profound conceptual problems which arise in approaches (such as Loop Quantum Gravity, for instance) where one tries to quantize gravity by starting from a vacuum state which amounts to {\it no spacetime at all}, let alone a spacetime manifold carrying some non-degenerate metric \cite{A,R,T}.
This advantage comes at a price, however, namely the intrinsic `bi-metric' nature of the formalism: to fully control the effective action functional, we must know its dependence on {\it two} rather than just one metric \cite{mr, elisa2}.

Assuming the corresponding variational principle is well defined, the effective action $\EAA[\flcb; \bg]$ gives rise to an effective field equation which governs the dynamics of $\flcb_{\mu\nu}(x)\equiv \flcb_{\mu\nu}[\bg](x)$ in dependence on the background metric:
\begin{align}
 \frac{\var}{\var \flcb_{\mu\nu}(x)}\EAA[\flcb;\bg]= 0 \label{eqn:01_03a}
\end{align}
For special, so-called `self-consistent' backgrounds $\bg_{\mu\nu}\equiv \bg_{\mu\nu}^{\text{selfcon}}$ it happens that eq. \eqref{eqn:01_03a} is solved by an identically vanishing fluctuation expectation value: $\flcb_{\mu\nu}[\bg^{\text{selfcon}}](x)\equiv 0$. Then the expectation value of the quantum metric $g_{\mu\nu}\equiv g_{\mu\nu}[\bg]$ equals exactly the background metric, $g_{\mu\nu}=\bg_{\mu\nu}$. The defining condition for a self-consistent background, 
\begin{align}
 \left. \frac{\var}{\var \flcb_{\mu\nu}(x)}\EAA[\flcb;\bg^{\text{selfcon}}]\right|_{\flcb=0}= 0 \label{eqn:01_03b}
\end{align}
is referred to as the {\it tadpole equation} since it expresses the vanishing of the fluctuation 1-point-function.
We may regard the tadpole equation as an `effective', i.e. quantum mechanically corrected analogue of the classical field equation. Only a single metric, $\bg^{\text{selfcon}}$, enters this equation. Interestingly enough, {\it the tadpole equation does not obtain as the stationarity condition of any action functional, i.e. not by a variation with respect to $\bg^{\text{selfcon}}$.}

In an analogous fashion we may define higher $n$-point 1PI Green's functions, for arbitrary $\bg$,
\begin{align}
 \left. \frac{\var}{\var \flcb_{\mu\nu}(x_1)}\cdots \frac{\var}{\var \flcb_{\rho\sigma}(x_n)}\EAA[\flcb;\bg]\right|_{\flcb=0} \label{eqn:01_03c}
\end{align}
An a priori different set of $n$-point functions is generated by differentiating the reduced functional $\EAA^{\text{red}}[\bg]\equiv \left.\EAA[\flcb;\bg]\right|_{\flcb=0}$ with respect to $\bg_{\mu\nu}$:
\begin{align}
 \frac{\var}{\var \bg_{\mu\nu}(x_1)}\cdots \frac{\var}{\var \bg_{\rho\sigma}(x_n)}\EAA^{\text{red}}[\bg] \label{eqn:01_03d}
\end{align}
It is an important theorem \cite{joos, DeWitt-books} that the sets of Green's functions \eqref{eqn:01_03c} and \eqref{eqn:01_03d} are {\it on-shell equivalent} if one uses a special type of gauge fixing condition (a `background gauge fixing term'), and if the quantization scheme respects split symmetry (see below) in the physical sector.

In standard applications of the background field method, say in perturbation theory on a classical flat spacetime, in Yang-Mills-type gauge theories, for instance, the doubling of fields is not much of a drawback usually. Using a `background'-type gauge fixing condition, the single-metric functional $\EAA^{\text{red}}[g]= \EAA[g,\bg=g]$ contains the same physics as $\EAA[g,\bg]$ in the sense it generates the same on-shell scattering matrix elements (and likewise for Yang-Mills theory). Moreover, in Yang-Mills theory on Minkowski space there are no deep conceptual reasons that would suggest the background approach. So, in situations where it does not simplify matters one just will not use it.

In quantum gravity, in particular in Asymptotic Safety, the bi-metric character must be taken much more seriously, for at least two reasons:
First of all, contrary to Yang-Mills theory on Minkowski space there is no obvious simple way of avoiding it, unless one is willing to embark on the profound difficulties of creating spacetime `from nothing'. 
Second, the background metric is indispensable in defining a coarse graining operation for the quantum fluctuations of the dynamical metric, at least as long as the notion of `coarse graining' is still meant to bear a certain resemblance to the classical one based upon Fourier analysis on flat space.

The gravitational average action $\EAA_k[g_{\mu\nu},\bg_{\mu\nu}]$ is formally derived from a gauge-fixed functional integral over $h_{\mu\nu}$, whereby the coarse graining is realized as a smooth infrared (IR) cutoff which suppresses the fluctuation modes of the covariant Laplacian built from the background metric, $-\bZ^2$, at the scale $k^2$. To this end one adds a bilinear term 
$\Delta_k S[h;\bg] \propto k^2 \int\md^d x \sqrt{\bg}\, h_{\mu\nu}\, R^{(0)}(-\bZ^2\slash k^2)\, h^{\mu\nu}$ 
to the bare action under the functional integral. The `shape function' $R^{(0)}(-\bZ^2\slash k^2)$ approaches zero (one) for arguments much larger (smaller) than one.
For the details of the construction we refer to \cite{mr}. Suffice it to say that $\EAA_k[g_{\mu\nu},\bg_{\mu\nu}]$ approaches the ordinary effective action $\EAA[g_{\mu\nu},\bg_{\mu\nu}]$ in the limit $k\rightarrow 0$, and the bare action $S$, up to a simple correction term, for $k\rightarrow \infty$. This interpolation is described by a functional RG equation (FRGE) which defines a flow on the theory space spanned by all diffeomorphism invariant action functionals depending on the two metrics $g_{\mu\nu}$ and $\bg_{\mu\nu}$, or equivalently on $\flcb_{\mu\nu}$ and $\bg_{\mu\nu}$, as well as on the ghosts $\Ghx,\, \GhAx$, and possibly also on  matter fields, $A$: $\EAA_k[\flcb,\Ghx,\GhAx,A;\bg]\equiv \EAA_k[g\equiv \bg+\flcb,\bg,\Ghx,\GhAx,A]$. Having to deal with an FRGE on this rather complicated theory space is the price we pay for background independence in this approach, and for avoiding an explicit `creatio ex nihilo' of spacetime with a non-degenerate metric on it.

The second argument of $ \EAA_k[g\equiv \bg+\flcb,\bg,\cdots]$ is referred to as the `extra $\bg$-dependence' \cite{elisa2} of the average action; it represents the dependence on the background metric which does not combine with $\flcb$ to form a full dynamical metric $g=\bg+\flcb$. If $\EAA_k$ has a non-trivial extra $\bg$-dependence it violates the `background-quantum field split symmetry' \cite{elisa2} given by $\var \flcb_{\mu\nu}=\epsilon_{\mu\nu}$, $\var \bg_{\mu\nu}=-\epsilon_{\mu\nu}$. The combination $\bg+\flcb$, and in fact the entire classical action, respects this invariance, but not the gauge fixing and the mode suppression term $\Delta_k \SW \propto \int h \, \RegX{-\bZ^2\slash k^2} h$. The latter is bilinear in $h_{\mu\nu}$, can have any dependence\footnote{The term $\Delta_k \SW[h;\bg]$ is however required  to be invariant under diffeomorphisms acting on $h_{\mu\nu}$ and $\bg_{\mu\nu}$ simultaneously.} on $\bg_{\mu\nu}$, however. As a consequence, the RG flow generates split symmetry violating terms and, at least from a certain level of precision onward, truncations of theory space must allow for this possibility \cite{MRS1,MRS2}. 

Dropping the requirement of split symmetry  leads to an infinite enlargement of theory space (even if one truncates at some fixed canonical dimension, for instance). To see this, consider an arbitrary split-symmetric and background gauge invariant monomial occurring in $\EAA_k$ with a corresponding running prefactor: $u(k)\,\Order{g}\equiv u(k)\,\Order{\bg+\flcb}$. Taylor expanding this monomial in powers of $\flcb_{\mu\nu}$ we obtain an infinity of terms proportional to $\OrderX{\flcb;\bg}{(p)}$ which is homogeneous in $\flcb$ of order $p$. Since coarse graining violates split symmetry, the parametrization of a  generic $\EAA_k$ must contain all $\mathcal{O}^{(p)}$'s with independent coefficients, i.e. as a sum of the type $\sum_{p=0}^{\infty} u^{(p)}(k) \, \OrderX{\flcb;\bg}{(p)}$. Moreover, there exist (background gauge invariant) monomials one can built from $\flcb_{\mu\nu}$ and $\bg_{\mu\nu}$ which one must include and do not arise in this way. 

The first explicit gravitational $\EAA_k$-flow was computed in \cite{mr} within a `single-metric truncation' which is to say that the only extra $\bg$-dependence allowed by the ansatz for $\EAA_k$ is the trivial one in the gauge fixing term. Until very recently, also all later computations of $\EAA_k$-flows took over this approximation. 
The technically much harder exploration of genuine `bi-metric truncations' which allow for a non-trivial extra $\bg$-dependence (beyond the gauge fixing term) is still in its infancy; the first bi-metric RG flows \cite{elisa2} were obtained in conformally reduced gravity \cite{creh1,creh2, creh3}, matter induced gravity \cite{MRS1}, and in the `two-fold Einstein-Hilbert truncation' of full fledged quantum gravity \cite{MRS2}.

It should also be emphasized that, even when one leaves numerical precision aside, essential conceptual properties of the gravitational average action can be understood only by appreciating its intrinsic bi-metric character. In the present paper we are confronted with an example of this kind. As we shall see, because of the unavoidable dependence of $\EAA_k$ on both $\flcb_{\mu\nu}$ and $\bg_{\mu\nu}$, or on two metrics, the issue of boundary corrections presents itself in a somewhat non-standard fashion.

In this paper we analyze two specific examples of truncations involving surface terms. The first one is of single-metric type and generalizes the Einstein-Hilbert truncation of pure gravity to manifolds with non-empty boundary. In the second example we consider a running bi-metric action induced by the quantum fluctuations of a scalar matter multiplet. In order to disentangle conceptual issues from the (rather severe) calculational difficulties in the bi-metric case we limit ourselves to the induced gravity approximation here and neglect the effect of the metric fluctuations. At least in an appropriate large $N$ limit this should be a reliable approximation.

The remaining sections of this paper are organized as follows. In Section 2 we study the single-metric Einstein-Hilbert truncation generalized by including a Gibbons-Hawking surface term in the FRG approach. In Section 3 we employ a more advanced, bi-metric truncation of a gravity and matter system in  the induced gravity approximation. Besides Newton type couplings on the boundary we also  include non-minimal matter couplings and analyze their RG behavior, discussing in particular the status of split symmetry and the indications for  Asymptotic Safety. In Section 4 we study the crucial conceptual issues and problems related to the different RG properties of the various, classically identical, Newton couplings on the bulk and the boundary. As an application, we describe their impact on  the thermodynamics of black holes. Section 5 contains the Conclusions and an outlook to future work generalizing the investigations initiated here.

\section{The single-metric Einstein-Hilbert truncation} \label{sec:03}

In this section we analyze a first example of a truncated RG flow in presence of a boundary, namely the $\partial \MaFs \neq \emptyset$ generalization of the Einstein-Hilbert truncation within the single-metric setting \cite{mr}.

\subsection{The truncation ansatz}

We are going to consider a scale dependent Euclidean action in $d$ dimensions
\begin{align}
 \EAA_k&=\EAAbulk_k+\EAAS_k \label{eqn:03_01}
\end{align}
which consists of a bulk part $\EAAbulk_k\left[\flcb_{\mu\nu},\Ghx^{\mu},\GhAx_{\mu};\bg_{\mu\nu}\right]$
and a boundary piece $\EAAS_k[\flcb_{\mu\nu};\bg_{\mu\nu}]$. The former has the same structure as in ref. \cite{mr}:
\begin{align}
\EAAbulk_k[\flcb,\Ghx,\GhAx;\bg]&= - \frac{1}{16\pi\Nk_k} \left.\int_{\MaFs}\md^d x \sqrt{g}\, \left\{\SR(g)-2\Kkbar_k\right\}\right|_{g=\bg+\flcb} \nonumber \\
&\quad +\left(\frac{1}{16\pi\Nk_k}\right) \frac{1}{2\alpha}\int_{\MaFs}\md^d x \sqrt{\bg}\, \bg^{\mu\nu}\left(\mathcal{F}_{\mu}^{\alpha\beta}\flcb_{\alpha\beta}\right)\left(\mathcal{F}_{\nu}^{\rho\sigma}\flcb_{\rho\sigma}\right) \nonumber \\
&\quad -\sqrt{2} \int_{\MaFs}\md^d x \sqrt{\bg}\, \,\GhAx_{\mu}\,\FPop[\flcb;\bg]\ud{\mu}{\nu}\,\Ghx^{\nu} \label{eqn:03_02}
\end{align}
The first term on the $\RHS$ of \eqref{eqn:03_02} is the Einstein-Hilbert action of the full metric $\bg_{\mu\nu}+\flcb_{\mu\nu}\equiv g_{\mu\nu}$ with a scale dependent prefactor, involving the `bulk Newton constant' $\Nk_k$, and the running `bulk cosmological constant' $\Kkbar_k$. The second and third term on the $\RHS$ of \eqref{eqn:03_02} are the gauge fixing and the ghost term, respectively. As in \cite{mr} we employ the background variant of the harmonic gauge which amounts to the choice 
$\mathcal{F}_{\mu}^{\alpha\beta}=\id_{\mu}^{\beta}\bg^{\alpha\gamma}\bZ_{\gamma}-\frac{1}{2}\bg^{\alpha\beta}\bZ_{\mu}$
with the corresponding Faddeev-Popov operator $\FPop[\flcb;\bg]$.\footnote{For its explicit form see eq. (2.11) of ref.  \cite{mr}.}
We shall neglect the RG running of the parameter $\alpha$ and set $\alpha=1$ henceforth.

As for specifying the domain on which the functionals $\EAAbulk_k$ and $\EAAS_k$ are defined, we impose Dirichlet boundary conditions for the fluctuation field: $\left.\flcb_{\mu\nu}\right|_{\partial\MaFs}=0$. Recall that, while $\flcb$ and its tangential derivatives vanish on $\partial\MaFs$, its normal derivative will be non-zero in general.

Furthermore, the argument $\bg_{\mu\nu}$ of $\EAA_k$ is allowed to vary over all (non-degenerate) Riemannian metrics on $\MaFs$ consistent with its topology. For concreteness we fix the topology to that of a $d$-dimensional disk.

In the boundary action $\EAAS_k$ we include a scale dependent Gibbons-Hawking term with a prefactor which is allowed to run independently from the one in the bulk:
\begin{align}
 \EAAS_k[\flcb;\bg]&=-\frac{1}{16\pi\NkEC_k}\int_{\partial\MaFs}\md^{d-1}x \sqrt{\iM}\, \left(2\EC-2\KkbarEC_k\right) \label{eqn:03_03}
\end{align}
Here $\iM_{\mu\nu}\equiv g_{\mu\nu}-n_{\mu}n_{\nu}$ denotes the boundary metric pertaining to the {\it full} metric $g_{\mu\nu}\equiv \bg_{\mu\nu}+\flcb_{\mu\nu}$. Furthermore, $n^{\mu}$ is the outward unit normal vector field of $\partial\MaFs$, and $\EC$ is the trace of the corresponding extrinsic curvature tensor:
\begin{align}
\EC&=\Z_{\nu}n^{\nu}= g^{\mu\nu}\Z_{\mu}n_{\nu} \label{eqn:03_03_1} 
\end{align}
Writing $\EC=\left(\iM^{\mu\nu}+n^{\mu}n^{\nu}\right)\Z_{\mu}n_{\nu}$ and exploiting the normalization condition $n^{\nu}n_{\nu}\equiv 1$ one obtains $\EC=\iM^{\mu\nu}\Z_{\mu}n_{\nu}$. This leads to the following representation which is often helpful:
\begin{align}
\EC=\iM^{\mu\nu}\left[\partial_{\mu}n_{\nu}-\Gamma^{\rho}_{\mu\nu}n_{\rho}\right]\label{eqn:03_03_2} 
\end{align}
Here $\iM^{\mu\nu}$ and the Christoffel symbol $\Gamma^{\rho}_{\mu\nu}$ refer to the metric $g_{\mu\nu}=\bg_{\mu\nu}+\flcb_{\mu\nu}$, which is also used to lower the index of the normal vector field: $n_{\mu}\equiv g_{\mu\nu}n^{\mu}$.

The ansatz for $\EAAS_k$ contains the `boundary Newton constant' $\NkEC_k$. If it happens to be equal to $\Nk_k$ from the bulk, the Einstein-Hilbert and the Gibbons-Hawking term have the correct relative renormalization for a well-posed variational problem. The ansatz \eqref{eqn:03_03} also includes a boundary cosmological constant $\KkbarEC_k$ which bears no special relationship to its bulk counterpart.

Note that the canonical mass dimensions of $\Nk_k$ and $\NkEC_k$ always agree, $\Mdim{\Nk_k}=\Mdim{\NkEC_k} =2-d$, those of the cosmological constants are different: $\Mdim{\Kkbar_k}=+2$, $\Mdim{\KkbarEC_k}=+1$. Correspondingly we introduce dimensionless couplings according to
\begin{align}
\tg_k\equiv k^{d-2} \Nk_k\, , & & \Kk_k &\equiv \Kkbar_k \slash k^2 \, ,\\
\tg^{\partial}_k\equiv k^{d-2} \NkEC_k\, ,&  &\KkEC_k & \equiv \KkbarEC_k \slash k 
\end{align}
We shall use $\left(\tg,\, \tg^{\partial},\, \Kk,\, \KkEC\right)$ as coordinates on the 4-dimensional theory space spanned by the truncation ansatz.

\subsection{The functional RG equation}

For truncations such as the one at hand in which the ghost term keeps its classical form the general flow equation for the gravitational average action reduces to \cite{mr}
\begin{align}
\partial_t \EAA_k[\flcb;\bg]&=\frac{1}{2}\Tr\left[\left(\kappa^{-2}\EAA_k^{(2)}[\flcb;\bg]+\RegGr_k[\bg]\right)^{-1} \partial_t \RegGr_k[\bg]\right] \nonumber\\
&\quad -\Tr\left[\left(-\FPop[\flcb;\bg]+\RegGh_k[\bg]\right)^{-1} \partial_t \RegGh_k[\bg]\right] \label{eqn:03_05}
\end{align}
Here $\EAA_k[\flcb;\bg]\equiv \EAA_k[\flcb,0,0;\bg]$ is the action functional for vanishing ghosts, $t=\ln k$ denotes the RG time, $\kappa^{-2}\equiv 32\pi\bar{\Nk}$ is a constant, and the matrix elements of the Hessian operator $\EAA^{(2)}_k$ are given by
\begin{align}
_{\mu\nu}\langle x\mid \EAA_k^{(2)}[\flcb;\bg]\mid y\rangle^{\alpha\beta}\equiv \frac{\bg_{\mu\rho}(x)\bg_{\nu\sigma}(x)}{\sqrt{\bg(x)}\sqrt{\bg(y)}}\,  \frac{\var^2\EAA_k[\flcb;\bg]}{\var\flcb_{\rho\sigma}(x)\var\flcb_{\alpha\beta}(y)} \label{eqn:03_06} 
\end{align}
where we use a self-explaining bra-ket notation. The FRGE of \eqref{eqn:03_05} is similar to the Wetterich equation of matter and Yang-Mills fields on flat space \cite{wett-mr}.

\subsection{Variation of the ansatz}

In practice $\EAA_k^{(2)}$ is computed most conveniently by performing two variations $\flcb_{\mu\nu}\mapsto \flcb_{\mu\nu}+\var \flcb_{\mu\nu}$ of the functional $\EAA_k[\flcb;\bg]$ at fixed $\bg_{\mu\nu}$, and `stripping off' the $\var\flcb_{\mu\nu}$'s then. In order to stay within the domain on which $\EAA_k$ is defined we must impose Dirichlet boundary conditions on the variations, too:
\begin{align}
 \left.\var\flcb_{\mu\nu}\right|_{\partial\MaFs}=0 \label{eqn:03_07}
\end{align}

Applying this procedure to $\EAAbulk_k$ we obtain the same result as in \cite{mr} where $\partial\MaFs=\emptyset$ had been assumed. Hereby one has to make essential use of the boundary condition \eqref{eqn:03_07} which eliminates potential surface terms arising from integrations by part. As a typical example, consider $\int_{\MaFs}\md^d x\sqrt{\bg}\, \var\flcb_{\alpha}^{\alpha}\bZ^{\mu}\bZ_{\mu}\var\flcb_{\beta}^{\beta}$ which equals $-\int_{\MaFs}\md^d x\sqrt{\bg}\, \left(\bZ^{\mu}\var\flcb_{\alpha}^{\alpha}\right)\left(\bZ_{\mu}\var\flcb_{\beta}^{\beta}\right)$ up to a surface term 
$\int_{\partial\MaFs}\md^{d-1} x\sqrt{\iM}\, n^{\mu}\var\flcb_{\alpha}^{\alpha}\left(\bZ_{\mu}\var\flcb_{\beta}^{\beta}\right)$. However, by virtue of $\var\flcb_{\mu\nu}=0$ on $\partial\MaFs$, this surface term vanishes.

Potentially dangerous are terms involving normal derivatives $n^{\rho}\partial_{\rho}\var\flcb_{\mu\nu}$ because they are non-zero in general, despite the Dirichlet conditions for $\var\flcb_{\mu\nu}$. However, since all integrals resulting from an application of Gauss' theorem consist of two $\var\flcb$'s but only one remaining derivative, they are bound to contain one undifferentiated factor of $\var\flcb$, vanishing on $\partial\MaFs$, and this causes the entire surface integral to vanish. Note, however, that this argument applies only to actions of at most {\it second order} in the derivatives; for higher order actions the situation will be more complicated \cite{barth}.

For the calculation of $\left(\EAAS_k\right)^{(2)}$ we must find out how the extrinsic curvature $\EC$ responds to a variation $\flcb_{\mu\nu}\mapsto \flcb_{\mu\nu}+\var \flcb_{\mu\nu}$ at fixed $\bg_{\mu\nu}$. It induces a change $\var g_{\mu\nu}=\var\flcb_{\mu\nu}$ of the full metric, whence
$ \left.\var g_{\mu\nu}\right|_{\partial\MaFs}=0$. This entails $\var\iM_{\mu\nu}=0$ since the normal vector field $n^{\mu}$ is unaffected by the variation. As a consequence, at most the Christoffel symbol in \eqref{eqn:03_03_2} can give rise to a non-zero change of $\EC$:
\begin{align}
 \var\EC&= -\iM^{\alpha\beta}\, n_{\rho}\,\var\Gamma_{\alpha\beta}^{\rho}\nonumber\\
&=-\frac{1}{2}\iM^{\alpha\beta}\, n^{\mu}\left[\partial_{\alpha}\var\flcb_{\beta\mu}+\partial_{\beta}\var\flcb_{\alpha\mu}-\partial_{\mu}\var\flcb_{\alpha\beta}\right] \nonumber \\
&=-\iM^{\alpha\beta}\left(\partial_{\beta}\var \flcb_{\alpha\mu}\right)n^{\mu}+\frac{1}{2}\iM^{\alpha\beta}\, n^{\mu}\, \partial_{\mu}\var\flcb_{\alpha\beta} \label{eqn:03_08}
\end{align}
The first term in the last line of \eqref{eqn:03_08} is zero since the projected derivative $\iM^{\alpha\beta}\partial_{\beta}$ acts tangentially to $\partial \MaFs$, with a vanishing result by \eqref{eqn:03_07}. In contrast, the second term, containing a normal derivative, is non-zero in general:
\begin{align}
 \var\EC&= \frac{1}{2}\iM^{\alpha\beta}\,n^{\mu}\partial_{\mu}\,\var\flcb_{\alpha\beta} \label{eqn:03_09}
\end{align}
Comparing \eqref{eqn:03_09} to \eqref{eqn:01_02} we see that the integral of $\var \EC$, with the correct prefactor, can indeed cancel the unwanted boundary terms in the variation of the bulk action, which is the raison d'\^etre of the Gibbons-Hawking term, of course.

However, applying a second variation to \eqref{eqn:03_09} we obtain zero, $\var^2\EC=0$, since $\iM^{\alpha\beta}$ (and $n^{\mu}$ clearly) does not change. With this result at hand it is now easy to see that 
\begin{align}
 \EAA^{(2)}_k\equiv \left(\EAAbulk_k + \EAAS_k\right)^{(2)}= \left(\EAAbulk_k \right)^{(2)}
\end{align}
So the overall conclusion is that, for the ansatz considered, the Hessian $\EAA^{(2)}_k$ receives actually no contributions from any boundary terms, neither from those potentially arising from the bulk action, nor from the boundary functional $\EAAS_k$. As a result, inserting the truncation ansatz into the RG equation \eqref{eqn:03_05} we encounter the same kinetic operator $\EAA_k^{(2)}\propto \left[-K\ud{\alpha\beta}{\mu\nu}\Z^2+U\ud{\alpha\beta}{\mu\nu}\right]$, with the tensors $K\ud{\alpha\beta}{\mu\nu}$ and $U\ud{\alpha\beta}{\mu\nu}$ defined in \cite{mr}, as in the case without boundary.

\subsection{The McKean-Singer heat kernel}

In order to derive the beta-functions of the 4 running coupling constants we must project the infinite dimensional flow on the subspace defined by the ansatz. To perform this projection we may insert special field configurations $\flcb,\,\bg$ in order to make the various field monomials, or linear combinations thereof, non-zero. Here it suffices to set $\flcb_{\mu\nu}(x)=0$  after the differentiations in \eqref{eqn:03_06}. For $\bg_{\mu\nu}$ we take a metric which is maximally symmetric at the interior points of $\MaFs$. We may think of $\MaFs$ as being cut out of a $d$-sphere along some `parallel' slightly `north' of its equator, say. Then both $\SR$ and $\EC$ are strictly positive on all, respectively, interior and boundary points of this manifold.

After inserting this manifold on both sides of the FRGE \eqref{eqn:03_05} the computation of the beta-functions for the bulk couplings proceeds along exactly the same lines as described in \cite{mr}. For the determination of $\partial_t \tg_k^{\partial}$ and $\partial_t \KkEC_k$ we need a generalization of the heat kernel expansion employed there which includes boundary contributions. In the case at hand the first few terms, known already from the pioneering work of McKean and Singer \cite{mckean-singer}, are sufficient:
\begin{align}
\Tr\left[e^{s\Z^2}\right]&=\frac{ \tr(\Id)}{\left(4\pi s\right)^{d\slash 2}} \left\{ \int_{\MaFs}\md^d x\, \sqrt{g} - \frac{1}{4}\sqrt{4\pi s} \int_{\partial\MaFs}\md^{d-1} x\,\sqrt{\iM} \right. \label{eqn:03_10}\\
&\left.\phantom{=\left(\frac{1}{4\pi s}\right)^{d\slash 2} \tr(\Id) \quad}%
+\frac{1}{6}s \left(\int_{\MaFs}\md^d x\, \sqrt{g}\, \SR +2 \int_{\partial\MaFs}\md^{d-1}x\, \sqrt{\iM}\EC\right)+\Order{s^{3\slash 2}}\right\} \nonumber
\end{align}
Terminating the asymptotic series at order $s^{3\slash 2}$ the terms retained match precisely those contained in the truncation ansatz. Recall that in the case with boundary there appear also Seeley-DeWitt coefficients of half-integer order \cite{mckean-singer,vassi-review}. Note also that the $\Order{s}$ term contains the Einstein-Hilbert and Gibbons-Hawking terms, respectively, in precisely the `preferred' combination with a relative coefficient of $+2$.

Employing the same Fourier transform-based method as in \cite{mr} we can use \eqref{eqn:03_10} in order to expand the trace of appropriate functions of the covariant Laplacian:
\begin{align}
\Tr\left[W\left(-\Z^2\right)\right]&=(4\pi)^{-d\slash 2}\tr(\Id) \left\{ Q_{d\slash 2}[W] \int_{\MaFs}\md^d x \, \sqrt{g}\right. \label{eqn:03_11}\\
 &\phantom{=(4\pi)^{-d\slash 2}\tr(\Id)}\quad- \frac{1}{2}\sqrt{\pi}\,Q_{(d-1)\slash 2}[W]\int_{\partial\MaFs} \md^{d-1}x\,\sqrt{\iM} \nonumber\\
& \left.\phantom{=(4\pi)^{-d\slash 2}}
+\frac{1}{6}\,Q_{d\slash 2 -1}[W] \left(\int_{\MaFs}\md^d x\, \sqrt{g}\, \SR+2\int_{\partial\MaFs}\md^{d-1}x \,\sqrt{\iM}\EC\right)+\cdots\right\}\nonumber 
\end{align}
The $Q_n$-functionals are given by
\begin{align}
Q_n[W]=\frac{1}{\Gamma(n)}\int_0^{\infty}\md z \, z^{n-1}\, W(z)\, , & & \text{for } n>0\, , \label{eqn:03_12}
\end{align}
along with $Q_0[W]=W(0)$, and $Q_n[W]=\left(-\partial_z\right)^{(-n)}W(0)$ for $n<0$.\footnote{For half-integer negative $n$ the fractional derivative is defined by an integral representation.}

Using \eqref{eqn:03_11} it is straightforward to evaluate the functional traces on the $\RHS$ of the flow equation \eqref{eqn:03_05} and to read off the scale derivatives of the 4 couplings $\tg_k,\, \Kk_k,\, \tg_k^{\partial}$ and $\KkEC_k$, respectively.

It turns out that the two differential equations for the bulk quantities close among themselves:
\begin{align}
\partial_t \tg_k=\left[d-2+\aDz(\tg_k,\Kk_k)\right]\tg_k\, , && \partial_t \Kk_k=\beta_{\Kk}(\tg_k,\Kk_k) \label{eqn:03_13}
\end{align}
The analogous pair of equations for the boundary couplings contain $\tg$ and $\Kk$ of the bulk, however:
\begin{align}
\partial_t \tg^{\partial}_k=\left[d-2+\aDzEC(\tg_k^{\partial},\tg_k,\Kk_k)\right]\tg_k\, , 
&& \partial_t \KkEC_k=\beta_{\KkEC}(\tg^{\partial}_k,\KkEC_k,\tg_k,\Kk_k) \label{eqn:03_14}
\end{align}
The anomalous dimensions $\aDz\equiv \partial_t \ln \Nk_k$ and $\aDzEC\equiv \partial_t \ln \NkEC_k$ are explicitly given by
\begin{align}
\aDz(\tg,\Kk)&=\frac{g B_1(\Kk)}{1-\tg B_2(\Kk)} \label{eqn:03_15}
\end{align}
with the same functions $B_1(\Kk)$ and $B_2(\Kk)$ as in \cite{mr}, and by
\begin{align}
 \aDzEC(\tg^{\partial},\tg,\Kk)&=\frac{1}{3}(4\pi)^{1-d\slash 2}\,\tg^{\partial}\left[d(d+1)\,\ThrfA{1}{d\slash 2- 1}{-2\Kk}-4d\,\ThrfA{1}{d\slash2-1}{0}\right. \nonumber \\
&\left. \phantom{=\frac{1}{3}(4\pi)^{1-d\slash 2}\tg^{\partial}}\qquad \quad -\frac{1}{2}d(d+1)\,\aDz(\tg,\Kk)\, \ThrfB{1}{d\slash 2-1}{-2\Kk}
\right]\label{eqn:03_16}
\end{align}
Here $\Phi_n^p$ and $\tilde{\Phi}_n^p$ are the standard threshold functions defined in \cite{mr}. They depend on the cutoff shape function $R^{(0)}$; for the optimized one \cite{litimgrav}, for example,
\begin{align}
\ThrfA{p}{n}{w}=\frac{1}{\Gamma(n+1)}\frac{1}{(1+w)^p}\, , && \ThrfB{p}{n}{w}=\frac{1}{(n+1)}\ThrfA{p}{n}{w} \label{eqn:03_17}
\end{align}
Finally, the beta-functions for the cosmological constants read
\begin{align}
 \beta_{\Kk}(\tg,\Kk)&=\left(\aDz-2\right)\Kk + \frac{1}{2}\,\tg\,(4\pi)^{1-d\slash2}\left[2d(d+1)\,\ThrfA{1}{d\slash2}{-2\Kk}\right. \label{eqn:03_18} \\
&\left.\phantom{\left(\aDz-2\right)\Kk + \frac{1}{2}\tg(4\pi)^{1-d\slash2}}\qquad \quad-8d\,\ThrfA{1}{d\slash2}{0}-d(d+1)\,\aDz(\tg,\Kk)\,\ThrfB{1}{d\slash2}{-2\Kk}
\right] \nonumber
\end{align}
for the bulk and, for its boundary analogue,
\begin{align}
 \beta_{\KkEC}(\tg^{\partial},\KkEC,\tg,\Kk)&=\left[\aDzEC(\tg^{\partial},\tg,\Kk)-1\right]\KkEC  \label{eqn:03_19}\\
&\quad\quad  -\frac{1}{8}\,\tg^{\partial}\,(4\pi)^{(3-d)\slash 2}\left[2d(d+1)\,\ThrfA{1}{(d-1)\slash2}{-2\Kk}-8d\,\ThrfA{1}{(d-1)\slash2}{0}\right. \nonumber \\
&\left.\phantom{ +\frac{1}{8}\tg^{\partial}(4\pi)^{(3-d)\slash 2}}\qquad \quad
-d(d+1)\,\aDz(\tg,\Kk)\,\ThrfB{1}{(d-1)\slash2}{-2\Kk}
\right] \nonumber
\end{align}
In the rest of this section we shall analyze this set of RG equations.

\subsection{The semiclassical regime, and beyond}
In order to get a first impression of what the above flow equations tell us we specialize them for the semiclassical regime where they can be solved easily. Here `semiclassical' stands for an approximation in which the RG equations are reduced to their one loop form by neglecting the `improvement' terms proportional to $\aDz$ and $\aDzEC$ on their $\RHS$, and by evaluating all threshold functions at zero cosmological constant.
In this manner we find the approximate anomalous dimensions
\begin{align}
\eta_{\text{N}}= - (d-2)\,\omega_d\, \tg\, , \qquad \eta^{\partial}_{\text{N}}= - (d-2)\,\omega^{\partial}_d\, \tg^{\partial}\, \label{eqn:03_19B}
\end{align}
with numerical constants $\omega_d$ and $\omega_d^{\partial}$, respectively. The dimensionful RG equations have the structure $\partial_t \Nk_k = \eta \Nk_k$ and admit the following simple but exact solutions:
\begin{subequations}\label{eqn:03_19CD}
\begin{align}
 \Nk_k &= \frac{\Nk_0}{1+\omega_d\, \Nk_0\, k^{d-2}} \label{eqn:03_19C} \\
\Nk^{\partial}_k &= \frac{\Nk^{\partial}_0}{1+\omega^{\partial}_d\, \Nk^{\partial}_0\, k^{d-2}} \label{eqn:03_19D}
\end{align}
\end{subequations}
If we additionally expand for small $k$ and solve also for the running cosmological constants in the same regime, we obtain
\begin{subequations}  \label{eqn:03_20}
\begin{align}
& \Nk_k=\Nk_0 \left[1-\omega_d \,\, \Nk_0\,\, k^{d-2}+ \cdots\right]\label{eqn:03_20a}\\
& \Nk^{\partial}_k=\Nk^{\partial}_0 \left[1-\omega^{\partial}_d \,\, \Nk^{\partial}_0\,\, k^{d-2} +\cdots\right] \label{eqn:03_20b}\\
& \Kkbar_k=\Kkbar_0 \,+\,\nu_d \, \,\Nk_0\,\, k^{d} +\cdots  \label{eqn:03_20c}\\
& \KkbarEC_k=\KkbarEC_0 \,+\,\nu^{\partial}_d \, \,\Nk^{\partial}_0\,\, k^{d-1} + \cdots \label{eqn:03_20d}
\end{align}
\end{subequations}
Here $\Nk_0,\, \Nk_0^{\partial},\, \Kkbar_0,$ and $\Kkbar_0^{\partial}$ are free constants of integration, and the dots represent higher orders in $\Nk_0 \, k^{d-2}$ and $\Nk^{\partial}_0 \, k^{d-2}$, respectively. 

The various $d$-dependent coefficients are given by the following expressions:
\begin{subequations} \label{eqn:03_21}
\begin{align}
\omega_d &= \frac{1}{3(d-2)(4\pi)^{d\slash 2-1}}\,\left[\vphantom{\frac{A}{B}}6\big(d(d-1)+4\big)\ThrfA{2}{d\slash2}{0} - d(d-3)\ThrfA{1}{d\slash 2 -1}{0}\right]  \label{eqn:03_21a} \\
\omega^{\partial}_d &= -\frac{d(d-3)}{3(d-2)(4\pi)^{d\slash 2-1}}\, \ThrfA{1}{d\slash 2 -1}{0} \label{eqn:03_21b} \\
\nu_d &= \frac{(d-3)}{(4\pi)^{d\slash 2-1}}\, \ThrfA{1}{d\slash 2}{0} \label{eqn:03_21c} \\
\nu^{\partial}_d &= -\frac{d(d-3)}{(d-1)\,2^{d-1}\,\pi^{(d-3)\slash 2}}\, \ThrfA{1}{(d-1)\slash 2 }{0}  \label{eqn:03_21d} 
\end{align}
\end{subequations}
It is instructive to look at the signs of these coefficients. To this end, let us specialize for $d=4$ dimensions and the `optimized' threshold functions \eqref{eqn:03_17}. This leads to
\begin{subequations} \label{eqn:03_22}
\begin{align}
\omega_4&= + \frac{11}{6\pi} >0 \label{eqn:03_22a}\\
\omega^{\partial}_4&= - \frac{1}{6\pi} <0 \label{eqn:03_22b}
\end{align}
\begin{align}
\nu_4&= + \frac{1}{8\pi} >0 \label{eqn:03_22c}\\
\nu^{\partial}_4&= - \frac{2}{9\pi} <0 \label{eqn:03_22d}
\end{align}
\end{subequations}

We observe that, at least within the approximation \eqref{eqn:03_20}, the bulk Newton constant $\Nk_k$ and the boundary counterpart $\NkEC_k$ run in opposite directions. The coefficient $\omega_4$ is positive, hence $\Nk_k$ {\it decreases} for increasing $k$, and this is precisely the hallmark of gravitational anti-screening \cite{mr}.
On the other hand, $\omega^{\partial}_4 <0$ implies that $\NkEC_k$ {\it increases} when $k$ is increased.  

As a result, the equality $\Nk_k=\NkEC_k$ is consistent with the RG evolution at most at a single scale $k$.
For instance, we might choose the two constants of integration, $\Nk_0$ and $\Nk_0^{\partial}$ to be equal. Then we have $\NkEC_k=\Nk_k$ at $k=0$, but going to a higher scale immediately destroys the equality. In this example, the average action $\EAA_k$ has the desired well posed variational principle at the `physical point' $k=0$.

It can be checked that the opposite running of $\Nk_k$ and $\NkEC_k$, i.e. the different signs of the anomalous dimensions $\aDz<0$ and $\aDzEC>0$, respectively, are robust with respect to changes of the cutoff shape function, and are realized for a wide range of dimensionalities.

Actually it is fairly easy to see that the positivity of $\aDzEC$ extends beyond the semiclassical approximation. Specializing \eqref{eqn:03_16} for $d=4$ and the threshold functions \eqref{eqn:03_17} we obtain
\begin{align}
\aDzEC&= \frac{\tg^{\partial}}{3\pi} \, \,\frac{1-\frac{5}{4}\aDz + 8\Kk}{(1-2\Kk)} \label{eqn:03_23}
\end{align}
This result is exact in the sense that the improvement term $\propto \aDz$ on the $\RHS$ of the FRGE was retained and $\Kk$ has not been set to zero in the arguments of the $\Phi$'s. It is obvious that the expression \eqref{eqn:03_23} is always positive in the regime of interest, $\tg^{\partial}>0$, $\aDz<0$, and $\Kk\in [0,1\slash2)$.

\begin{SCfigure}[50][htbp]
\hspace{1.5cm}
\psfrag{tagk}{$k$}
\psfrag{tagA0}[bc]{$0$}
\psfrag{tagB0}[tl]{$0$}
\psfrag{tagC}{$C_k$}
\psfrag{tagG}{$1 \slash G^{\partial}_k$}
\includegraphics[width=0.3\textwidth]{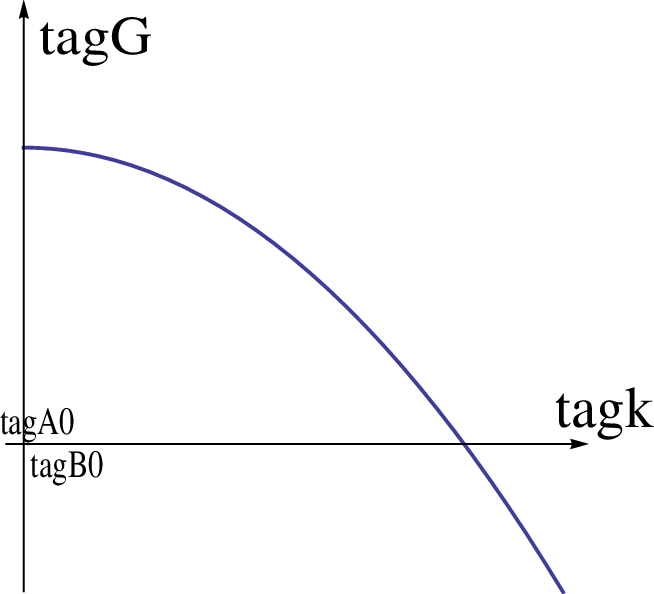} \hspace{1.5cm}
\caption{Schematic plot of the scale dependence of the Gibbons-Hawking coupling $1\slash \Nk_k^{\partial}$ in the single-metric Einstein-Hilbert truncation. The level-(0) coupling $1\slash \Nk_k^{(0,\partial)}$ appearing in the matter induced bi-metric action of Section \ref{sec:04} enjoys a qualitatively identical $k$-dependence.}\label{fig:parabola}
\end{SCfigure}

We may conclude therefore that the simple formula \eqref{eqn:03_19D} provides us with a fairly reliable parametrization of the running Gibbons-Hawking coupling%
\footnote{The actual coordinate on theory space is taken to be the {\it inverse} (dimensionless) Newton constant, $1\slash \tg_k^{\partial}$, the true prefactor of the Gibbons-Hawking term. It is well behaved at the zero displayed by \eqref{eqn:03_23B} where $\tg_k^{\partial}$ itself has an artificial and inconsequential pole.}
 under very general conditions:
\begin{align}
 \frac{1}{\Nk_k^{\partial}}&=\frac{1}{\Nk_0^{\partial}}+\omega_4^{\partial} k^2 \equiv \frac{1}{\Nk_0^{\partial}}- |\omega_4^{\partial}| k^2 \label{eqn:03_23B}
\end{align}
The function $1 \slash \Nk_k^{\partial}$ is depicted in Fig. \ref{fig:parabola}. 
If, for instance, $\Nk_0^{\partial}=\Nk_0$, it starts out positive, decreases for increasing $k$, has a zero near the Planck scale $k=m_{\text{Pl}}\equiv \Nk_0^{-1\slash2}$, and turns negative then. (Recall that the formulae \eqref{eqn:03_19CD} do not assume $\Nk_0 k^{d-2}$ to be small.)

The well known {\it decrease} of the bulk Newton constant in the UV has been interpreted as an indication for the {\it anti-screening} character of QEG at short distances \cite{mr}. Now we find that, in the same theory and within essentially the same truncation, the corresponding boundary Newton constant {\it increases} in the UV. One might be worried therefore about whether the conjecture of a gravitational anti-screening is really correct. 
In order to get a picture as clear as possible it is helpful to re-consider this issue in an explicit bi-metric context; this will be the topic of the next section.

Finally we mention that the bulk and boundary cosmological constants, too, display an opposite RG running. While $\Kkbar_k$ increases with increasing $k$, its counterpart $\KkbarEC_k$ decreases.

\subsection{The fixed points}
Next we search for fixed points of the flow equations \eqref{eqn:03_13} - \eqref{eqn:03_19}, i.e. for common zeros of all four beta-functions:
\begin{subequations} \label{eqn:03_30}
\begin{align}
 \beta_{\tg}\left(\tg_*,\Kk_*\right)&=0=\beta_{\Kk}\left(\tg_*,\Kk_*\right)\label{eqn:03_30a} \\
\beta_{\tg^{\partial}}\left(\tg_*,\Kk_*,\tg^{\partial}_*\right)&=0=\beta_{\KkEC}\left(\tg_*,\Kk_*,\tg^{\partial}_*,\Kk^{\partial}_*\right)\label{eqn:03_30b}
\end{align}
\end{subequations}
As the system decouples partially we can first solve the $\tg$ - $\Kk$ subsystem \eqref{eqn:03_30a} separately, insert its solution $\left(\tg_*,\Kk_*\right)$ into the equations \eqref{eqn:03_30b}, and solve them for the remaining fixed point coordinates $\tg^{\partial}_*$ and $\KkEC_*$. 
We shall restrict ourselves to $d=4$ and the `optimized' cutoff here.

The $\tg$-$\Kk$ subsystem possesses the well-known Gaussian and non-Gaussian fixed points, respectively:
\begin{subequations} \label{eqn:03_31}
\begin{align} 
 \text{GFP}_{\tg\text{-}\Kk} &: \qquad\tg_*=0, && \Kk_*=0 && \label{eqn:03_31a} \\
\text{NGFP}_{\tg\text{-}\Kk} &:\qquad\tg_*=\tg^{\text{NG}}_*>0, && \Kk_*=\Kk^{\text{NG}}_*>0 &&\label{eqn:03_31b}
\end{align}
\end{subequations}
Numerically we have
\begin{align}
 \tg^{\text{NG}}_* \approx 0.707 \, , && \Kk^{\text{NG}}_*\approx 0.193 \label{eqn:03_32}
\end{align}

Each one of the two fixed points in the subsystem, when inserted into \eqref{eqn:03_30b}, gives rise to 2 fixed points of the full system. In the case of the $\text{GFP}_{\tg\text{-}\Kk}$ ($\text{NGFP}_{\tg\text{-}\Kk}$) they are denoted G-G-FP and G-NG-FP (NG-G-FP and NG-NG-FP), respectively. In this notation the labels `Gaussian' (G) and `non-Gaussian' (NG) refer to the two options for satisfying $\beta_{\tg}=0$, namely by setting either $\tg=0$ or $\aDz=-2$, and analogously for $\beta_{\tg^{\partial}}=0$.

It turns out that the second set of equations, \eqref{eqn:03_30b}, can easily be solved analytically in terms of $\tg^{\text{NG}}$, $\Kk^{\text{NG}}$.
In Table \ref{tab:03_01} we summarize the resulting coordinates of all 4 fixed points.
\begin{table}[h]
\begin{center}
\begin{tabular}{|c||c|c|c|c|}\hline
 & $\tg_*$ & $\Kk_*$ & $\tg_*^{\partial}$ & $\KkEC_*$\\
\hline \hline
G-G-FP & $0$ & $0$ & $0$ & $0$\\
G-NG-FP & $0$ & $0$ & $-6\pi$ & $+\frac{4}{3}$\\
NG-G-FP & $\tg_*^{\text{NG}}$ & $\Kk_*^{\text{NG}}$ & $0$ & $0$\\
NG-NG-FP & $\tg_*^{\text{NG}}$ & $\Kk_*^{\text{NG}}$ & $-12\pi\,\frac{1-2\Kk_*^{\text{NG}}}{7+16\Kk_*^{\text{NG}}}$ &$+\frac{4}{3}\,\frac{6+16\Kk_*^{\text{NG}}}{7+16\Kk_*^{\text{NG}}}$\\
\hline
\end{tabular}
\caption{The coordinates of the four fixed points}
\label{tab:03_01}
\end{center}
\end{table}

The $\text{NGFP}_{\tg\text{-}\Kk}$ fixed point in the subsystem is precisely the one which is usually considered a candidate for the Asymptotic Safety construction on spacetime manifolds without boundary \cite{mr,oliver, frank1}. Here we find that in presence of boundaries it gets lifted to 2 fixed points of the full system, the NG-G-FP with $\tg^{\partial}_*=\KkEC_*=0$, and the NG-NG-FP with $\tg^{\partial}_*=-2.292<0$, $\KkEC_*=1.201>0$.

Concerning the possibility of choosing $\Nk_k$ and $\NkEC_k$ equal, we observe that the relation $\Nk_k=\NkEC_k$, or $\tg_k=\tg_k^{\partial}$, is inconsistent with the RG evolution in the asymptotic scaling regime of both the NG-G-FP and the NG-NG-FP.

Linearizing the RG flow about the latter two fixed points we find the following set of critical exponents:
\begin{align}
&\text{NG-G-FP:} && \Theta_{1,2} = 1.475 \pm 3.043 \Ii\, , & & \Theta_3= -2 \, , & &\Theta_4 =1 \label{eqn:03_33}  \\
&\text{NG-NG-FP:} & & \Theta_{1,2} = 1.475 \pm 3.043 \Ii\, , & & \Theta_3= 3 \, , & &\Theta_4 =2 \label{eqn:03_34}
\end{align}
At the NG-NG-FP all four scaling fields are relevant, i.e. the corresponding values of $\Re \Theta$ are all positive. In the case of the NG-G-FP one scaling field is irrelevant. As a consequence, the dimensionalities of the respective UV critical hypersurfaces are $\dim \cUV(\text{NG-G-FP})=3$ and $\dim \cUV(\text{NG-NG-FP})=4$. 

  \section{Bi-metric average action induced by \\ matter field fluctuations} \label{sec:04}

In this section we discuss our second explicit calculation, a truncation of the bi-metric type. Since analyses of this kind are rather involved technically we are not going to consider the fluctuations of full Quantum Einstein Gravity here, but only those of a multiplet of scalar fields $A_j$, $j=1,\cdots, \ns$, coupled to gravity. Invoking the limit of large $\ns$, we retain only the scalar contributions to the gravitational beta-functions, discarding the more complicated ones stemming from quantum fluctuations of the gravitational field itself.

\subsection{A bi-metric truncation with boundary terms} \label{sec:04_01}

We start by fixing a spacetime manifold $\MaFs$ with $\partial \MaFs\neq \emptyset$, a space of non-dynamical metrics $\bg_{\mu\nu}$ compatible with the topology of $\MaFs$, and corresponding spaces from which the dynamical fields $\flcb_{\mu\nu}$ and $A$ are to be taken. The latter are required to satisfy Dirichlet boundary conditions:
\begin{align}
 \left.\flcb_{\mu\nu}\right|_{\partial\MaFs}=0\,, && \left. A \right|_{\partial \MaFs}=\AS \label{eqn:04_00a}
\end{align}
where $\AS$ is an arbitrary but fixed multiplet of scalar fields defined on the submanifold $\partial \MaFs$. Variations of $A$ thus vanish on the boundary, i.e. $\left. \var A\right|_{\partial\MaFs}=0$.

The truncation ansatz comprises the first two terms of the $\flcb_{\mu\nu}$-expansion. We write
\begin{align}
 \EAA_k[\flcb,A;\bg]&=\EAA^{\background}_k[A;\bg]+\EAA^{\lin}_k[\flcb,A;\bg] \label{eqn:04_01}
\end{align}
whereby the `background' and `linear' pieces $\EAA^{\background}_k$ and $\EAA^{\lin}_k$, respectively, are of zeroth and first order in $\flcb_{\mu\nu}$. In the purely background dependent functional $\EAA^{\background}_k$ we include the Einstein-Hilbert and the Gibbons-Hawking term with $k$-dependent coefficients, as in the previous section, along with a globally $O(\ns)$-invariant matter field action:
\begin{align}
 \Gamma_k^{\background}[A;\bg]
&= - \frac{1}{16\pi \nkA} \int_{\MaF}\md^d x \sqrt{\bg}\, \left(\SRb - 2\kkA\right) \nonumber\\
&\quad - \frac{1}{16\pi \nkAB} \int_{\partial\MaF}\md^{d-1} x \sqrt{\biM}\, \left(2\bEC - 2\kkAB\right) \nonumber \\
& \quad +\int_{\MaFs} \md^d x \sqrt{\bg} \left\{\frac{1}{2}\,\bg^{\mu\nu} \partial_{\mu}A\partial_{\nu} A +\frac{1}{2} \xiA \SRb A^2 + \PotAkA\right\}\nonumber \\
&\quad + \xiAS\int_{\partial\MaFs}\md^{d-1} x \sqrt{\biM}\,  \bEC A^2 \label{eqn:04_02}
\end{align}
In the two last lines of \eqref{eqn:04_02} appropriate sums over the `flavor index' $j=1,\cdots,\ns$ are understood. The scalars feel the background metric via a standard kinetic term, non-minimal couplings $\SRb \, A^2$ and $\bEC\,A^2$ with running coefficient $\xiA$ and $\xiAS$, respectively, and a $O(\ns)$-invariant, but otherwise arbitrary $k$-dependent potential $\PotAkA$.

For the terms linear in $\flcb_{\mu\nu}$ we make the following ansatz:
\begin{align}
\Gamma^{\lin}_k[\flcb,A;\bg]&=
 \frac{1}{16 \pi \nkB}\int_{\MaFs}\md^dx\sqrt{\bg}\, \, \calE_k^{\mu\nu}[\bg,A]\, \,\flcb_{\mu\nu} \label{eqn:04_03}
 \\
&\quad  +\int_{\partial \MaFs}\md^{d-1}x\sqrt{\biM}\, \left\{ \frac{1}{16\pi } \left(\frac{1}{\nkB}-\frac{1}{\nkBB}\right)- \frac{1}{2}\left(\xiBR-\xiBS\right)A^2    \right\} n^{\lambda}\partial_{\lambda}\flcb^{\mu}_{\phantom{\mu}\mu} \nonumber
\end{align}
In the bulk term of \eqref{eqn:04_03} we employ the convenient abbreviation
\begin{align}
 \calE_k^{\mu\nu}[\bg,A]&\equiv{\bar{G}}^{\mu\nu}-\frac{1}{2}\, \eka \,\bg^{\mu\nu} \SRb + \kkB\,\bg^{\mu\nu} - 8\pi \nkB{\cal T}_k^{\mu\nu}[A;\bg]\label{eqn:04_04}
\end{align}
Hereby $\bar{G}^{\mu\nu}=\Ricb^{\mu\nu}-\frac{1}{2}\bg^{\mu\nu}\SRb$  is the usual Einstein tensor of the background metric, and ${\cal T}_k^{\mu\nu}$ is an energy momentum tensor defined as 
\begin{align}
{\cal T}_k^{\mu\nu}[A;\bg] &\equiv%
 ( \partial^{\mu}A)(\partial^{\nu}A) - \frac{1}{2}\, \bg^{\mu\nu}\bg^{\rho\sigma}(\partial_{\rho}A)( \partial_{\sigma}A )- \frac{1}{2}\, \bg^{\mu\nu}\xiBg\,\SRb A^2 -  \bg^{\mu\nu}\PotAkB \label{eqn:04_05}  \\
&\quad  +\xiBR \left\{\bg^{\mu\nu}\bZ^2(A^2)-\bZ^{\mu}\bZ^{\nu}(A^2) +\Ricb^{\mu\nu}A^2   \right\} \nonumber
\end{align}
Obviously, if it were not for the surface terms\footnote{This difficulty is well known from perturbation theory, see for instance \cite{barr-solo}.}, stationarity under a variation $\flcb_{\mu\nu} \mapsto \flcb_{\mu\nu}+\var \flcb_{\mu\nu}$ would imply the effective Einstein equation $ \calE_k^{\mu\nu}[\bg,A]=0$.

The bulk terms of  $\EAA^{\background}_k+\EAA^{\lin}_k$ are exactly those considered in ref. \cite{MRS1} for the special case where $\partial\MaFs =\emptyset$ and $\xiA=\xiBR=\xiBg=0$. Among them there are the purely gravitational terms without matter fields. In particular there exist two $\flcb_{\mu\nu}$-independent field monomials with zero and two derivatives, respectively, namely $\int \sqrt{\bg}$ and $\int \sqrt{\bg}\, \SRb$, and their prefactors define the running quantities $\kkA \slash \nkA$ and $1 \slash \nkA$. In the linear sector there are three possible tensor structures with at most two derivatives, namely $\flcb_{\mu\nu}$ contracted with ${\bar{G}}^{\mu\nu}$, $\bg^{\mu\nu}\SRb$, and $\bg^{\mu\nu}$, respectively. They give rise to the first three terms on the $\RHS$ of \eqref{eqn:04_04}. Their scale dependence defines the `level-$(1)$' running couplings $1\slash \nkB$, $\eka \slash \nkB$, and $\kkB \slash \nkB$, respectively. (Here the superscripts $(0),(1),\dots$ indicate the {\it level}, i.e. the $\flcb_{\mu\nu}$-order at which the couplings occur.) The level-$(0)$ matter field sector is a $\bg$-dependent, but otherwise standard scalar action along with a non-minimal coupling contribution on the surface  in the last two  lines of \eqref{eqn:04_02}, and at level-(1) we include the set of terms contained in $\flcb_{\mu\nu} {\cal T}_k^{\mu\nu}[A;\bg] $ and the $\xi$ dependent surface term of \eqref{eqn:04_03}. The motivation for the form \eqref{eqn:04_05} of the tensor $ {\cal T}_k^{\mu\nu}$ is that it gives rise to precisely those field monomials which are known to occur {\it when the split symmetry is intact.}

In fact, let us assume for a moment that the running couplings satisfy the following set of relations (in general they do not):
\begin{subequations}
\begin{align}
 &\nkA=\nkB\,, & &\nkAB = \nkBB \label{eqn:04_05Ba} \\
&\xiA=\xiBg=\xiBR\,, & &\xiAS=\xiBS \label{eqn:04_05Bb} \\
&\eka=0\,,  &&\PotAkA=\PotAkB\label{eqn:04_05Bc} 
\end{align}\label{eqn:04_05B}
\end{subequations}%
\noindent If \eqref{eqn:04_05B} holds then $\EAA^{\background}_k+\EAA^{\lin}_k$ has no extra background dependence. That is, to first order in $\flcb_{\mu\nu}$ it can be rewritten as a functional of the sum $\bg_{\mu\nu}+\flcb_{\mu\nu}$ only:
\begin{align}
 \EAA^{\background}_k[A;\bg]+\EAA^{\lin}_k[\flcb,A;\bg]=\EAA^{\background}_k[A;\bg+\flcb] + \Order{\flcb^{2}} \label{eqn:04_05C}
\end{align}
Stated differently, in this case, $\EAA_k^{\lin}$ happens to be the first variation of $\EAA_k^{\background}[A;\bg]$ with respect to $\bg$, and this explains the similarity of ${\cal T}_k^{\mu\nu}$ with a conventional energy momentum tensor.

This argument should merely be regarded a motivation for a sensible truncation ansatz which is as simple as possible in the sense that it covers the split-symmetric limiting case at least. In general, the split symmetry is violated, and we would have to consider many more running couplings in order to be `complete' in an appropriate sense (retaining all matter terms with two derivatives and one power of $\flcb$, for instance). In the present paper we are mostly interested in conceptual rather than precision issues and so we shall not refine the truncation in this way, in particular as we are not interested here in the running of the matter sector itself, but rather the gravitational action it induces.

As for the boundary terms included in the truncation ansatz, the purely gravitational part consists of the standard Gibbons-Hawking term at level-$(0)$, i.e. in the background functional, and the term $\propto \int_{\partial\MaFs}\md^{d-1}x \sqrt{\biM}\, \bg^{\mu\nu} n^{\alpha} \partial_{\alpha} \flcb_{\mu\nu}$ at the linear level. The inclusion of precisely this term, too, is motivated by the fact that in the split-symmetric case it combines with the Gibbons-Hawking term to a functional of $\bg+\flcb$ alone. Indeed, when the relations \eqref{eqn:04_05B} hold true we have the Taylor expansion
\begin{align}
& \left.-\frac{1}{16\pi}\left[\frac{1}{\nkA}\int_{\MaFs}\md^d x \sqrt{g} \, \SR(g)+ \frac{1}{\nkAB} \int_{\partial\MaFs}\md^{d-1}x \sqrt{\iM} \cdot 2 \EC\right]\right|_{g=\bg+\flcb} \nonumber \\
&\qquad\qquad =
-\frac{1}{16\pi}\left[\frac{1}{\nkA}\int_{\MaFs}\md^d x \sqrt{\bg} \, \SRb+ \frac{1}{\nkAB} \int_{\partial\MaFs}\md^{d-1}x \sqrt{\biM} \cdot 2 \bEC\right] \nonumber \\
& \qquad\qquad \quad
+\frac{1}{16\pi}\left(\frac{1}{\nkA}-\frac{1}{\nkAB}\right)  \int_{\partial\MaFs}\md^{d-1}x \sqrt{\biM} \bg^{\mu\nu} n^{\alpha} \partial_{\alpha} \flcb_{\mu\nu}
+\Order{\flcb^2} \label{eqn:04_05D}
\end{align}
Thanks to \eqref{eqn:04_05D}, eq. \eqref{eqn:04_05C} is satisfied also in the boundary sector, and this is in fact what motivates the specific boundary term we included into the level-$(1)$ truncation ansatz, eq. \eqref{eqn:04_03}.

Likewise, we arranged the couplings multiplying the monomials of the matter fields on the boundary in such a way that for $\xiAS=\xiBS$ the split-symmetric case is recovered. The additional term proportional to $\xiBR$ originates from an integration by parts which is necessary to convert the level-$(1)$ scalar field action to its above form involving the energy momentum tensor of eq. \eqref{eqn:04_05}. 

At this point it is important to stress that the issue of split symmetry being intact or violated has a priori nothing to do with problems of boundary terms  possibly being maladjusted to the corresponding bulk terms. In a more general truncation where we retain arbitrary orders in $\flcb$, split symmetry implies that the (boundary) Newton constants, (boundary) cosmological constants, $\cdots$ at the various levels are all equal:
\begin{align}
& \Nk_k^{(0)}=\Nk_k^{(1)}=\Nk_k^{(2)}=\Nk_k^{(3)}=\cdots \, , \nonumber \\
&\Nk_k^{(0,\partial)}=\Nk_k^{(1,\partial)}=\Nk_k^{(2,\partial)}=\Nk_k^{(3,\partial)}=\cdots \, . \label{eqn:04_05E}
\end{align}
Here $\Nk^{(p)}$ and $\Nk^{(p,\partial)}$ are defined via the running prefactors of the $p^{\text{th}}$ terms in the $\flcb_{\mu\nu}$- expansion of $\sqrt{g}\SR$ and $\sqrt{\iM}\EC$, respectively.

Note that \eqref{eqn:04_05E} does not imply any relation among surface and bulk couplings, such as $\nkA=\nkAB$, for instance. (This would result in a `good' $\var \bg_{\mu\nu}$-variational principle for $\EAA_k^{\background}$, which is actually {\it not} what one is aiming at.) Moreover, it is perfectly possible that the level-$(1)$ surface term in the truncation ansatz is non-zero even with exact split symmetry since there is no general reason why $\nkB=\nkBB$ should hold. 

Analogous considerations apply to the $\xi$-parameters. Split symmetry requires $\xi_k^{(0)}=\xi_k^{(1)}=\cdots$ and $ \xi_k^{(0,\partial)}=\xi_k^{(1,\partial)}=\cdots$ but has no implications for the relative magnitude of $\xi_k^{(p)}$ in the bulk and $\xi_k^{(p,\partial)}$ on the boundary.

\subsection{The RG equations}

In the following, for the sake of simplicity, we restrict ourselves to a scalar potential containing only a quadratic and a quartic term at both the background level, 
$\PotAkA=\frac{1}{2} \bmk^{(0)\, 2} A^2 + \frac{1}{24}\bomg^{(0)} A^4$, and the linear level:
$\PotAkB= \frac{1}{2}\bmk^{(1)\, 2} A^2 + \frac{1}{24}\bomg^{(1)} A^4$. Our ansatz for $\EAA_k$ contains a total of 17 running couplings then.

The truncation ansatz is now substituted into the FRGE and its $\RHS$ is then projected onto the truncated theory space by means of a Taylor expansion in $\flcb_{\mu\nu}$, $\SRb$, and $A^2$, respectively. Hereby we retain only zeroth and first orders in these quantities, except for the expansion in $A^2$ where second order contributions are also relevant. We then evaluate the resulting traces over functions of $\bZ^2$ using heat kernel techniques. Whereas in the level-$(0)$ sector the heat kernel expansion in eq. \eqref{eqn:03_10} is sufficient to project out all relevant field monomials, in the level-$(1)$ sector the fluctuation field $\flcb_{\mu\nu}$ makes its appearance under the traces, and we have to use the following generalization \cite{vassi-review}:
\begin{align}
\Tr\left[f\,e^{s\bZ^2}\right]&=\frac{\tr(\Id)}{\left(4\pi s\right)^{d\slash 2}}  \left\{ \int_{\MaFs}\md^d x\, \sqrt{\bg}\, f - \frac{1}{4}\sqrt{4\pi s} \int_{\partial\MaFs}\md^{d-1} x\,\sqrt{\biM}\, f \right.\nonumber\\
&\phantom{=\frac{\tr(\Id)}{\left(4\pi s\right)^{d\slash 2}} \quad}%
+\frac{1}{6}s \left(\int_{\MaFs}\md^d x\, \sqrt{\bg}\, \SRb\, f +2 \int_{\partial\MaFs}\md^{d-1}x\, \sqrt{\biM}\left[\bEC\,f + \frac{3}{2} n^{\lambda}\bZ_{\lambda} f\right]\right)\nonumber\\
&\left.\phantom{=\frac{\tr(\Id)}{\left(4\pi s\right)^{d\slash 2}} \quad}%
\qquad\qquad
+\Order{s^{3\slash 2}}\right\}  \label{eqn:04_05X}
\end{align} 
Especially interesting is the last surface contribution  in \eqref{eqn:04_05X} that will give rise to a non-trivial running of $\Nk_k^{(1,\partial)}$. 

The operator of the functions under the traces are expressed by their Mellin transforms $Q_n$ which multiply the corresponding field monomials. When applying the heat kernel expansion we neglect all contributions that yield invariants outside the truncated theory space.
Finally we equate the coefficients of equal basis monomials on both sides of the FRGE and read off the beta-functions for the various dimensionful couplings. We refer to Appendix \ref{sec:B} of this paper, and to Appendix A of ref. \cite{MRS1} for the details of this calculation.

Instead of using the so extracted RG equations for the dimensionful couplings, it is more convenient to employ dimensionless couplings. All Newton type couplings, bulk and boundary, of level-$(0)$ and level-$(1)$,  are converted according to $\tg_k \equiv \Nk_k\, k^{d-2}$.

For the dimensionless cosmological constants we must distinguish  the bulk couplings $\Kk^{(0)}_k\equiv \Kkbar^{(0)}_k \slash k^2$,  $\Kk^{(1)}_k\equiv \Kkbar^{(1)}_k \slash k^2$, and the boundary one: $\Kk^{(0,\partial)}_k\equiv \Kkbar^{(0,\partial)}_k \slash k$. 

Clearly, masses scale with one power of $k$, and so the dimensionless couplings are given by $\mk^{(0)}\equiv \bmk^{(0)} \slash k$ and $\mk^{(1)}\equiv \bmk^{(1)} \slash k$ for level-$(0)$ and level-$(1)$, respectively.

Except for $ \omgA^{(0)} \equiv\bomg^{(0)}  k^{d-4}$ and  $ \omgA^{(1)} \equiv\bomg^{(1)}  k^{d-4}$ all remaining couplings of the matter sector are dimensionless already.

In the sequel of this section we present all 17  beta-functions for the dimensionless couplings. In the next section we shall then discuss those of their properties which are important to analyze various conceptual questions concerning the gravitational sector. In the forthcoming companion paper \cite{daniel-prep} we shall describe a detailed numerical analysis of the RG flow defined by these beta-functions.

\paragraph{(A) Matter couplings at level-(0).} In our purely matter induced approximation, the potentials $\PotAkA$ and $\PotAkB$ are important sources for the scale dependence of the various couplings. The mass and four-vertex coefficients of $\PotAkA$ obey a closed subsystem given by
\begin{subequations}\label{eqn:04_06}
\begin{align}
 \partial_t \left(\mkB\right) &=-2\,\mkB - (4\pi)^{-\frac{d}{2}}\ns   \, \omgA^{(0)}\,\,
   \ThrfA{2}{d\slash 2}{\mkB}	\label{eqn:04_06a} \\
\partial_t \omgA^{(0)} &= \left(d-4\right) \omgA^{(0)}+ 6\,(4\pi)^{-\frac{d}{2}}\ns \,   
\, \ThrfA{3}{d\slash 2}{\mkB}\, \omgA^{(0)\, 2}	 \label{eqn:04_06b}
\end{align}
\end{subequations}
The running of $\mkB$ and $\omgA^{(0)}$ influences  the flow of all remaining couplings. 

\paragraph{(B) Matter couplings at level-(1).} In the level-$(1)$ sector, the scalar potential $\PotAkB$ contains the couplings $\mkC$ and $\omgA^{(1)}$. Once a solution to the equations \eqref{eqn:04_06} is fixed
we can solve for  the scale dependence of $\mkC$ and $\omgA^{(1)}$. They are found to satisfy the following system of differential equations:
\begin{subequations}\label{eqn:04_06-B}
\begin{align}
&\partial_t  \left(\mkC\right) =-2\mkC+(4\pi)^{-\frac{d}{2}}\ns\,  \,\left\{\,\omgA^{(0)} \big[\left(d-2\right)\,
 \ThrfA{3}{d \slash 2 +1}{\mkB}
+2\, \mkC\, \ThrfA{3}{d\slash 2}{\mkB}\big]
\right. \nonumber \\
&\phantom{\partial_t  \left(\mkC\right) =+(4\pi)^{-\frac{d}{2}}\ns\, k^{d-2}  \qquad}\left. \qquad
-  \,\omgA^{(1)} \, \ThrfA{2}{d\slash 2}{\mkB} \right\} 
\label{eqn:04_06-Ba} \\
&\partial_t \omgA^{(1)}=(d-4)\omgA^{(1)}+(4\pi)^{-\frac{d}{2}}\ns\,  \,\left\{\, 9\,\omgA^{(0)\, 2} \big[\left(2-d\right)\,
 \ThrfA{4}{d \slash 2 +1}{\mkB}-2\,\mkC\,  \ThrfA{4}{d\slash 2}{\mkB} \big] \right.
\nonumber \\
&\phantom{\partial_t \omgA^{(1)}=+(4\pi)^{-\frac{d}{2}}\ns\, k^{d-4} \qquad} \left. \qquad \quad
+ 12\,\omgA^{(0)}\, \omgA^{(1)}\,\, \ThrfA{3}{d\slash 2}{\mkB}\, \right\} 
\, \label{eqn:04_06-Bb}
\end{align}
\end{subequations}
As one can see, all non-canonical scale dependence of $\omgA^{(1)}$ disappears when $\omgA^{(0)}$ vanishes.

\paragraph{(C) The \boldmath{$\xi$}-parameters.} 
In total we have five non-minimal coupling parameters, namely $\xiA$ and $\xiAS$ at level-(0), as well as $\xiBg$, $\xiBR$, and $\xiBS$ at level-(1). Three of them, those pertaining to the bulk, enter the trace on the $\RHS$ of the flow equation and thus potentially affect the evolution of the other couplings, in particular of the Newton constants.
The RG equations for the $\xi$-parameters themselves, in  the background sector, are given by
\begin{subequations} \label{eqn:04_07A}
\begin{align}
 \partial_t  \xiA&=-\frac{1}{6}\,  (4\pi)^{-\frac{d}{2}}\ns\,  \omgA^{(0)} \left\{ \ThrfA{2}{d \slash 2 - 1}{\mkB}-12\,\xiA \ThrfA{3}{d \slash 2}{\mkB}  \right\} \label{eqn:04_07a} \\
\partial_t \xiAS  &=  -\frac{1}{6}\,(4\pi)^{-\frac{d}{2}}\ns\,\omgA^{(0)} \,\, \, \ThrfA{2}{d \slash 2 - 1}{\mkB} \label{eqn:04_07aS}
\end{align}
\end{subequations}%
Furthermore, the non-minimal couplings of the linear level are determined by the following set of differential equations which to some extent resembles the structure of (\ref{eqn:04_07A}a,b):\addtocounter{equation}{-1}
\begin{subequations}\label{eqn:04_07} \stepcounter{equation} \stepcounter{equation}
\begin{align} 
\partial_t \xiBg  &=+ (4\pi)^{-\frac{d}{2}}\ns\,  \left\{\,2\,\left[\xiA\,\omgA^{(1)}+\left(\xiBg+\tfrac{d-4}{12}\right)\, \omgA^{(0)}\right] \ThrfA{3}{d\slash 2}{\mkB} \right. \label{eqn:04_07b} \\
&\phantom{=+ (4\pi)^{-\frac{d}{2}}\ns\, k^{d-4} \qquad}\left.	
- 3\left(d-2\right)\,\xiA\, \omgA^{(0)}\, \ThrfA{4}{(d\slash2)+1}{\mkB} \right. \nonumber \\
&\phantom{=+ (4\pi)^{-\frac{d}{2}}\ns\, k^{d-4} \qquad}\left.
 - 6\,\xiA \,\mkC\,\omgA^{(0)} \,\,\,\ThrfA{4}{d\slash 2}{\mkB}\right. \nonumber \\
&\phantom{=+ (4\pi)^{-\frac{d}{2}}\ns\, k^{d-4} \qquad}\left.
+  \frac{1}{3}\mkC\, \omgA^{(0)}\,\, \ThrfA{3}{d\slash 2 -1}{\mkB}
 - \frac{1  }{6}\,\omgA^{(1)} \,\,\ThrfA{2}{d\slash 2 -1}{\mkB}
\right\} \nonumber
\end{align}
\begin{align}
 \partial_t \xiBR &=+2 \,(4\pi)^{-\frac{d}{2}}\ns\,  \,\left(\xiBR - \frac{1}{6}\right)\omgA^{(0)}  \,\, \ThrfA{3}{d\slash2}{\mkB},\label{eqn:04_07c} \\
\partial_t \xiBS&= -\frac{1}{4}\,(4\pi)^{-\frac{d}{2}}\ns\, \left\{ 
\, \left(\tfrac{10-3d}{3}\right)\, \omgA^{(0)}\,   \ThrfA{3}{d\slash2}{\mkB}
\right.\label{eqn:04_07d} 
  \\
&\left. \phantom{+\frac{1}{4}\, (4\pi)^{-\frac{d}{2}} \ns\, \qquad } \quad
 - 2\, \mkC\,\omgA^{(0)} \, \ThrfA{3}{d\slash2-1}{\mkB}+\omgA^{(1)} \,\ThrfA{2}{d\slash2-1}{\mkB } \,
\right\} \nonumber
\end{align}
\end{subequations}

The couplings we consider next enjoy a special status: none of them appears in the Hessian on the RHS of the FRGE. As a result, we could simplify the appearance of their beta-functions by forming appropriate combinations of them such that the coefficient of each field monomial consists of one such combination only. In order to keep the interpretation of the various couplings simple, we shall not do this here, however.

\paragraph{(D) The Newton type couplings.} The truncation contains four Newton type couplings $\Nk_k^a$ with $a=(0),(0,\partial),(1),(1,\partial)$. The general form of their RG equation is $\partial_t \tg^a_k = \left(d-2+\aDzG^a\right)\tg^a_k$ which 
contains the anomalous dimension
$\aDzG^a = \partial_t \ln \Nk^a_k$. The anomalous dimensions describe the non-canonical contributions to the RG running of the Newton couplings. They read as follows:
\begin{subequations} \label{eqn:04_09}
\begin{align}
 \aDzG^{(0)}\left(\tg^{(0)},\mka,\xia\right) &= \frac{2}{3}
(4\pi)^{-\frac{d}{2}+1}\ns\,\tg^{(0)}_k\, \ThrfA{1}{d \slash 2 - 1}{\mkB} \cdot \label{eqn:04_09a}\\
&\phantom{
(\pi)^{-\frac{d}{2}+1}\ns \ThrfA{1}{d \slash 2 - 1}{\mkB}}\cdot 
\left\{1- 6\,\xiA \frac{\ThrfA{2}{d \slash 2}{\mkB}}{\ThrfA{1}{d \slash 2 - 1}{\mkB}}  \right\}	\nonumber \\
 \aDzG^{(0,\partial)}\left(\tg^{(0,\partial)},\mka\right)&  = \frac{2}{3}(4\pi)^{-\frac{d}{2}+1}\ns\,\tg^{(0,\partial)}
\,
 \ThrfA{1}{d \slash 2 - 1}{\mkB } \label{eqn:04_09b}\\
  \aDzG^{(1)}\left(\tg^{(1)},\mka,\xibR\right)&= \frac{2}{3} (4\pi)^{-\frac{d}{2}+1}\ns\, \tg_k^{(1)}  \,\left(1-6\,\xiBR  \right)\ThrfA{2}{d\slash2}{\mkB}  \label{eqn:04_09c} \\
 \aDzG^{(1,\partial)}{\left(\tg^{(1,\partial)},\mka,\xibR\right)}&=  \frac{2}{3}(4\pi)^{-\frac{d}{2}+1} \ns\,\tg^{(1,\partial)}_k \left\{\left(1-\tfrac{3}{2}(\tfrac{d-2 }{2})\right)\, \ThrfA{2}{d\slash 2 }{\mkB} \vphantom{ \frac{3}{2}} \right.\nonumber \\
&\left. \phantom{=+  \frac{2}{3}(4\pi)^{-\frac{d}{2}+1} \ns\,\tg^{(1,\partial)}_k\qquad} -  \frac{3}{2}\mkC\,\ThrfA{2}{d\slash2-1}{\mkB }\right\}  \label{eqn:04_09d}
\end{align}
\end{subequations}

\paragraph{(E) The cosmological constant type couplings.} We only have three cosmological constant type couplings due to the boundary conditions imposed on the fluctuation field $\flcb_{\mu\nu}$. For the present truncation we find their flow equations to be
\begin{subequations} \label{eqn:04_10}
\begin{align}
\partial_t \lambda_k^{(0)}&=
( \aDzG_k^{(0)}-2) \Kk_k^{(0)}
+  2\,(4\pi)^{-\frac{d}{2}+1}\ns  \,\ThrfA{1}{d\slash 2}{\mkB } \tg^{(0)}_k  \\
\partial_t \lambda_k^{(0,\partial)} &=
(\aDzG_k^{(0,\partial)}-1)\Kk_k^{(0,\partial)}
-  \sqrt{\pi} \,(4\pi)^{-\frac{d}{2}+1}\ns\,  \,  
\ThrfA{1}{(d-1)\slash 2}{\mkB }\tg^{(0,\partial)}_k 
\end{align}
\begin{align}
\partial_t\lambda^{(1)}&=(\aDzG_k^{(1)}-2 ) \Kk_k^{(1)}\nonumber \\
&\qquad
 +    (4\pi)^{-\frac{d}{2}+1}\ns\, \left\{
 \left(2-d\right)\,
 \ThrfA{2}{d \slash 2 +1}{\mkB}-  2\,\mkC\,\ThrfA{2}{d\slash 2}{\mkB} \right\}\,\tg_k^{(1)}
\end{align}
\end{subequations}
As expected, the beta-functions depend on the corresponding anomalous dimensions as well as on the related Newton type couplings. 

\paragraph{(F) The coupling \boldmath{$\eka$}.} Finally, the deviation of the relative coefficient of $\Ricb^{\mu\nu}$ and $\bg^{\mu\nu}\SRb$ from $-1\slash 2$, appearing in the Einstein field equations, is described by $\eka$. Its flow behavior is contained in the following differential equation:
\begin{align}
 \partial_t \eka&=\aDzG_k^{(1)}\eka   -4\, (4\pi)^{-\frac{d}{2}+1}\ns\, \left\{\left(\tfrac{2-d}{12}-\xiBg+\xiBR\right)\, \ThrfA{2}{d\slash 2}{\mkB}
\right. \nonumber \\
&\quad\phantom{= -4\, (4\pi)^{-\frac{d}{2}+1}\ns\, k^{d-2}  \qquad}
+\,\mkC\left[2\, \xiA \,\ThrfA{3}{d\slash 2}{\mkB}-\tfrac{1}{6}\ThrfA{2}{d\slash 2 -1}{\mkB}  \right] \nonumber \\
&\quad\phantom{= -4\, (4\pi)^{-\frac{d}{2}+1}\ns\, k^{d-2}  \qquad}\left.
+ \left(d-2\right)\,\xiA \,\ThrfA{3}{(d\slash2)+1}{\mkB}
\right\} \tg_k^{(1)} \label{eqn:04_08}
\end{align}
The importance of $\eka$ resides in the fact that, if non-vanishing, $\eka$ leads to an effective field equation for the metric which could never arise in a single-metric setting. This makes the violation of the split symmetry manifest. 

All of the above beta-functions referring to boundary couplings are new. Special cases of those for the level-(0) bulk couplings and the level-(1) gravity couplings can already be found in the literature \cite{Narain,MRS1}.\footnote{Ref. \cite{Narain} contains the beta-functions of, in their notation, $\tilde{\lambda}_2 \mathrel{\widehat{=}} \frac{1}{2}m^{(0)\,2}$, $\tilde{\lambda}_4 \mathrel{\widehat{=}} \frac{1}{24}u^{(0)}$, $\tilde{\xi}_0\mathrel{\widehat{=}}(16\pi \tg^{(0)})^{-1}$, and $\tilde{\xi}_2\mathrel{\widehat{=}}-\frac{1}{2}\xia$ for $d=4$. Ref. \cite{MRS1} calculates the beta-functions for  $\tg_k^{(0)}$, $\tg_k^{(1)}$, $\lambda_k^{(0)}$, $\lambda_k^{(1)}$, $\eka$ in the case of vanishing $\xi$-parameters.
}

\subsection{Properties of the RG flow}
We will focus in the following on  the running of the Newton constants and the non-minimal coupling parameters. For this purpose it is sufficient to neglect the running in the matter sector and to  treat $m_k^{(0)}\equiv m^{(0)}$, $m_k^{(1)}\equiv m^{(1)}$, $\omgA^{(0)}\equiv u^{(0)}$, and $\omgA^{(1)}\equiv u^{(1)}$ as constants. 
Furthermore, to make the results more transparent we will sometimes display them for the optimized shape function only, for which the threshold functions assume the simple form \eqref{eqn:03_17}. 

\subsubsection{The couplings of the matter sector}

\paragraph{(A) The \boldmath{$\xiA$}-parameter.}
The beta-function of the coupling that multiplies the $\int \sqrt{\bg}\,\SRb A^2$ monomial, i.e. the $\xiA$-parameter, is proportional to $u^{(0)}$ and thus is switched on by a non-vanishing four-vertex. In the approximation of 
 $m^{(0)}_k\equiv m^{(0)}$ and $\omgA^{(0)}\equiv u^{(0)}$ the differential equation \eqref{eqn:04_07a} can be solved analytically: 
\begin{align}
 \xiA&=\left(\xi^{(0)}_{k_0}- \xi^{(0)}_{*}\right)  \left(\frac{k}{k_0}\right)^{\alpha  \,  u^{(0)} } + \xi^{(0)}_{*}\,, && \text{with   } \quad \alpha\equiv \frac{2\,\ns}{(4\pi)^{d\slash 2}} \ThrfA{3}{d\slash 2}{m^{(0)\, 2}} >0 \label{eqn:04_11}
\end{align}
Here the constant $\xi^{(0)}_{*}\equiv\tfrac{1}{12} \ThrfA{2}{d\slash 2-1}{m^{(0)\, 2}}\slash \ThrfA{3}{d\slash 2}{m^{(0)\, 2}}$ is a fixed point  for $\xiA$. Its beta-function \eqref{eqn:04_07a} vanishes at this point, $\partial_t \xiA=0$. Since $\ThrfA{p}{n}{m^{(0)\, 2}}>0$ for any $p$ and $n$, the exponent $\alpha$ is positive. Therefore an initial value $\xi^{(0)}_{k_0}$ larger (smaller) then $\xi^{(0)}_{*}$ will lead to a monotonically increasing (decreasing) function $k \mapsto \xiA$, as long as $u^{(0)}>0$. Hence, the parameter $\xiA$ runs away from its fixed point value for increasing $k$. 

If we start at some $k_0$ with a minimally coupled scalar, $\xi^{(0)}_{k_0}=0$, the term $\int \sqrt{\bg}\,\SRb A^2$ is induced by the RG flow, and its coefficient stays strictly negative during the entire  evolution. The four vertex  drives the theory away from minimal coupling.

For the optimized cutoff shape function and arbitrary dimension $d$ the fixed point is at $\xi^{(0)}_{*}=\tfrac{d}{24} (1+m^{(0)\, 2})$. 

For $m^{(0)}=0$, and $d=4$, we find the cutoff independent result  $\xi^{(0)}_{*}=1\slash 6$; it coincides with the value of $\xi$ in the case of a conformally coupled scalar field in $d=4$. 

\paragraph{(B) The \boldmath{$\xiAS$}-parameter.}
Whereas the bulk coupling $\xiA$ enters the RHS of the FRGE via the Hessian,  its counterpart on the surface, i.e. $\xiAS$, does not and has a much simpler RG equation therefore. 
In fact, its beta-function \eqref{eqn:04_07aS} has the same form as the bulk equation \eqref{eqn:04_07a} in the minimally coupled case, $\xiA=0$. 

We are able to match $\xiA$ and $\xiAS$ at most at one scale, say $k_0$; above or below $k_0$ the couplings differ unavoidably. For $m^{(0)}_k\equiv m^{(0)}$ and $\omgA^{(0)}\equiv u^{(0)}$  the differential equation  \eqref{eqn:04_07aS} has the solution:
\begin{align}
 \xiAS=  -\frac{1}{6}\,(4\pi)^{-\frac{d}{2}}\ns\,\omgA^{(0)} \,\, \, \ThrfA{2}{d \slash 2 - 1}{m^{(0)\, 2}} \ln\left(\frac{k}{k_0}\right) +\xi^{(0,\partial)}_{k_0}\label{eqn:04_12}
\end{align}
Apparently, the scale dependence of $\xiAS$ is driven by the level-(0) four vertex, $u^{(0)}$, too. It decreases (increases) for positive (negative) $u^{(0)}$ and increasing $k$.

\paragraph{(C) The \boldmath{$\xiBg$}-parameter.}
At level-$(1)$, the beta-function of the parameter $\xiBg$, given in eq. \eqref{eqn:04_07b}, depends on $m^{(0)}$, $m^{(1)}$, $u^{(0)}$, $u^{(1)}$, $\xiA$, and on $\xiBg$ itself. By fixing the couplings of the two scalar potentials, and inserting \eqref{eqn:04_11} we can analytically solve for $\xiBg$. This yields 
\begin{align}
 \xiBg&=  -c_1\, (d-4)
+  \left(c_2\,m^{(1)\, 2}-c_3\,m^{(1)\, 2}\, \xia_*  -c_4 \, \xia_* \right)\left[\left(\frac{k}{k_0}\right)^{\alpha\,u^{(0)}} -1\right] \label{eqn:04_13} \\
&\quad \nonumber 
+c_5 \left(12 \xibg_{k_0} +d-4\right) \left(\frac{k}{k_0}\right)^{\alpha\,u^{(0)}}  \\
&\quad 
+\big[c_6\, u^{(0)}+c_7\, m^{(1)\, 2}\, u^{(0)} - c_8\, u^{(1)}\big]  \left(\xia_*-\xia_{k_0}\right)\left(\frac{k}{k_0}\right)^{\alpha\,u^{(0)}}\ln \left(\frac{k}{k_0}\right) \nonumber
\end{align}
The quantities $c_1,\cdots,c_8$ are positive, $k$-independent constants which depend on the spacetime dimension $d$ and shape function $R^{(0)}$, however. 
The solution \eqref{eqn:04_13} shows a complicated structure that is sensitive to a large number of parameters. In the following we consider various special cases which reveal the generic features of \eqref{eqn:04_13} and shed light on certain aspects of split symmetry breaking.

Equation \eqref{eqn:04_13} shows that generically it is the level-(0) coupling $u^{(0)}$ which triggers the dominant running of $\xiBg$, leading to a power law, or log-power growth (decay) for positive (negative) $u^{(0)}$. There is one exception,  namely when $\xiA$ is {\it not} tuned to its fixed point value $\xia_*$. Then, even for $u^{(0)}=0$, the four vertex of level-$(1)$, i.e. $u^{(1)}$, induces a logarithmic running of $\xiBg$.

The beta-function of $\xiBg$ given in \eqref{eqn:04_07b} allows for a fixed point $\xiBg=\xibg_*$. Substituting $\xiA=\xia_*=\tfrac{1}{12} \ThrfA{2}{d\slash 2-1}{m^{(0)\, 2}}\slash \ThrfA{3}{d\slash 2}{m^{(0)\, 2}}$ into this beta-function, $\partial_t \xiBg=0$ has the unique solution\footnote{Here we assume $u^{(0)}\neq 0$. If the level-(0) quartic coupling vanishes, and $\xiA=\xia_*$ is substituted, the beta-function of $\xiBg$ is zero and we have a family of fixed point values $\xibg_*\in \noR$.}
\begin{align}
 \xibg_*&=
- \frac{(d-4)}{12}
- \frac{1}{6} m^{(1)\, 2}\,\frac{\ThrfA{3}{d\slash 2-1}{m^{(0)\, 2}}}{\ThrfA{3}{d\slash 2}{m^{(0)\, 2}}}	\nonumber\\
& \phantom{=}\, + \,\xia_* \,\frac{1}{\ThrfA{3}{d\slash 2}{m^{(0)\, 2}}}\left[ \frac{3(d-2) }{2 } \ThrfA{4}{d\slash 2+1}{m^{(0)\, 2}} +3\, m^{(1)\, 2}\ThrfA{4}{d\slash 2}{m^{(0)\, 2}}\right]\label{eqn:04_14}  
\end{align}
Consider now the case of vanishing masses $m^{(0)}=m^{(1)}=0$, in $d=4$. The first, second and the last term on the RHS of \eqref{eqn:04_14} disappear then, and the remaining threshold-functions all have the structure $\ThrfA{n+1}{n}{0}$, which for every choice of $R^{(0)}$ yields $\ThrfA{n+1}{n}{0} = 1\slash \GF{n+1}$. Thus, eq. \eqref{eqn:04_14} reduces to 
\begin{align}
\xibg_* = \xia_* =\frac{1}{6} \label{eqn:04_14X}
\end{align}
This equality of the two $\xi$-parameters reflects the intact split symmetry in the case of a conformally coupled scalar in $d=4$.
However, for general spacetime dimensions or non-conformal actions, comparing equation \eqref{eqn:04_13} to eq. \eqref{eqn:04_11} shows that $\xibg_* \neq \xia_*$, and so {\it generically the split symmetry is violated.} (Recall also the discussion of eq. \eqref{eqn:04_05Bb}.)

\paragraph{(D) The \boldmath{$\xiBR$}-parameter.}
The beta-function of $\xiBR$ in eq. \eqref{eqn:04_07c} is proportional to  $u^{(0)}$ which triggers the renormalization group effects for $\xiBR$. If we leave the mass $m^{(0)}$ and the four vertex $u^{(0)}$ constant, the differential equation \eqref{eqn:04_07c} decouples and yields the following solution for $\xiBR$, with the same constant $\alpha$ as in \eqref{eqn:04_11},
\begin{align}
\xiBR&= \left(\xibR_{k_0}-\frac{1}{6}\right)\left(\frac{k}{k_0}\right)^{\alpha\, u^{(0)}}	+\frac{1}{6} \label{eqn:04_14B}  
\end{align}
There exists a non-trivial fixed point for general spacetime dimensions $d$ and arbitrary masses $m^{(0)}$, $m^{(1)}$, namely $\xibR_*=1\slash 6$. For positive (negative) quartic coupling $u^{(0)}$ this corresponds to a UV-repulsive (attractive) fixed point. 

\paragraph{Remarks:} Now that we have solved for the $k$-dependence of all bulk $\xi$-parameters, several remarks are in order.

The eqs. \eqref{eqn:04_11} and \eqref{eqn:04_14B} for $\xiA$ and $\xiBR$ show a very similar structure. In both cases the crucial exponent is $\alpha= \frac{2\,\ns}{(4\pi)^{d\slash 2}} \ThrfA{3}{d\slash 2}{m^{(0)\, 2}} >0$, and below or above the fixed point value $\xi_*$ the respective coupling monotonically decreases or increases, respectively. The dependence of the $\xi_k$'s on the corresponding fixed points $\xi_*$ shows  however that, in general, $\xiA$ and $\xiBR$ have a different evolution with the scale $k$. Whereas  $\xibR_*=1\slash 6$  is a fixed number, $\xia_*$ is a monotonically increasing function of the mass $m^{(0)}$. 

As an example, consider the initial values $\xibR_{k_0}=\xia_{k_0}=1\slash 6 + \epsilon$, $u^{(0)}>0$, and $m^{(0)\, 2}=12\, \epsilon$, for a small $\epsilon>0$. Then $\xiBR>\xibR_*$ for all $k$ and asymptotically $\xibR_{k\rightarrow \infty} \rightarrow + \infty$. However, $\xia_{k_0}=1\slash 6 + \epsilon <  1\slash 6 \,(1+m^{(0)\, 2})=\xia_*$ and so $\xiA$ runs from the initial value towards negative infinity, $\xia_{k\rightarrow \infty} \rightarrow - \infty$. This amounts to a considerable violation of split symmetry.

In fact, the  $\xi$-parameters of the bulk multiply the various terms occurring in the $\flcb_{\mu\nu}$-expansion of the monomial $\int \sqrt{g}\,\SR(g)$. In our notation, $\xiA$ describes the running at level-$(0)$, i.e. $\int \sqrt{\bg}\,\SRb(\bg)$, whereas $\xiBg$ and $\xiBR$ are the couplings associated to the basis invariants of level-$(1)$, i.e. $\int (\var \sqrt{g})\SRb(\bg)$ and $\int \sqrt{\bg} \var(\SR(g))$, respectively. To restore split symmetry, the scale dependence of all three couplings would have to coincide, see eq, \eqref{eqn:04_05Bb}. This is clearly not the case for general RG trajectories and spacetime dimensions. It is $\xiBg$ that mainly spoils this symmetry.

A special case in which the split symmetry in the bulk $\xi$-sector remains intact is the conformally coupled action (having $m^{(0)}=m^{(1)}=0$ and $\xi=(d-2)\slash (4d-4)$ in general), in exactly {\it four} spacetime dimensions, $d=4$. For this particular setting, all $\xi$-parameters on the bulk 
remain at their common fixed point value $\xi_*=1\slash 6$ if this value is assigned to them by the initial conditions:\footnote{Note that the $u^{(1)}$ dependence in $\partial_t\xiBg$ vanishes at $\xiA=\xia_*$ and that $\ThrfA{2}{1}{0}=2\ThrfA{3}{2}{0}$ holds.}
\begin{align}
 \xiA=\xiBg=\xiBR=\frac{1}{6} \qquad \Leftrightarrow \qquad \partial_t\xiA=\partial_t\xiBg=\partial_t\xiBR=0 \nonumber
\end{align}
However, already if we relax the assumption of $d=4$, $\xiBg$ will deviate from $\xiBR$ and $\xiA$ upon a small change in $k$.

\paragraph{(E) The \boldmath{$\xiBS$}-parameter.}
Finally, we consider the non-minimal coupling parameter of level-$(1)$ on the boundary. In the approximation  of constant masses and quartic couplings the RHS of its RG equation \eqref{eqn:04_07d} is $k$-independent and we may simply integrate $\partial_t \xiBS\equiv \beta_{\xibS}(m^{(0)},u^{(0)},m^{(1)},u^{(1)};d)=\const$ in order to obtain $\xiBS$:
\begin{align}
 \xiBS&= \beta_{\xibS}(m^{(0)},u^{(0)},m^{(1)},u^{(1)};d) \, \ln \left(\frac{k}{k_0}\right)+ \xibS_{k_0}  \label{eqn:04_15}  
\end{align}
The entire RG flow of $\xiBS$ is completely determined by matter couplings. We will mainly focus on the first term in its beta-function of eq. \eqref{eqn:04_07d}, and therefore set $m^{(1)}=0=u^{(1)}$:
\begin{align}
  \xiBS&= 
 (4\pi)^{-\frac{d}{2}}\ns\,  
\, \left(\tfrac{3d-10}{12}\right)\,    \ThrfA{3}{d\slash2}{m^{(0)\, 2}}\,\omgA^{(0)}
\, \ln \left(\frac{k}{k_0}\right)+ \xibS_{k_0}  \label{eqn:04_16}  
\end{align}
The following features of the eqs. \eqref{eqn:04_15} and \eqref{eqn:04_16}  are noteworthy:

Due to the Dirichlet boundary conditions for $\flcb_{\mu\nu}$ and $\var A$, the matching of $\xiBS$ with $\xiBR$ is not necessary to have a well defined variational problem for the  scalar  but it is necessary if we want the equation  of motion for  $\flcb_{\mu\nu}$ to follow from the stationarity of the action \eqref{eqn:04_03}. Even if we set by hand  $m^{(1)}=u^{(1)}=0$ there is no RG trajectory  along which $\xiBS$ equals $\xiBR$ for all  $k$. The couplings run entirely differently, with a logarithmic and a power law behavior for $\xiBS$ and $\xiBR$, respectively. Furthermore, notice that (except for $u^{(0)}=0$) the coupling $\xibS$ has no fixed point at all. 

Concerning split symmetry, we find that even if we restore the symmetry on the {\it bulk} by fixing $d=4$ and $\xia=\xibg=\xibR=1\slash 6$ which is stable under the flow, along with $m^{(0)}=m^{(1)}=0$,  split symmetry is still violated. The reason is the mismatch of the  boundary couplings at level-(0) and level-(1), $\xiAS$ and $\xiBS$. Actually, for $u^{(1)}=0=m^{(1)}$ it can be easily deduced from equations  \eqref{eqn:04_12} and \eqref{eqn:04_16} that both couplings display an opposite running in case of $d\geq 4$.

\clearpage
\paragraph{Summary ($\xi$-parameters):}

\paragraph{(i)} On the one hand, contributions to the beta-functions of the $\xi$'s arise from the expansion of the functional trace in the FRGE with respect to $\xiA\, \SRb$ and the level-$(1)$ part of the Hessian, $\xiBg\, \SRb\, \flcb^{\mu}_{\phantom{\mu}\mu}$ and $\xiBR\, \var \SR$. On the other hand, the heat kernel expansion generates $\frac{1}{6} \,\SRb$ terms in the beta-functions `from nothing', i.e. they are not proportional to one of the $\xi$'s itself. This results in a relative coefficient of $1\slash 6$ of these two types of contributions to the beta-functions governing the bulk $\xi$ couplings. While the same holds true for $\xiAS$ multiplying $\bEC$ 
at level-(0), its analog at level-$(1)$, the coefficient $\xiBS$ of $n^{\lambda}\bZ_{\lambda}  \flcb^{\mu}_{\phantom{\mu}\mu}$, comes with an additional factor of $3 \slash 2$ in the heat kernel expansion, \eqref{eqn:04_05X}.

\paragraph{(ii)} Even if a trajectory hits the point in theory space corresponding to a minimally coupled action, $\xi=0$, a non-vanishing quartic coupling $u_k^{(0)}$ will cause the non-minimal couplings to grow again.

\paragraph{(iii)} Considering the differential equations \eqref{eqn:04_07} for the five $\xi$-parameters, we found that in four, and only four, dimensions a conformally coupled matter sector remains conformally coupled over all RG scales.
In dimensions other than four, the non-minimal coupling in a conformal theory is $\xi=(d-2)\slash 4(d-1)$ which we found to have  no distinguished meaning compared to other values of $\xi$.

\paragraph{(iv)} Split symmetry can be restored in the bulk sector, in four dimensions, and for conformally coupled matter. However, on the boundary, the level-(0) and level-(1) couplings can be tuned to match {\it at most at one scale } and split symmetry is violated at any other.

\subsubsection{Newton type couplings}
In the remainder of this section we take a closer look at the RG running of the Newton type couplings $\tg_k^a$, $a=(0), (0,\partial), (1), (1,\partial)$.
The anomalous dimensions which we listed in equation \eqref{eqn:04_08} show  a common pattern, but at the same time differ in certain aspects considerably. 
To study them we shall dispose of all inessential complications. In particular we neglect the running of the $\xi$'s and approximate 
\begin{align}
\xiA=\xia=\const \qquad\quad \text{ and } \qquad \quad \xiBR=\xibR=\const  \label{eqn:04_16Xa}
\end{align}
in the sequel. With this restriction, the differential equations for all Newton couplings reduce to the form\footnote{For concreteness we consider $d> 2$ in the following.} 
\begin{align}
\partial_t \tg_k^{a}=(d-2)\big(1- \omega_d^a\, \,\tg_k^a\big)\, \tg_k^a \, , \label{eqn:04_16Xb}
\end{align}
where the constants $\omega_d^a$ are defined by the anomalous dimensions $ \eta^a\equiv -(d-2)\,\omega_d^a \tg_k^a$. The equation \eqref{eqn:04_16Xb} has the same form as in the single-metric case and so the general solution is given by \eqref{eqn:03_19CD}: 
\begin{align}
\Nk_k^a&= \tg_k^a\, k^{2-d}= \frac{\Nk_0^a}{1+ \omega_d^a\, \Nk_0^a\, k^{d-2}} \label{eqn:04_16Xc}
\end{align}
Within this approximation, all dimensionless Newton constants have a Gaussian fixed point at $\tg_*^a=0$, and a non-trivial one at
\begin{align}
 \tg_*^a&= \frac{1}{\omega_d^a}\, , \qquad \quad a=(0),(0,\partial),(1),(1,\partial). \label{eqn:04_g-ngfp}
\end{align}
The special properties of the Newton couplings are listed in the following.

\paragraph{(F) Level-(0), bulk: \boldmath{$\tg_k^{(0)}$}.}
At level-$(0)$ we re-encounter the basis invariants we investigated in the single-metric truncation for pure gravity. Here we study the  running of the couplings $\tg_k^{(0)}$ and $\tg_k^{(0,\partial)}$ induced by the matter sector. The anomalous dimension \eqref{eqn:04_09a} related to $\tg^{(0)}$ depends on $\xia$ that determines the sign of $\omega_d^{(0)}$ and thus the behavior of $\tg_k^{(0)}$ under the RG flow:
\begin{align}
 \omega_d^{(0)}&=-  \gamma_d \, 
\left(1- 6\,\xia \frac{\ThrfA{2}{d \slash 2}{m^{(0)\, 2}}}{\ThrfA{1}{d \slash 2 - 1}{m^{(0)\, 2}}}  \right)\, \ThrfA{1}{d \slash 2 - 1}{m^{(0)\, 2}}	\label{eqn:04_18} 
\end{align}
Here we abbreviated $\gamma_d\equiv\tfrac{2}{3} \,(4\pi)^{-\frac{d}{2}+1}\ns\slash   (d-2)\,>0$. 

The running of $\Nk_k^{(0)}$ can be `switched off' by setting $\xia=\xia_{\,\rm z}$ where $\xia_{\,\rm z}$ is the zero of $\omega_d^{(0)}$. It is given by $\xia_{\,\rm z}= \frac{1}{6}\, \ThrfA{1}{d \slash 2 - 1}{m^{(0)\, 2}}\slash\ThrfA{2}{d \slash 2}{m^{(0)\, 2}} >1\slash 6$. At $\xia=\xia_{\,\rm z}$ the coefficient $\omega_d^{(0)}$, hence $\eta^{(0)}\equiv - (d-2)\omega_d^{(0)}\tg_k^{(0)}$, changes its sign. Assuming $\tg_k^{(0)}>0$ for a moment, the anomalous dimension of $\tg_k^{(0)}$ is negative (positive) for $\xia$ larger (smaller) than $\xia_{\,\rm z}$.  In fact, if we fix $\xia$ such that $\eta^{(0)}$ is {\it negative} at some IR scale $k=k_0$, it stays negative at all higher scales even if we allow $\xia$ to run. This is due to the fact that $\xia$ monotonically increases above its fixed point value $1\slash 6$, and thus the second term in the brackets of \eqref{eqn:04_18} dominates the overall sign of $\eta^{(0)}$ for all $k>k_0$.

The fact that $\eta^{(0)}$ can become negative is also essential for having a {\it positive} fixed point value $\tg_*^{(0)}$.  From \eqref{eqn:04_g-ngfp} with \eqref{eqn:04_18}:
\begin{align}
 \tg_*^{(0)}&=  \frac{ (4\pi)^{d \slash 2-1}}{\ThrfA{2}{d \slash 2}{m^{(0)\, 2}}  } \frac{(d-2)}{4\,\ns}\cdot \left(\xia-\xia_{\,\rm z}\right)^{-1} \label{eqn:04_18B} 
\end{align}
Obviously, $\tg_*^{(0)}$ is positive for $\xia>\xia_{\,\rm z}$. 

The combination of the two points made above shows that {\it for any fixed value $\xia > \xia_{\,\rm z}$, the RG flow of the matter induced truncation mimics the key property of full fledged Quantum Einstein Gravity: Newton's constant has the negative anomalous dimension characteristic of anti-screening, and it admits a non-trivial fixed point at a positive value $\tg_*^{(0)}$.} 
For this reason we may use the simpler scalar system with a non-minimal coupling as a qualitatively correct toy model for full QEG.

\paragraph{(G) Level-(0), boundary: \boldmath{$\tg_k^{(0,\partial)}$}.}
The boundary Newton type coupling on level-$(0)$ obeys the same RG equation as  $\tg_k^{(0)}$ of the bulk if we set $\xia=0$ there:
\begin{align}
 \omega_d^{(0,\partial)}&= - \gamma_d\,\ThrfA{1}{d \slash 2 - 1}{m^{(0)\, 2}}\, < \, 0 \label{eqn:04_19} 
\end{align}
As a consequence, $\eta^{(0,\partial)}=-(d-2)\omega_d^{(0,\partial)}\tg^{(0,\partial)}$ is always positive (negative) for $\tg^{(0,\partial)}$ positive (negative).
Correspondingly the fixed point is situated at negative $\tg^{(0,\partial)}$, namely at
\begin{align}
 \tg_*^{(0,\partial)}&=  -\frac{ (4\pi)^{d \slash 2-1}}{\ThrfA{1}{d \slash 2-1}{m^{(0)\, 2}}  }\frac{3(d-2)}{2\,\ns }  \,\,\,<\,0 \label{eqn:04_20} 
\end{align}

Except for $\xia=0$, the boundary and bulk Newton couplings on level-$(0)$ have different beta-functions. For $\xia < \xia_{\,\rm z}$ they run in the same direction, for $\xia>\xia_{\,\rm z}$ in opposite directions.

\paragraph{(H) Level-(1), bulk: \boldmath{$\tg_k^{(1)}$}.}
In the anomalous dimension $\eta^{(1)}$ of the level-$(1)$ Newton coupling in the bulk, the non-minimal parameter $\xiBR$ enters:
\begin{align}
 \omega_d^{(1)}&= -\gamma_d  \,\left(1-6\,\xibR  \right)\ThrfA{2}{d\slash2}{m^{(0)\, 2}}  \label{eqn:04_21} 
\end{align}
The $\xibR$-parameter can flip the sign of $\eta^{(1)}$ and thereby qualitatively change the flow behavior of $\tg^{(1)}$.

The $\xibR$ value that leads to a vanishing  $\omega_d^{(1)}$ in \eqref{eqn:04_21}, and marks the transition from positive to negative $\eta^{(1)}$, coincides with its  fixed point value  $\xibR_*=1\slash 6$. 
For $\xibR\neq 1\slash 6$, there exists a non-zero fixed point for $\tg^{(1)}$ at 
\begin{align}
\tg_*^{(1)}&=  \frac{ (4\pi)^{d \slash 2-1}}{\ThrfA{2}{d \slash 2}{m^{(0)\, 2}}  } \frac{(d-2)}{4\, \ns}\, \left(\xibR-\frac{1}{6}\right)^{-1} \label{eqn:04_22}  
\end{align}

 Intact split symmetry at the NGFP would require both $\tg_*^{(0)}=\tg_*^{(1)}$ and $\xia_*=\xibR_*$. These two conditions are mutually exclusive, however. Indeed, by eqs. \eqref{eqn:04_18B} and  \eqref{eqn:04_22}, the first condition implies $\xibR_*-\frac{1}{6}=\xia_*-\xia_{\,\rm z}$ which upon using the second condition becomes $\xia_{\,\rm z}=1\slash 6$. This latter constraint is impossible to satisfy since $\xia_{\,\rm z}$ is strictly larger than $1\slash 6$.

This is a special case of a more general argument. Let us ask if we can achieve the split symmetric evolution $\partial_t \tg_k^{(0)}=\partial_t \tg_k^{(1)}$ along some RG trajectory. This requires that  $\omega_d^{(0)}=\omega_d^{(1)}$ which by eqs. \eqref{eqn:04_18B} and  \eqref{eqn:04_21}  corresponds to 
\begin{align}
 \left(\xiBR-\frac{1}{6}\right)&= \left(\xiA-\xia_{\,\rm z}\right)\label{eqn:04_23}  
\end{align}
At the same time we have to fulfill $\xiA=\xiBR$, yielding a contradiction, since $\xia_{\,\rm z}>1 \slash 6$. Thus, split symmetry is unavoidably violated.

\paragraph{(I) Level-(1), boundary: \boldmath{$\tg_k^{(1,\partial)}$}.}
Finally, consider the RG flow of the  level-$(1)$ Newton coupling on the boundary. In order to have a well-defined variational principle for $\flcb_{\mu\nu}$ it is desirable that $\tg_k^{(1)}=\tg_k^{(1,\partial)}$. From equation \eqref{eqn:04_09d} we deduce
\begin{align}
 \omega_d^{(1,\partial)}&= -\gamma_d \, \left[\left(1-\tfrac{3}{2}(\tfrac{d-2 }{2})\right)\, \ThrfA{2}{d\slash 2 }{m^{(0)\, 2}} \vphantom{ \frac{3}{2}}  -  \frac{3}{2}m^{(1)\, 2}\,\ThrfA{2}{d\slash2-1}{m^{(0)\, 2} }\right]\label{eqn:04_24}  
\end{align}
Looking at this equation together with eq. \eqref{eqn:04_21} it is evident that a parallel RG evolution $\partial_t \tg_k^{(1)}=\partial_t \tg_k^{(1,\partial)}$ which requires $\omega_d^{(1)}=\omega_d^{(1,\partial)}$ is a matter of fine tuning $\xibR$ and $m^{(1)}$ rather than a universal property.

 In contrast to the other $\omega_d^a$ functions, $\omega_d^{(1,\partial)}$ is strictly positive for $d>2$ and all $m^{(1)}$. For this reason the required $\xibR$ is strictly larger than the conformal value $1\slash 6$ if we enforce the bulk-boundary matching $\tg_k^{(1)}=\tg_k^{(1,\partial)}$ by setting $\omega_d^{(1)}=\omega_d^{(1,\partial)}$.

 In the special case of $m^{(1)}=0$ the requirement  $\omega_d^{(1)}=\omega_d^{(1,\partial)}$ fixes the  parameter $\xibR$ to be $\xibR_{\text{match}}\equiv (d-2)\slash 8$.
\footnote{In four dimensional spacetimes $\xibR_{\text{match}}=1\slash 4$. As compared to the conformal point the `defect' of a relative coefficient of $2 \slash 3$ originates directly from the heat kernel expansion where the relevant monomial comes with the coefficient $1 \slash 4$ instead of $1 \slash 6$ as $\SRb$ and $\bEC$ do.}
Clearly the condition $\xibR=\xibR_{\text{match}}$ is unstable under RG evolution; at best it can hold approximately, in a truncation which neglects the running of $\xibR$.
Though we may arrange for  matching boundary and bulk Newton constants in level-$(1)$ at a certain scale $k_0$, a running $\xiBR$ will immediately destroy it since $\xiBR$ rapidly evolves towards larger values near $\xibR_{k_0}=\xibR_{\text{match}}$ for $d>3$. 

The above {\it boundary-bulk matching} is not to be confused with {\it split symmetry}. While the former refers to a fixed level, the latter links different levels. For the boundary Newton constants, exact split symmetry would require $\tg_k^{(0,\partial)}=\tg_k^{(1,\partial)}$ at all scales. This relationship, too, is unstable under RG evolution. The anomalous dimensions $\eta^{(0,\partial)}$ and $\eta^{(1,\partial)}$ have opposite signs for initially coinciding couplings,  $\tg_{k_0}^{(0,\partial)}=\tg_{k_0}^{(1,\partial)}$, which drives them apart immediately.

The coupling $\tg^{(1,\partial)}$ has a non-trivial  fixed point at
\begin{align}
\tg_*^{(1,\partial)}&=  \frac{ (4\pi)^{d \slash 2-1}}{\ThrfA{2}{d \slash 2}{m^{(0)\, 2}}  }\frac{(d-2)}{2\,\ns} \, \left(\frac{3 d-10 }{12}  +   \frac{m^{(1)\, 2}}{2}\,\frac{\ThrfA{2}{d\slash2-1}{m^{(0)\, 2} }}{\ThrfA{2}{d\slash 2 }{m^{(0)\, 2}}}  \right)^{-1} \label{eqn:04_25}  
\end{align}
For $d>3$ and any value of $m^{(0)}$ and $m^{(1)}$ we have  $\tg_*^{(1,\partial)}>0$.

\subsubsection{Non-Gaussian fixed points of the full system}
Finally we ask whether in the present 17 dimensional bi-metric truncation the scalar fields induce a Non-Gaussian fixed point in the entire gravitational sector. In \cite{MRS1} it was found, in a similar setting but without surface terms, that in fact such a Non-Gaussian fixed point occurs. It is not difficult to see that the latter generalizes to spacetimes with a boundary which is further  evidence for the Asymptotic Safety of quantum gravity. 

We focus on $d=4$ in the following. Then, in  the matter sector, we recover the trivial fixed point at
\begin{subequations}
\begin{align}
m^{(0)}_*=m^{(1)}_* =0\,, && u^{(0)}_*=u^{(1)}_*=0 \label{eqn:04_26}  
\end{align}
\end{subequations}
By virtue of \eqref{eqn:04_26} also the beta-functions of all five non-minimal couplings vanish automatically, i.e. without imposing further conditions on the couplings. As a consequence, the fixed point is 5-fold degenerate with respect to the values of the $\xi$-parameters. Although the fixed point values of the remaining couplings depend on the $\xi$-parameters, they do not restrict them,  and we may choose $\xia_*$, $\xiaS_*$, $\xibg_*$, $\xibR_*$, and $\xibS_*$ arbitrarily.  

However, there are two exceptions: If $\xia_*=\xi_{\,\rm z}$, we have only the trivial fixed point value for the level-(0) Newton coupling on the bulk, i.e. $\tg_*^{(0)}=0$. The same holds for $\xibR_*=1\slash 6$ and the level-(1) Newton coupling on the bulk, $\tg_*^{(1)}=0$. 

Thus, a Non-Gaussian fixed point in the complete  gravitational sector is given by \addtocounter{equation}{-1}
\begin{subequations}\stepcounter{equation}
\begin{align}
 &\xiaS_*=\arbit \, , && \xibS_*=\arbit \, , \\
& \xibg_*=\arbit\, , \\
&\xia_*\neq  \frac{1}{6}\, \frac{\ThrfA{1}{1}{0}}{\ThrfA{2}{2}{0}} \, , && \xibR_*\neq \frac{1}{6}\, , \\
& \tg_*^{(0,\partial)}=-\frac{12 \pi}{\ns \ThrfA{1}{1}{0}} \, , &&  \tg_*^{(0)}=-\frac{12 \pi}{\ns \left(\ThrfA{1}{1}{0} - 6 \ThrfA{2}{2}{0} \xia_*\right) }\,, \\
& \tg_*^{(1,\partial)}=+\frac{24 \pi}{\ns \ThrfA{2}{2}{0}} \, , &&  \tg_*^{(1)}=+\frac{24 \pi}{\ns \left(2 - 12\,  \xibR_*\right)\ThrfA{2}{2}{0} }\,, \\
& \Kk_*^{(0,\partial)}=\sqrt{\pi} \,\frac{\ThrfA{1}{3\slash 2}{0}}{\ThrfA{1}{1}{0}}\, , &&  \Kk_*^{(0)}= \frac{3\ThrfA{1}{2}{0}}{ 2\left(6\ThrfA{2}{2}{0}\xia_* -\ThrfA{1}{1}{0}\right)}\, , \\
 & \Kk_*^{(1)}= \frac{3 \ThrfA{2}{3}{0}}{ 2\left(6\,\xibR_* -1\right) \ThrfA{2}{2}{0}}\, ,
& &E_*=\frac{\left(12 \frac{\ThrfA{3}{3}{0}}{\ThrfA{2}{2}{0}}\, \xia_* - 6\,\xibg_*\right)}{\left(1-6\,\xibR_*\right)} -1
\end{align}
\end{subequations}

Remarkably, on the manifold of fixed points parametrized by the $\xi$'s, there exists a distinguished point, namely  $\xia_*=0=\xiaS_*$ and $\xibR_*=1\slash 12 = \xibS_*$: There we obtain a well defined standard variational principle for $\flcb_{\mu\nu}$ and $A$, since  $\tg_*^{(1)}=\tg_*^{(1,\partial)}$ from \eqref{eqn:04_26} so that the $\partial \MaFs$-term in \eqref{eqn:04_03} vanishes. (At this point, $\tg_*^{(0)}=\tg_*^{(0,\partial)}$ is satisfied, too, so that even the variation with respect to $\bg_{\mu\nu}$ is well defined.) 

In particular this distinguished point seems perfectly suitable for constructing a continuum limit. Here we shall not embark on a detailed discussion of the RG trajectories and the critical manifold, however; this will be done in a companion paper \cite{daniel-prep} using numerical methods.
\section{Variational problems, the counting of field modes, and black hole thermodynamics}

In this section we discuss various notions of effective, i.e. $k$-dependent field equations implied by the gravitational average action. 
As a first application we exploit that they provide us with a universal tool for `counting' the number of field modes having covariant momenta between infinity and the IR cutoff $k$ which, in $\EAA_k$, are integrated out already.
We shall describe several explicit examples, emphasizing in particular the r\^{o}le of boundary terms. The main application will be the thermodynamics of black holes and the underlying statistical mechanics of fluctuation modes.

\subsection{Effective field equations and on-shell actions}

We begin our discussion at the exact level, i.e. no truncation is involved, but we assume $\partial\MaFs=\emptyset$ and leave the complications due to a non-empty boundary aside for a moment.

\subsubsection{Running stationary points}

It follows directly from the definition of the average action \cite{wett-mr, mr} that the $k$-dependent field-source relationship which governs the expectation values $g_{\mu\nu}=\langle \gamma_{\mu\nu}\rangle$, $\flcb_{\mu\nu}\equiv \langle h_{\mu\nu}\rangle$, and similarly for the matter fields, is the stationarity condition of the functional $\widetilde{\EAA}_k\equiv \EAA_k + \Delta_k \SW$, rather than of $\EAA_k$ itself.
On an arbitrary theory space involving dynamical, i.e. non-background fields $\Phi_i$ coupled to sources ${\cal J}_i$ it reads
\begin{align}
\frac{\var \widetilde{\EAA}_k[\Phi]}{\var \Phi_i}\equiv \frac{\var \EAA_k[\Phi]}{\var \Phi_i}+\frac{\var \Delta_k \SW[\Phi]}{\var \Phi_i} = {\cal J}_i 	\label{eqn:05_01}
\end{align}
Eq. \eqref{eqn:05_01} gives rise to a first notion of a scale dependent `effective field equation': the running stationary point condition of $\widetilde{\EAA}_k$. It is a complicated system of equations, involving $\flcb_{\mu\nu}$, $\bg_{\mu\nu}$, and all matter fields; it allows for the computation of all expectation values of the dynamical fields {\it in dependence on both the sources and the background fields}.

For a gravity-scalar system, for instance, the set $\Phi_i$ consists of $A$ and $\flcb_{\mu\nu}$. The effective field equations read therefore, with vanishing sources, say,
\begin{align}
\frac{\var \EAA_k[A,\flcb;\bg]}{\var A(x)}+\sqrt{\bg}\,{\cal R}_k[\bg] A(x) =0 \label{eqn:05_02}
\end{align}
A similar equation, involving a term ${\cal R}_k \flcb_{\mu\nu}$, holds for the metric fluctuation. This coupled system yields the solutions $A(x)\equiv A[\bg](x)$ and $\flcb_{\mu\nu}(x)\equiv \flcb_{\mu\nu}[\bg](x)$ as functionals of the fixed but arbitrary background metric $\bg$.

\subsubsection{Running tadpole equation and self-consistent backgrounds}
We can now ask under what conditions $\flcb_{\mu\nu}=0$ is a solution to the above coupled system, together with an appropriate configuration of $A$. In general this will happen only for very  special, `self-consistent' backgrounds. They are determined by the $\flcb$-analogue of \eqref{eqn:05_02}, without the ${\cal R}_k \flcb_{\mu\nu}$ term, however, since we set $\flcb_{\mu\nu}=0$ after the first variation:
\begin{align}
 \left.\frac{\var}{\var\flcb_{\mu\nu}(x)} \EAA_k[A,\flcb;\bg_k^{\text{selfcon}}]\right|_{\flcb=0}=0 \label{eqn:05_03}
\end{align}
We call $\bg^{\text{selfcon}}_k$ a self-consistent background metric for the scale $k$ if there exists a scalar field configuration $A(x)$ such that the coupled system of equations \eqref{eqn:05_02} and \eqref{eqn:05_03} is satisfied.

This latter system constitutes a second natural notion of an effective field equation generalizing the classical Einstein equation; it is a scale dependent version of the tadpole condition mentioned in the Introduction. Contrary to the running stationary point condition, it involves only one rather than two metric variables, namely $\bg_{\mu\nu}$.

Let us assume we are given the (exact) average action as a Taylor series in $\flcb$,
\begin{align}
 \EAA_k[A,\flcb;\bg]= \EAA_k^{\background}[A;\bg]+\EAA_k^{\lin}[A,\flcb;\bg]+\EAA_k^{\text{quad}}[A,\flcb;\bg]+\cdots
\label{eqn:05_03a}
\end{align}
where the various contributions on the $\RHS$ of \eqref{eqn:05_03a} contain $0,1,2,\cdots,$ factors of $\flcb_{\mu\nu}$. If we insert \eqref{eqn:05_03a} into \eqref{eqn:05_03} then all terms except the linear one will drop out:
\begin{align}
 \frac{\var}{\var \flcb_{\mu\nu}(x)} \EAA_k^{\lin}[A,\flcb;\bg^{\text{selfcon}}_k]=0 \label{eqn:05_03b}
\end{align}

To avoid any misunderstanding later on we emphasize that \eqref{eqn:05_03b} has nothing to do with a truncation at the linear level. It is an exact statement expressing the vanishing of the 1-point function. Instead, truncations of \eqref{eqn:05_03a} at some order of $\flcb_{\mu\nu}$ would influence only the precision with which $\EAA_k^{\lin}$ can be computed from the flow equation since contributions coming from the quadratic term $\EAA_k^{\text{quad}}$, say, affect the running of $\EAA_k^{\lin}$.

As we assume $\partial \MaFs=\emptyset$ at the moment, $\EAA_k^{\lin}$ has the structure
\begin{align}
 \EAA_k^{\lin}[A,\flcb;\bg]=\int_{\MaFs}\md^d x \sqrt{\bg} \, \Upsilon_k^{\mu\nu}[\bg,A](x) \flcb_{\mu\nu}(x) \label{eqn:05_03c}
\end{align}
and the functional derivative in \eqref{eqn:05_03b} is not problematic, whence the tadpole equation becomes
\begin{align}
 \Upsilon_k^{\mu\nu}[\bg,A]=0 \label{eqn:05_03d}
\end{align}

Two important facts are noteworthy about the running tadpole equation: First, since it is an equation for $\bg$ but the differentiation in \eqref{eqn:05_03} is with respect to another field, namely $\flcb$, {\it the tadpole equation cannot be written as the $\bg$-derivative of any diffeomorphism invariant functional $F[\bg,A]$ in general.} Second, the integrability of this system is not automatic, but under special circumstances solutions can exist. (For a detailed discussion of the second issue we refer to \cite{MRS1}.)

Note also that generically, i.e. when $\EAA_k$ is not split symmetric, the tadpole and the stationary condition are not equivalent.

Both the {\it running stationarity condition}  and the {\it running tadpole equation} generalize the field equations of classical gravity theory in a way which goes far beyond replacing the Einstein-Hilbert action $\SW_{\text{EH}}[g]$ with some other diffeomorphism invariant single-metric functional $\SW[g]$. In the first case one deals with field equations involving two metrics $\bg_{\mu\nu}$ and $\bg_{\mu\nu}+\flcb_{\mu\nu}\equiv g_{\mu\nu}$; their structure is constrained only by the requirement that they must be representable as the $\flcb$-derivative of some invariant functional. In the second case the effective field equations contain only one metric, namely $\bg$, but there is no longer the requirement to be the $\bg$-derivative of an invariant action.\footnote{An example of this freedom is the $\eka$-term in the bi-metric truncation.} Thus the gravitational average action allows for a great variety of potential modifications of the classical Einstein equations. A purely phenomenological analysis of such bi-metric actions should therefore be a worthwhile complement to the flow equation studies.

\subsubsection{Running on-shell actions}

The effective average action is known to satisfy an exact functional integro-differential equation \cite{wett-mr}. For quantum gravity with quantized scalar matter it reads \cite{mr}:
\begin{align}
 \exp\{-\EAA_k[\flcb,A,\Ghx,\GhAx;\bg]\}  &=\int {\cal D} h \,{\cal D} \hat{A}\, {\cal D} \hat{\Ghx}\, {\cal D} \hat{\GhAx} \, \exp\left[ \vphantom{\int} -\widetilde{\SW}[h,\hat{A},\hat{\Ghx},\hat{\GhAx};\bg] \right.\label{eqn:05_04} \\ &
\quad \phantom{=\int {\cal D} h  }
 +\int \md^d x\ \left\{ (h_{\mu\nu}-\flcb_{\mu\nu})\frac{\var}{\var \flcb_{\mu\nu}} + (\hat{A}-A) \frac{\var}{\var A}  \right. \nonumber \\ 
&\quad \phantom{=\int {\cal D} h  } \qquad \qquad \left. \left. \quad
+ (\hat{\Ghx}^{\mu}-\Ghx^{\mu}) \frac{\var}{\var \Ghx^{\mu}} +   (\hat{\GhAx}_{\mu}-\GhAx_{\mu}) \frac{\var}{\var \GhAx_{\mu}}  \right\} \EAA_k[\flcb,A,\Ghx,\GhAx;\bg]\right] 
\nonumber \\ &\quad \phantom{=\int {\cal D} h   }
\cdot \exp\big\{-\Delta_k \SW[h-\flcb,\hat{A}-A,\hat{\Ghx}-\Ghx,\hat{\GhAx}-\GhAx;\bg]\big\} \nonumber
\end{align}
Here we wrote $(h_{\mu\nu},\hat{A},\hat{\Ghx}^{\mu},\hat{\GhAx}^{\mu})\equiv \hat{\Phi}$ for the quantum fields (integration variables) and $(\flcb_{\mu\nu},A,\Ghx^{\mu},\GhAx_{\mu})\equiv \Phi$ for their expectation values. Furthermore, the action $\widetilde{\SW}\equiv \SW + \SW_{\text{gf}}+\SW_{\text{gh}}$ contains the gauge fixing and Faddeev-Popov ghost terms besides the diffeomorphism invariant bare action $\SW$.

Let us assume $\Phi(x)\equiv \Phi^{\text{SP}}_k[\bg](x)$ is a running stationary point of $\EAA_k[\Phi;\bg]$. If we insert it on both sides of the integro-differential equation, the functional derivatives on the RHS of  \eqref{eqn:05_04} all vanish, and we obtain
\begin{align}
 e^{-\EAA_k[\Phi^{\text{SP}}_k[\bg];\,\bg]}= \int {\cal D} \hat{\Phi}\,\, e^{-\widetilde{\SW}[\hat{\Phi};\,\bg]}\,e^{-\Delta_k\SW[\hat{\Phi}-\Phi^{\text{SP}}_k[\bg]]} \label{eqn:05_05}
\end{align}
The simplest, and most important, special case of \eqref{eqn:05_05} is realized when $\bg\equiv \bg^{\text{selfcon}}_k$ is a running self-consistent background $(\flcb_{\mu\nu}=0)$, the scalar has no expectation value $(A=0)$, and the ghost configuration\footnote{Since $\EAA_k$ preserves ghost number this is always a solution to the corresponding field equations.}  $\Ghx^{\mu}=0=\GhAx_{\mu}$ is picked. Then $\Phi^{\text{SP}}_k\equiv 0$, and we have
\begin{align}
 e^{-\EAA_k[0;\,\bg^{\text{selfcon}}_k]}= \int {\cal D} \hat{\Phi}\,\, e^{-\widetilde{\SW}[\hat{\Phi};\,\bg^{\text{selfcon}}_k]} \,e^{-\Delta_k\SW[\hat{\Phi}]} \label{eqn:05_06}
\end{align}
This equation shows that the quantity
\begin{align}
\ParFD_k \equiv e^{-\EAA_k [0;\,\bg^{\text{selfcon}}_k]} \label{eqn:05_07}
\end{align}
has the following very interesting interpretation: It is the partition function, cut off at the IR scale $k$, of a certain statistical mechanical system, with Boltzmann factor $e^{-\widetilde{\SW}}$, and defined on a classical spacetime with metric $\bg^{\text{selfcon}}_k$.
This system has vanishing fluctuation averages, $\langle \hat{\Phi} \rangle \equiv \Phi = 0$, and the relative contribution of the various field modes to the statistical sum is weighted by the  suppression factor $e^{-\Delta_k\SW}$.
In this sense the gravitational average action, evaluated on a running self-consistent background, is a tool for `counting' the states (field modes) integrated out between infinity and the IR scale $k$.

In a non-gauge theory without fermions an integral of the type \eqref{eqn:05_06} would imply that $\ln \ParFD_k$ is a monotonic function%
\footnote{Clearly $\EAA_k[0;\bg^{\text{selfcon}}_k]$ is reminiscent of a c-function. We shall come back to this aspect elsewhere \cite{daniel-prep}.} 
 which decreases for increasing $k$:
\begin{align}
 \frac{\partial}{\partial k} \ln \ParFD_k < 0 \label{eqn:05_07AA}
\end{align}
In the present case where some of the fields in $\hat{\Phi}$ are Grassmann odd the inequality \eqref{eqn:05_07AA} might be violated, at least for some RG trajectories and some intervals of $k$. The same conclusion can also be drawn from the FRGE for $\EAA_k$ where the potential minus signs implicit in the supertrace will spoil \eqref{eqn:05_07AA} in general.

\subsection{Variational principle in presence of a boundary}
In the previous subsection we excluded spacetime  manifolds which have a boundary so that the formal functional derivative $\var \slash \var \flcb_{\mu\nu}(x)$ has a clear meaning in the sense of a well-defined variational principle for which it is a shorthand notation.

Now we return to manifolds with a boundary $\partial \MaFs \neq \emptyset$. The problem then consists in giving a precise meaning to the variations in presences of boundary terms. Here we restrict the discussion to the truncated theory space of the gravity-scalar model of Section \ref{sec:04}, with $\EAA_k$ given by the sum of the background terms \eqref{eqn:04_02} and the linear part \eqref{eqn:04_03}. As our truncation stops at  linear order the two types of effective field equations discussed previously are equal in this case since the `running stationary point' happens to be independent of $\flcb_{\mu\nu}$.

Let us consider the response of $\EAA_k$ to a variation $\flcb_{\mu\nu} \rightarrow \flcb_{\mu\nu} + \var \flcb_{\mu\nu}$ where we require
\begin{align}
 \left.\var \flcb_{\mu\nu} \right|_{\partial\MaFs} = 0 \label{eqn:05_08}
\end{align}
in order not to leave the domain of $\EAA_k$. (Recall that $\flcb_{\mu\nu}$ itself satisfies Dirichlet conditions.) From eq. \eqref{eqn:04_03} we obtain for $\var \EAA_k=\var \EAA_k^{\lin}$:
\begin{align}
 \var \EAA_k[\flcb,A;\bg]&=%
\frac{1}{16\pi \nkB}\int_{\MaFs}\md^d x \sqrt{\bg}\,\, \calE_k^{\mu\nu}[\bg,A] \var\flcb_{\mu\nu}(x) \label{eqn:05_09}\\
& +\int_{\partial \MaFs}\md^{d-1}x\sqrt{\biM}\, \left\{ \frac{1}{16\pi } \left(\frac{1}{\nkB}-\frac{1}{\nkBB}\right)- \frac{1}{2}\left(\xiBR-\xiBS\right)A^2    \right\} n^{\alpha}\partial_{\alpha}\var\flcb^{\mu}_{\phantom{\mu}\mu} \nonumber
\end{align}
If the $\partial \MaFs$-integral in \eqref{eqn:05_09} is absent, the requirement of stationarity leads to an effective Einstein equation without problems. With \eqref{eqn:04_04} it reads\footnote{See ref. \cite{MRS1} for  a discussion of the integrability issues related to \eqref{eqn:05_10}.}
\begin{align}
{\bar{G}}^{\mu\nu}-\frac{1}{2}\, \eka \,\bg^{\mu\nu} \SRb + \kkB\,\bg^{\mu\nu}  = 8\pi \nkB{\cal T}_k^{\mu\nu}[A;\bg]
\label{eqn:05_10}
\end{align}
However, as \eqref{eqn:05_08} does not imply the vanishing of the normal derivative $n^{\alpha}\partial_{\alpha} \var \flcb_{\mu\nu}$ the surface term of \eqref{eqn:05_09} is non-zero generically. 

There are several relevant issues here.

\paragraph{(A)} First of all note that the difficulty of the `disturbing' surface term concerns only one of several irreducible components of the metric fluctuation. If one performs a transverse-traceless (York) decomposition of $\flcb_{\mu\nu}$, ref. \cite{oliver}, it is only its trace part $\flcb_{\mu\nu}^{\text{tr}}\equiv \phi \bg_{\mu\nu} \slash d$, with $\phi \equiv \bg^{\mu\nu}\flcb_{\mu\nu}$, which is affected by the surface term in $\EAA_k^{\lin}$. Thanks to the Dirichlet boundary conditions for $\flcb_{\mu\nu}$ we may write, on $\partial\MaFs$,
\begin{align}
n^{\alpha}\partial_{\alpha}\flcb^{\mu}_{\phantom{\mu}\mu}\equiv \bg^{\mu\nu} n^{\alpha}\partial_{\alpha}\flcb_{\mu\nu}= \bg^{\mu\nu} n^{\alpha}\bZ_{\alpha}\flcb_{\mu\nu} = n^{\alpha}\bZ_{\alpha}(\bg^{\mu\nu}\flcb_{\mu\nu}) = n^{\alpha}\partial_{\alpha} \phi\, . \nonumber
\end{align}
Hence the surface integral in \eqref{eqn:04_03} reads $\int_{\partial\MaFs}\md^{d-1}x \sqrt{\biM}\, n^{\alpha}\partial_{\alpha}\phi$. Obviously the other (i.e. TT, TL, and LL) parts of the York decomposition do not contribute so that those irreducible components, at fixed $\phi$, enjoy a standard variational principle. Note that $\phi$ amounts to a fluctuation of the conformal factor of $g_{\mu\nu}$.

\paragraph{(B)} There is the possibility that the surface term vanishes as a consequence of 
\begin{align}
\nkB = \nkBB\quad \Leftrightarrow \quad \tg^{(1)}_k = \tg^{(1,\partial)}_k \qquad \text{and} \qquad \xiBR=\xiBS \label{eqn:05_11} 
\end{align}
For a single scale $k$, at the `physical point' $k=0$, for instance, this can always be arranged for, presumably, by picking an appropriate RG trajectory (i.e. by choosing suitable constants of integration). An exact equality of the level-(1) bulk and boundary Newton constant is non-trivial, however. It requires their respective anomalous dimensions to agree,
\begin{align}
 \aDzG^{(1)}=\aDzG^{(1,\partial)} \quad \text{when}\quad \tg^{(1)}_k = \tg^{(1,\partial)}_k .\label{eqn:05_12} 
\end{align}
In a randomly chosen truncation the (generalization of the) condition \eqref{eqn:05_11} will not be met in general. 

However, there do exist distinguished `perfect truncations' where the bulk and boundary Newton constants are equal on all scales. In Section \ref{sec:04} we found an (admittedly somewhat artificial) example of this kind: If we focus on the subsector of the average action \eqref{eqn:04_01} in which $m^{(0)}=m^{(1)}=0=u^{(0)}=u^{(1)}$, and $\xibR=(d-2)\slash 8=\xibS$ are kept fixed {\it by the very definition of the ansatz,} the flow on the smaller theory space will respect \eqref{eqn:05_11}. It is an intriguing conjecture that such perfect truncations possess an enhanced degree of self-consistency and reliability.

\paragraph{(C)} A more speculative possibility which goes beyond the concrete matter system of the present paper is the following. It is conceivable that on very particular theory spaces, with carefully chosen field contents and symmetries, the desired bulk-boundary matching occurs automatically and universally on all scales without any further ado. 

This possibility is not as far fetched as it might seem perhaps. In fact, there is an example where almost precisely this `miracle' is known to happen: From a technical point of view the trace computations needed for the FRGE projections parallel exactly the evaluation of the spectral action in non-commutative geometry \cite{spect-act}. However, within the latter setting, it has been shown \cite{cham-connes-surf} that the spectral triple encoding the standard model of particle physics automatically gives rise to the correctly adjusted surface term, a very remarkable result indeed.

\paragraph{(D)} Up to now we tried to follow the standard variational method of General Relativity, except that we employed the decomposition $g_{\mu\nu}=\bg_{\mu\nu}+\flcb_{\mu\nu}$ and regarded $\flcb_{\mu\nu}$ the dynamical field which carries the entire variation, $\var g_{\mu\nu}=\var \flcb_{\mu\nu}$. However, we were conservative in our choice of the function space, ${\cal F}$, in which $\flcb_{\mu\nu}$ and $\var \flcb_{\mu\nu}$ are supposed to live,
\begin{align}
 {\cal F}\equiv \left\{ f_{\mu\nu} \text{ tensor on }\MaFs\, ;\quad f_{\mu\nu}=0 \text{ on } \partial \MaFs \right\}\, \label{eqn:05_13}
\end{align}
and this has led to the problematic surface term in $\var\EAA_k$.

So, might  it be possible to be more modest and choose a smaller function space, requiring that also the normal derivative $\bZ_n$ of $\var \flcb_{\mu\nu}$ vanishes on the boundary:
\begin{align}
 \var \flcb_{\mu\nu} \in {\cal F}^{\prime}\equiv \left\{ f_{\mu\nu} \text{ tensor on }\MaFs\, ;\quad f_{\mu\nu}=0 \text{ and } \bZ_n f_{\mu\nu}=0\text{ on } \partial \MaFs \right\}\qquad ?\label{eqn:05_14}
\end{align}
For this new choice the surface terms in $\var\EAA_k$ vanish {\it always}, and the effective field equation is unambiguously given by \eqref{eqn:05_10}.

The conventional answer to the above equation is a clear `no', in particular when the underlying functional integral is intended to represent a transition amplitude between 3-geometries; $\MaFs$ is the portion of spacetime between `initial' and `final' time slices, making up $\partial\MaFs$, then.
Fixing the field and its normal derivative on both the initial and final slice amounts to imposing twice too many boundary conditions for a second order field equation.  So we would loose most, or perhaps all solutions when we try to base the variational principle on ${\cal F}^{\prime}$ rather than ${\cal F}$.

However, in the bi-metric setting with its background split  there are cases where this answer does not apply, or is much less convincing at least:

\paragraph{(i)} Consider an arbitrary (second order in the derivatives) bi-metric action $\EAA_k[\flcb;\bg]$. Let us leave aside higher functional derivatives $(\var \slash \var \flcb)^n \EAA_k[\flcb;\bg]$ and consider the first one only. Furthermore, regarding $n=1$, let us sacrifice the possibility of identifying {\it arbitrary} stationary points, but let us be content  with self-consistent backgrounds. Then, loosely speaking, all that needs to have a precise meaning is the first $\flcb_{\mu\nu}$-derivative of $\EAA_k$, not for all $\flcb$, but only near $\flcb_{\mu\nu}\equiv 0$.
The essential observation is that the self-consistent solution $\flcb_{\mu\nu}\equiv 0$ on all of $\MaFs$, if it exists, is {\it not} lost when we restrict ${\cal F}$ to ${\cal F}^{\prime}$, the trivial reason being that the zero solution has vanishing derivatives everywhere on $\MaFs$ and, by continuity, a vanishing normal derivative on $\partial\MaFs$.

\paragraph{(ii)} Let us consider an, otherwise arbitrary, bi-metric action which happens to be linear in $\flcb_{\mu\nu}$, either as the result of some exact calculation or, as in Section \ref{sec:04}, because theory space has been truncated in this way. Then $\flcb_{\mu\nu}$ has the character of an {\it auxiliary field}. No variational principle whatsoever could yield a dynamical equation for $\flcb_{\mu\nu}$, but only the tadpole constraint depending on $\bg_{\mu\nu}$ and the matter fields.
As for the interpretation of the results for the truncated gravity-scalar system in Section \ref{sec:04}, this last argument suggests that the tadpole equation \eqref{eqn:05_10} can be taken seriously even when the boundary correction is maladjusted to the corresponding bulk term, i.e. when the surface integral in \eqref{eqn:05_09} has a non-zero prefactor, $\nkB \neq \nkBB$ or $\xiBR\neq\xiBS$. In this case we restrict the variations of the `auxiliary field' $\flcb_{\mu\nu}$ to ${\cal F}^{\prime}$.

\paragraph{Discussion:}
In this paper we shall not try to definitely resolve the issue of the bulk-boundary matching required by the variational principle. Presumably this is anyhow only possible by fully appreciating that $\EAA_k$ is not a classical but an effective action containing arbitrarily high derivatives acting on the metric and correspondingly complicated surface terms. This will require a major structural generalization of the FRGE, for the following reason. 

In the setting of the present paper the boundary couplings do not enter the Hessian $\EAA_k^{(2)}$ on the RHS of the flow equation; hence they cannot back-react on the RG evolution which rather is fully determined by the bulk couplings. In more complicated truncations, and at the exact level this situation will change; $\EAA_k^{(2)}$ will consist of bulk-bulk, bulk-boundary, and boundary-boundary blocks, and also the cutoff operator $\CutO_k$ has an analogous block structure. At this point one must take a decision about how to coarse-grain fields living on the boundary. A priori there is a considerable freedom in choosing a cutoff for them, and clearly this choice will be crucially important for the bulk-boundary matching.

Returning to the more restricted scope of the present paper we shall consider it legitimate to extract effective field equations from the bulk action at level-(1) and assume that no surface terms interfere with that. This is justified by either invoking the restriction from  ${\cal F} $ to ${\cal F}^{\prime} $ or, very conservatively, by narrowing down the truncation to the `perfect' one of (B) above; this will not take anything away from the non-trivial results of the next section, in particular on black hole thermodynamics.

 \subsection{Counting field modes}

In this subsection we present a number of examples,  for various truncations and regimes along the  RG trajectory, which illustrate the `state counting' property of $\ln \ParFD_k = - \EAA_k[0;\,\bg_k^{\text{selfcon}}]$.
Here we focus on the relevance of surface terms.
For simplicity we fix $d=4$ in this subsection.

\subsubsection{Single-metric truncation, no boundary}
Let us consider pure gravity and the single-metric ansatz $\EAA_k\equiv \EAA_k^{\text{bulk}}$ of eq. \eqref{eqn:03_02}, assuming $\partial \MaFs = \emptyset$ for a moment. With vanishing ghosts, self-consistent backgrounds are solutions of the (conventional looking, but `running') Einstein equation
\begin{align}
 G_{\mu\nu}(\bg_k^{\text{selfcon}})= - \Kkbar_k\,(\bg_k^{\text{selfcon}})_{\mu\nu} \label{eqn:05_15}
\end{align}
For every given solution to \eqref{eqn:05_15}, eq. \eqref{eqn:03_02} leads to the running on-shell -action
\begin{align}
 \EAA_k^{\text{bulk}}[0,0,0;\,\bg_k^{\text{selfcon}}]&= - \frac{\Kkbar_k}{8\pi  G_k} \left. \int_{\MaFs} \md^4 x \sqrt{g}\right|_{g=\bg_k^{\selfcon}}	\label{eqn:05_16}
\end{align}
This quantity is strictly negative, and $\ln \ParFD_k$ positive therefore. (We assume $\Kkbar_k$ and $G_k$ positive here.)

As an example, consider the maximally symmetric solution to \eqref{eqn:05_15} for $\Kkbar_k >0$, namely the 4-sphere $S^4(L)$ with radius $L_k= (3\slash \Kkbar_k)^{1\slash 2}$.
It has scalar curvature $\SR= 12 \slash L^2_k=4 \Kkbar_k$ and the volume%
\footnote{Here and in the following we write $s_n\equiv \text{vol}\, S^n(1)= 2 \pi^{(n+1)\slash 2}\slash \Gamma\big((n+1)\slash 2 \big)$  and $b_n\equiv \text{vol} \, B^n(1)= \pi^{n\slash 2}\slash \Gamma\big(n\slash 2 +1 \big)$ for the volume of the unit $n$-sphere and $n$-ball, respectively.}
 $\int \md^4 x \sqrt{g}= s_4 L_k^4 = 9 {s_4} \slash \Kkbar_k^2$. Hence
\begin{align}
 \ln \ParFD_k &= \frac{9 {s_4}}{8\pi} \, \frac{1}{G_k \Kkbar_k} = \frac{9{s_4}}{8\pi} \, \frac{1}{g_k \lambda_k}	\label{eqn:05_17}
\end{align}

This is a very intriguing and important result. It shows that the weighted number of modes integrated out between infinity and the IR cutoff $k$ depends only on the properties of the  dimensionless combination of couplings  $\Nk_k \Kkbar_k=g_k \lambda_k$ if we employ the single-metric Einstein-Hilbert truncation. 

Along a RG trajectory of type \Rmnum{3}a, for instance \cite{frank1}, this product decreases from its fixed point value $\lim_{k\rightarrow \infty} g_k \lambda_k= g_* \lambda_*= \Order{1}$ to the infrared value $G_{\text{obs}} \Kkbar_{\text{obs}}$ which is observed at low scales; in real Nature it is of the order $10^{-120}$. 

It is known \cite{frank-cosmo, entropy} that for all type \Rmnum{3}a trajectories admitting a long classical regime there is a huge hierarchy $G_{\text{obs}} \Kkbar_{\text{obs}} \ll g_* \lambda_*$. Hence, as expected, and consistent with \eqref{eqn:05_07AA},
\begin{align}
 \ln \ParFD_{k\rightarrow 0} \gg	\ln \ParFD_{k\rightarrow \infty} \label{eqn:05_18}
\end{align}

Along the hypothetical trajectory realized in Nature \cite{h3, entropy} the Boltzmann weighted number of modes integrated out when the cutoff approaches zero is about $\ln \ParFD_{k\rightarrow 0} \approx 10^{120}$, while the NGFP value $\ln \ParFD_{k\rightarrow \infty}$ is basically zero.

\subsubsection{Bi-metric truncation: flat space with boundary}\label{sec:05_03B}

As a second example we consider the matter induced bi-metric action $\EAA_k= \EAA_k^{\background}+\EAA_k^{\lin}$ of Section \ref{sec:04_01} on a 4D Euclidean spacetime with a non-empty boundary. We consider the associated tadpole equation \eqref{eqn:05_09} justified now and explore its contents. For simplicity we specialize for a regime of the underlying RG trajectory in which the cosmological constant in this equation, the level-(1) coupling $\Lambda_k^{(1)}$, is negligible.  As a result, since $\mathcal{T}_k^{\mu\nu}[A=0;\bg]=0$, there exists a class of special solutions to the coupled gravity + scalar system with
\begin{align}
 \Ricb_{\mu\nu}=0 \, ,\qquad A=0 \qquad (\Lambda_k^{(1)}=0) \label{eqn:05_19}
\end{align}
On the boundary, $\left.  A \right|_{\partial \MaFs}=0$.

The simplest Ricci flat solution is flat space clearly. So let us assume $\MaFs$ is a subset of $R^4$, with $\partial \MaFs\neq \emptyset$, and equipped with a flat metric. Inserting the configuration
\begin{align}
 \bg_{\mu\nu}^{\selfcon}= \delta_{\mu\nu}\, ,\quad  A=0 \, ,  \quad \text{with}\quad \Lambda_k^{(1)}=0 \, ,\label{eqn:05_20}
\end{align}
into $\EAA_k[\flcb=0,A;\bg_{\mu\nu}]=\EAA_k^{\background}[A;\bg_{\mu\nu}]$ given by \eqref{eqn:04_01} with \eqref{eqn:04_02} we obtain
\begin{align}
 -\ln \ParFD_k= \EAA_k[0,0;\delta_{\mu\nu}]&=  \frac{\Lambda_k^{(0)}}{8\pi G_k^{(0)}}\,\text{vol}(\MaFs)+  \frac{\Lambda_k^{(0,\partial)}}{8\pi G_k^{(0,\partial)}}\,\text{vol}(\partial\MaFs) \nonumber \\
& \quad - \frac{1}{8\pi G_k^{(0,\partial)}}\int_{\partial \MaFs}\md^3 x \sqrt{\biM}\,\bEC \label{eqn:05_21}
\end{align}

A perhaps surprising property of this equation is that it involves a non-zero (bulk) cosmological constant term even though it applies to flat space. However, the condition for flat space to be a self-consistent background is that the cosmological constant {\it at level-(1)} is negligible. Its counterpart at level-(0), the one appearing in \eqref{eqn:05_21} may have any value.
Only when split symmetry happens to be intact we have $\Lambda_k^{(0)}=\Lambda_k^{(1)}$ so that the bulk term on the RHS of \eqref{eqn:05_21} indeed vanishes. 
This results in a kind of `holographic' property of the function $\ParFD_k$ which then is expressed by surface terms only. While the Gibbons-Hawking term is the most important contribution, the condition $\Lambda_k^{(0)}=\Lambda_k^{(1)}=0$ still leaves room for a non-zero boundary cosmological constant $\Lambda_k^{(0,\partial)}$, leading to a term proportional to $\text{vol}(\partial\MaFs)$.

In the language of thermodynamics equation \eqref{eqn:05_21} defines a certain `free energy'. The terms proportional to $\Lambda_k^{(0)}\, \text{vol}(\MaFs)$ and $\Lambda_k^{(0,\partial)}\,\text{vol}(\partial\MaFs)$ amount to a homogeneously distributed bulk and surface energy density reminiscent of the volume and surface energy of a liquid droplet. 
In this picture the last term in \eqref{eqn:05_21}, the Gibbons-Hawking contribution, is equally natural and describes how the droplet gains or looses energy by developing a curved or crumpled surface.

What is the statistical mechanics, and what are the pertinent degrees of freedom which underlie this thermodynamics at the microscopic level?

In the context of the effective average action the answer is clear: It is the statistical mechanics of the matter and geometry {\it fluctuations} about their respective backgrounds. The generalized harmonic modes of those fluctuations are `counted' by the partition function $\ParFD_k$ when the IR cutoff $k$ is lowered from infinity to zero. The various running coupling constants contained in $\EAA_k$ and $\ln \ParFD_k$ 
parametrize how the number of fluctuation modes, contributing to the functional integral and weighted with the `Boltzmann factor' $e^{-\widetilde{S}}$,  decreases when we `zoom' deeper and deeper into the microscopic structure of spacetime by increasing $k$.

Another remark is in order at this point. We stress that the occurrence of surface terms in the partition function $\ParFD_k$ on empty flat space is both {\it unavoidable} and {\it natural} from the physics point of view. It is unavoidable because the RG flow generates such terms when $\partial \MaFs \neq\emptyset$. Therefore, contrary to the classical action underlying the variational principle of General Relativity we may not subtract any terms on an ad hoc basis `by hand' from $\EAA_k$.
The surface terms are also natural because the number of field modes counted by $\ParFD_k$ will depend on the shape of $\partial\MaFs$ in general, and this dependence can lead to observable effects, the most famous example being the Casimir effect.

In classical relativity, in order to obtain a finite action for asymptotically flat spacetimes, one usually replaces $\EC \rightarrow \EC-\EC_0$ in the Gibbons-Hawking action $S_{\text{GH}}$. Here $\EC_0$ is the extrinsic curvature of $\partial\MaFs$ when embedded into a {\it flat} spacetime. From the above remarks it should be clear that within the average action approach this procedure would not only be unmotivated but wrong since we might loose essential physics.

To be more concrete about $\MaFs$ let us consider two examples.

\paragraph{(A)} Let us take $\MaFs$ to be a 4-ball, $\MaFs= B^4(L)$, with arbitrary radius $L$. Embedding the boundary $\partial \MaFs=S^3(L)$ into $R^4$ its extrinsic curvature equals $\bEC= 3 \slash L$. As a result,
\begin{align}
 -\ln \ParFD_k &= \frac{b_4}{8\pi}\frac{\Lambda^{(0)}_k L^4}{G_k^{(0)}} + \frac{s_3}{8\pi}\frac{\Lambda^{(0,\partial)}_k L^3}{G_k^{(0,\partial)}}  - \frac{3 s_3}{8\pi}\frac{L^2}{G_k^{(0,\partial)}} \label{eqn:05_22}
\end{align}
Here it is particularly obvious that the asymptotic series for the heat kernel gives rise to a systematic expansion in powers of $1\slash L$.

\paragraph{(B)} The next example is a simple Euclidean caricature of the foliated cylinder type spacetimes one considers in relation with the initial or boundary value problem of Lorentzian gravity. Again we embed $\MaFs$ into $(R^4,\delta_{\mu\nu})$. We fix a foliation of $R^4$ in terms of  flat 3-dimensional hypersurfaces labeled by a parameter $t$ referred to as `Euclidean time'. This gives rise to a corresponding foliation on $\MaFs$.
We consider $\MaFs$ foliated by hypersurfaces $\Sigma_t$, $t\in[t_1,t_2]$ which are bounded by closed 2-surfaces $S_t$.
Thus $\partial\MaFs$ consists of the hypersurfaces $\Sigma_{t_1}$, $\Sigma_{t_2}$, and the union of all $S_t=\partial \Sigma_t$. 

For simplicity we take $\MaFs$ to be the direct product of a 3-ball of radius $\ell$, $B^3(\ell)$, with the time interval. Thus $\Sigma_t=B^3(\ell)$ and $S_t=S^2(\ell)$ for any $t\in[t_1,t_2]$ so that we have $\text{vol}(\MaFs)=b_3 \ell^3 (t_2-t_1)$ and $\text{vol}(\partial \MaFs)=2\, b_3 \ell^3+ s_2 \ell^2 (t_2-t_1)$. The extrinsic curvature of the flat $t=\const$ surfaces in $R^4$ vanishes and so $\Sigma_{t_1}$ and $\Sigma_{t_2}$ do not contribute to the surface integral over $\bEC$. The only contribution comes from the union of all 2-spheres $S_t$. Since the trace of the extrinsic curvature of $S^2(\ell)$ embedded in flat $R^3$ is given by $\EC=2 \slash \ell $ we therefore get $\int \md^3 x \sqrt{\biM}\, \bEC=(t_2-t_1) \cdot (2\slash\ell)\cdot \text{vol}\, S^2(\ell)= 2 s_2 (t_2-t_1) \ell$. 

This brings us to the final result
\begin{align}
 -\ln \ParFD_k&=\left[\frac{\Lambda^{(0)}_k }{6\,G_k^{(0)}}\ell^3 + \frac{\Lambda^{(0,\partial)}_k }{2\,G_k^{(0,\partial)}}\ell^2 - \frac{\ell}{G_k^{(0,\partial)}}\right] (t_2-t_1)+ \frac{\Lambda_k^{(0,\partial)}}{3 G_k^{(0,\partial)}}\ell^3 \label{eqn:05_23}
\end{align}
The coefficient of $(t_2-t_1)$ on the RHS of \eqref{eqn:05_23} may be thought of as a certain energy  associated to the empty flat 3-dimensional space (not `spacetime') interior to a 2-sphere of radius $\ell$.

\subsubsection{Bi-metric truncation: Thermodynamics of the Schwarzschild black hole}

We continue to restrict ourselves to the self-consistent backgrounds of the type \eqref{eqn:05_19}: a vanishing scalar field together with a Ricci-flat metric, $\Ric_{\mu\nu}(\bg)=0$. Perhaps the most prominent representative of this class is the Euclidean Schwarzschild solution $\md s^2= f(r)\md t^2+ f(r)^{-1}\md r^2  + r^2\md \Omega^2$ with
\begin{align}
 f(r)= 1-\frac{R_{{\rm S}}}{r} \label{eqn:05_26}
\end{align}
Here $r\in (R_{{\rm S}},\infty)$, $t\in[0,\beta]$, and time is now compactified to a circle of circumference $\beta\equiv 4\pi R_{{\rm S}}$. 

\paragraph{(A) The constant of integration.} Note that the Schwarzschild radius $R_{{\rm S}}$ has the status of a free constant of integration with the dimension of a {\it length}. It is usually re-expressed in terms of a mass, $M$, so as to recover Newtonian gravity asymptotically. Then $R_{{\rm S}}=2\, G\, M$ where $G$ is the Newton constant of the classical theory. Since in quantum gravity it is not a priori obvious which constant $G_k^{(0)}$, $G_k^{(0,\partial)}$, $G_k^{(1)}$, $\cdots$ should be used to convert $R_{{\rm S}}$ to a mass we shall refrain from doing this and continue to label the family of Schwarzschild metrics $g_{\mu\nu}^{\text{Sch}}$ by the length parameter $R_{{\rm S}}$.

\paragraph{(B) The running on-shell action.}  We consider the Euclidean Schwarzschild manifold $\MaFs$ foliated by 3D hypersurfaces $\Sigma_t$ of constant $t$. They carry the metric $\md s^2_{\Sigma_t}= f(r)^{-1}\md r^2 + r^2 \md \Omega^2$.
With the time compactified and the period chosen as%
\footnote{If $\beta\neq 4\pi R_{{\rm S}}$ there is another boundary at the horizon, however. We shall not consider this generalization here.}
 $\beta=4\pi R_{{\rm S}}$, the only boundary of $\MaFs$ is the union of the asymptotic 2-spheres $\partial \Sigma_t = S^2$ on which $r=\const\equiv \hat{r}$, $\hat{r}\rightarrow \infty$.

Let us insert this background into the truncation ansatz \eqref{eqn:04_01} with \eqref{eqn:04_02} where, by assumption, $\Lambda_k^{(1)}=0$. All that needs to be evaluated is $-\ln \ParFD_k= \EAA_k[\flcb=0,A=0;\bg=g^{\text{Sch}}]=\EAA_k^{\background}[0;g^{\text{Sch}}]$.
Since in the equation \eqref{eqn:04_02} for $\EAA_k^{\background}$ all terms containing $A$, and also the bulk term involving $\SRb$, will vanish we are  left only with the bulk and boundary cosmological constant terms, respectively, together with the extrinsic curvature term:
\begin{align}
 -\frac{2}{16\pi G_k^{(0,\partial)}} \int_{\partial\MaFs}\md^3 x \sqrt{\biM}\, \left(\bEC-\bEC_0\right)- \frac{2}{16\pi G_k^{(0,\partial)}}\int_{\partial\MaFs}\md^3 x \sqrt{\biM}\, \bEC_0 \label{eqn:05_27}
\end{align}
Here we rewrote the extrinsic curvature of $\MaFs$ in $\partial\MaFs$ by adding and subtracting $\bEC_0$, the curvature of $\partial\MaFs$ embedded in flat space, $\bEC\equiv (\bEC-\bEC_0)+\bEC_0$. As a result, the first integral of \eqref{eqn:05_27} is the usual subtracted Gibbons-Hawking term, while the second, to leading order in $R_{{\rm S}}\slash r \rightarrow 0$, becomes independent of the spacetime curvature caused by the black hole. For very large $\hat{r}$  it corresponds to the surface term of flat spacetime considered in Section \ref{sec:05_03B}. When added to the bulk and boundary cosmological constant terms it yields the free energy of flat spacetime. Thus the on-shell average action boils down to
\begin{align}
 -\ln \ParFD_k= - \frac{1}{8\pi G_k^{(0,\partial)}} \int_{\partial\MaFs}\md^3 x \sqrt{\biM}\, \left(\bEC-\bEC_0\right)+ \cdots \label{eqn:05_28}
\end{align}
where the dots stand for the contributions of flat space. 
The evaluation of the integral in \eqref{eqn:05_28} is standard; it yields $\int \md^3 x\sqrt{\biM}\left(\bEC-\bEC_0\right)=-2\pi R_{{\rm S}}\, \beta$ when $\hat{r}\rightarrow \infty$. For the partition function this leads us to
\begin{align}
 -\ln \ParFD_k=  \frac{\beta R_{{\rm S}}}{4 G^{(0,\partial)}}+ \cdots = \frac{\mathcal{A}}{4 G_k^{(0,\partial)}}+ \cdots \label{eqn:05_29}
\end{align}
where $\mathcal{A}\equiv 4\pi R_{{\rm S}}^2$ denotes the area of the event horizon (in the Lorentzian interpretation).

Eq. \eqref{eqn:05_29} is a very instructive formula. Structurally it coincides with the familiar semiclassical Gibbons-Hawking result \cite{BH-Therm}. However, the (unique and truly constant) classical Newton constant appearing there got replaced by a specific member of the various infinite families of running Newton-type couplings which parametrize a general bi-metric average action. Thus we see that {\it the scale dependence of the partition function $\ParFD_k$ and the 
derived thermodynamical quantities  are governed by  the boundary Newton constant at level zero}.

\paragraph{(C) Thermodynamics at finite scale.} 
Note that the Bekenstein-Hawking temperature is scale independent,
\begin{align}
 T=\frac{1}{\beta}=\frac{1}{4\pi R_{{\rm S}} }\, , \label{eqn:05_temp}
\end{align}
while the free energy $F_k\equiv -\beta^{-1} \ln \ParFD_k$ inherits its $k$-dependence from $\Nk_k^{(0,\partial)}$:
\begin{align}
 F_k&= \frac{R_{{\rm S}}}{4 \Nk_k^{(0,\partial)}}= \frac{1}{16\pi \Nk_k^{(0,\partial)}} \frac{1}{T} \label{eqn:05_free}
\end{align}
If we apply the standard relations $U=-T^2 \tfrac{\partial}{\partial T} (F\slash T)$ and $S= - \partial F \slash \partial T$ at every {\it fixed value of $k$} we obtain for the internal energy and entropy, respectively:
\begin{align}
 U_k&=\frac{R_{{\rm S}}}{2 \Nk_k^{(0,\partial)}}= \frac{1}{8\pi \Nk_k^{(0,\partial)}}\frac{1}{T}=2 F_k \label{eqn:05_int} \\
S_k &= \pi \frac{R_{{\rm S}}^2}{\Nk_k^{(0,\partial)}}= \frac{\mathcal{A}}{4 \Nk_k^{(0,\partial)}} = \frac{1}{16\pi \Nk_k^{(0,\partial)}}\frac{1}{T^2} \label{eqn:05_entro}
\end{align}
Likewise $C=\partial U \slash \partial T$ yields the specific heat capacity
\begin{align}
 C_k&= -2\pi \frac{R_{{\rm S}}^2}{\Nk_k^{(0,\partial)}}= - \frac{1}{8\pi\Nk_k^{(0,\partial)}}\frac{1}{T^2} \label{eqn:05_capac}
\end{align}
Obviously the RG running of all thermodynamical functions of interest is governed by a single running coupling, namely $1\slash \Nk_k^{(0,\partial)}$.

\paragraph{(D) Running ADM mass.} 
Up to now we never ascribed any mass to the Schwarzschild spacetime.  As we mentioned already, it is more natural to characterize it by a length such as $R_{{\rm S}}$. Its conversion  to a mass is a matter of convention, strictly speaking, which in quantum gravity becomes particularly ambiguous.
Nevertheless, our result \eqref{eqn:05_29} suggests that a natural way of relating a mass to a black hole with a given parameter $R_{{\rm S}}$ is by means of $G_k^{(0,\partial)}$:
\begin{align}
 M_k\equiv \frac{R_{{\rm S}}}{2\,G_k^{(0,\partial)}} \label{eqn:05_30}
\end{align}
We emphasize again that $R_{{\rm S}}$ has no $k$-dependence; it labels different solutions of the truncated tadpole equation, $\Ricb_{\mu\nu}=0$, which happens to be independent of any running coupling. So the $k$-dependence of the running mass $M_k$ is entirely due to $G_k^{(0,\partial)}$. It can be seen as a {\it scale dependent generalization of the classical ADM mass}.\footnote{In more general backgrounds it becomes $M_k=-(8\pi G_k^{(0,\partial)})^{-1}\oint (\EC-\EC_0)$ where the integral is over an asymptotic sphere.}
The definition  \eqref{eqn:05_30} is motivated by observing that all relations of semiclassical black hole thermodynamics  retain their form when quantum gravity effects are included via the average action {\it provided we replace the classical mass by the running mass $M_k$.} The mass \eqref{eqn:05_30} controls the partition function of a single black hole,
\begin{align}
 -\ln \ParFD_k=\,\, 4\pi G^{(0,\partial)}_k M_k^2= \,\, \frac{1}{2}\,\beta\,M_k\,, \label{eqn:05_31}
\end{align}
and the ensuing thermodynamical relations
\begin{align}
 \frEny_k=\frac{1}{2}M_k\, \qquad U_k = M_k \, , \qquad S_k=4\pi \Nk_k^{(0,\partial)}M_k^2\, , \label{eqn:05_31B}
\end{align}
look like their semiclassical counterparts with the replacement $M\rightarrow M_k$. In its other r\^{o}les, the classical mass $M$ might possibly get replaced by running masses with a different $k$-dependence.

\paragraph{(E) Explicit \boldmath{$k$}-dependence of the boundary Newton constant.} 
Let us see now what we obtain for $M_k$ from the matter induced beta-functions. The running of $\Nk_k^{(0,\partial)}$ is governed by the anomalous dimension $\eta^{(0,\partial)}$ in eq. \eqref{eqn:04_09b}. For simplicity we neglect the scalar mass here and set $m_k^{(0)}\equiv 0$. Then \eqref{eqn:04_09} becomes
\begin{align}
 \eta^{(0,\partial)}&= - (d-2)\, \omega_d^{(0,\partial)} \tg^{(0,\partial)} \label{eqn:05_32}
\end{align}
with the crucial coefficient
\begin{align}
 \omega_d^{(0,\partial)}&=- \frac{2\,\ns}{3(d-2)\,(4\pi)^{\frac{d}{2}-1}}\,
 \ThrfA{1}{d \slash 2 - 1}{0}  <0 \label{eqn:05_33} 
\end{align}
The anomalous dimension \eqref{eqn:05_32} has the same general structure as  its single-metric analogue in eq. \eqref{eqn:03_19B}. The latter equation contains a coefficient $\omega_d^{\partial}$ which was found to be negative in Section \ref{sec:03}. In the matter induced bi-metric context the corresponding quantity is negative, too: $\omega_d^{(0,\partial)}<0$. 
This leads us to the following exact solution for the RG equation $\partial_t \left(1 \slash \Nk_k^{(0,\partial)}\right)=-\eta^{(0,\partial)} \left( 1 \slash \Nk_k^{(0,\partial)}\right)$, in 4 dimensions:
\begin{align}
 \frac{1}{\Nk_k^{(0,\partial)}}=\frac{1}{\Nk_0^{(0,\partial)}} + \omega_4^{(0,\partial)}k^2 \label{eqn:05_34}
\end{align}
Remarkably, this is the same result as eq. \eqref{eqn:03_23B} obtained with the single-metric truncation of full QEG. This fairly robust prediction for the behavior of $1\slash \Nk_k^{(0,\partial)}$ is sketched in Fig \ref{fig:parabola}.

\paragraph{(F) Explicit \boldmath{$k$}-dependence of the ADM mass.} 
Using \eqref{eqn:05_34} in \eqref{eqn:05_30} we associate the following running mass to the black hole with Schwarzschild radius $R_{{\rm S}}$:
\begin{align}
 M_k&= \left[1 + \omega_4^{(0,\partial)}\Nk_0^{(0,\partial)}k^2\right] M_0 \qquad \text{where}\qquad M_0\equiv \frac{R_{{\rm S}}}{2\Nk_0^{(0,\partial)}} \label{eqn:05_35}
\end{align}
Let us match the surface and bulk Newton constants at $k=0$ and identify this quantity with the standard Newton constant, $\Nk_0^{(0,\partial)}=\Nk_0^{(0)}\equiv \Nk \equiv m_{\text{Pl}}^{-2}$.
Then we recover the ordinary relationship $M_0=R_{{\rm S}}\slash (2 \Nk)$ in the extreme infrared (at $k=0$), but at higher scales the mass associated to the very same geometry is {\it smaller} than $M_0$:
\begin{align}
 M_k&= \left[1 - |\omega_4^{(0,\partial)}| \left(\frac{k}{m_{\text{Pl}}}\right)^2\right] M_0  \label{eqn:05_36}
\end{align}
In writing down \eqref{eqn:05_36} we made it manifest that the coefficient $\omega_4^{(0,\partial)}$ turned out negative. As a consequence, {\it  $M_k$ decreases for increasing $k$, reaches zero at a scale near $k=m_{\text{Pl}}$, and becomes negative for even larger $k$-values.}

An attempt at interpreting this behavior could be as follows. It is known that in QEG the bulk Newton constant $\Nk_k$ decreases for increasing $k$, and this was interpreted as an indication of  gravitational anti-screening due to the energy and momentum of the virtual particles surrounding every massive body; because of the attractivity of gravity, they are pulled towards this body, adding positively to its bare mass, whence the virtual cloud leads to an effective mass that increases with increasing distance \cite{mr}.

Now we have seen that the boundary Newton constant $\Nk_k^{\partial}$, or $\Nk_k^{(0,\partial)}$, {\it increases} for increasing $k$.
Interestingly enough, what at first sight might seem to contradict the picture of gravitational anti-screening, in view of $M_k=\frac{1}{2} R_{{\rm S}} \slash \Nk_k^{(0,\partial)}$, at least heuristically, actually {\it confirms} it: According to this definition of `mass', {\it the running mass of any material body decreases with increasing $k$, or decreasing distance.} Somewhere near $k=m_{\text{Pl}}$ it even seems to vanish, indicating probably that in this regime a more elaborate treatment is necessary.

\begin{figure}[htbp]
\begin{center}
\psfrag{tagk}{$k$}
\psfrag{tagA0}[bc]{$0$}
\psfrag{tagB0}[tl]{$0$}
\psfrag{tagC}{$C_k$}
\psfrag{tagG}{$F_k, U_k, S_k$}
\subfigure{\includegraphics[width=0.35\textwidth]{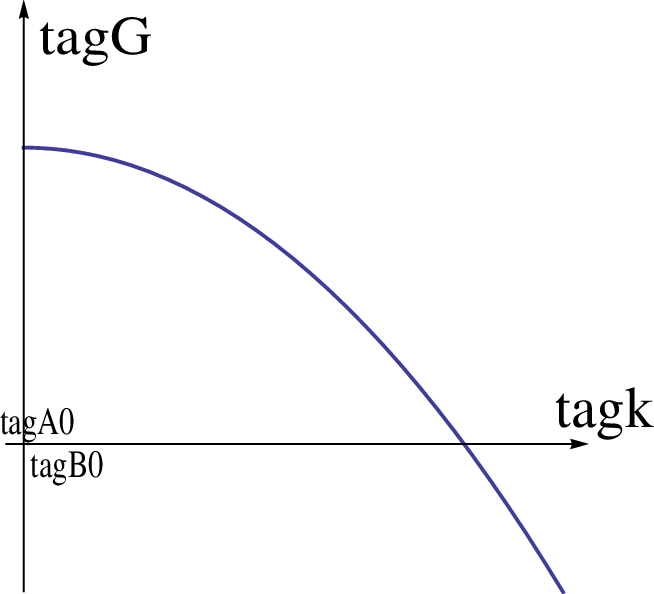}}\hspace{0.1\textwidth}
\subfigure{\includegraphics[width=0.35\textwidth]{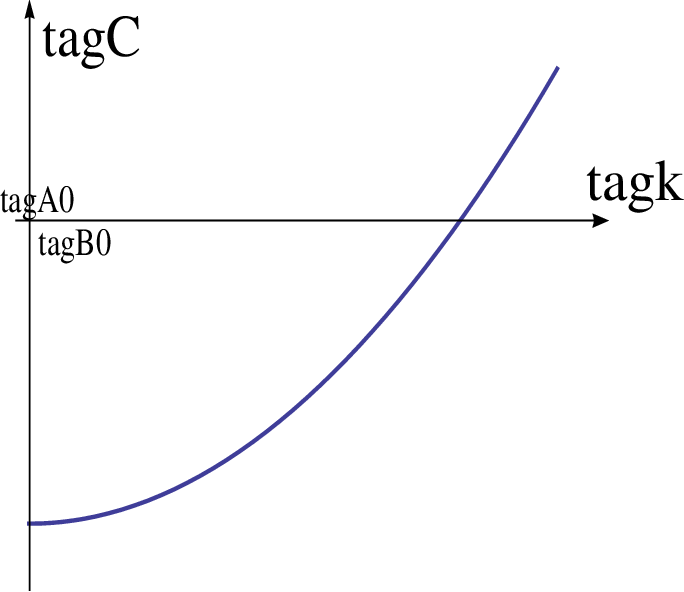}}
\caption{Scale dependence of free energy, internal energy, and entropy (left panel), as well as the specific heat capacity (right panel) of the Schwarzschild black hole.}\label{fig:thermo}
\end{center}
\end{figure}

\paragraph{(G) Running thermodynamic quantities.} 
The running free and internal energy, the entropy and the specific heat capacity are governed by the same function of $k$ as $M_k$ in \eqref{eqn:05_36}:
$ F_k , \, U_k ,\, S_k ,\, C_k\, \propto \left[1 - |\omega_4^{(0,\partial)}| (k \slash  m_{\text{Pl}})^2 \right]$.
Note, however, that the specific heat has a negative IR value, $C_0=-2\pi R_{{\rm S}}^2\slash G_0^{(0,\partial)}$, and switches its sign in the opposite direction, from negative to positive, at the zero in the Planck regime, see Fig. \ref{fig:thermo}.

\paragraph{(H) A natural cutoff identification.} 
Even though the shortcut to extracting physical information from the running couplings in $\EAA_k$ by identifying $k$ with some physical scale (`RG improvement') is notoriously ambiguous in general, it is clear that the black hole spacetime specified by a given $R_{{\rm S}}$ has a distinguished intrinsic mass scale associated to it that does not rely on any artificial conversion factor, namely $1 \slash R_{{\rm S}}$, or the temperature $T=\left(4\pi R_{{\rm S}}\right)^{-1}$. 

If we tentatively adopt the cutoff identification $k\approx 1\slash R_{{\rm S}}$, and go from macroscopic astrophysical black holes to microscopic ones with a Planckian Schwarzschild radius, we find that $M_k$ decreases monotonically, heading for $M_k=0$ near $R_{{\rm S}}=\ell_{\text{Pl}}$. Remarkably, by eqs. \eqref{eqn:05_31}, the thermodynamical quantities, the entropy in particular, all vanish in this limit: $F_k \rightarrow 0$, $U_k\rightarrow 0$, $S_k\rightarrow 0$  for  $R_{{\rm S}}\searrow \ell_{\text{Pl}} $.

It is particularly intriguing that {\it the specific heat capacity $C_k$ changes its sign from negative to positive near the Planck scale.} This suggests that near $\ell_{\text{Pl}}$ the notorious instability of classical gravity possibly gets tamed in a dynamical way: The system no longer can lower its energy by accreting further mass and the gravitational collapse might come to a halt.

This picture based on the boundary Newton constant is surprisingly similar to what we found in \cite{bh,erick2,erick1} by a rather different reasoning, namely the `RG improvement' of the {\it bulk} Newton constant in the classical formula $f(r)=1-2GM \slash r$. Keeping $M$ fixed we replaced $G\rightarrow \Nk_k$ and identified $1\slash k$ with the radial proper distance.

Future work will have to clarify the precise relationship between the two treatments, in particular whether they are different pictures of the same phenomenon or should be superimposed rather \cite{daniel-prep}.

\section{Conclusion}
In this paper we studied functional RG flows of quantum gravity on spacetime mani\-folds with boundary. We considered two specific truncations which, as a new feature, contained various surface terms such as the Gibbons-Hawking term for instance. The first example was the single-metric Einstein-Hilbert  truncation of full fledged (pure) gravity, the second an induced gravity scenario based on quantized scalar fields. The latter example is simple enough to be treated in a bi-metric truncation which helped in obtaining a conceptually clear picture. We derived and analyzed the beta-functions describing its RG flow on a 17-dimensional theory space. The discussion focused on the breaking of the background-quantum field  split symmetry, and on the violation of the bulk-boundary matching among the various Newton-like constants. We found that the two phenomena are intertwined in a complicated way.
Leaving the structural generalization of the FRGE which is needed for arbitrary (untruncated) action functionals to future work, we justified the variational procedure for the second derivative actions considered here.

For a proper interpretation of the surface terms and the variational principle it was crucial to take the bi-metric character of gravitational average action into account. In an expansion with respect to $\flcb_{\mu\nu}$ a generic functional $\EAA_k$ contains `towers' of bulk Newton constants $\Nk_k^{(p)}$, boundary Newton constants $\Nk_k^{(0,\partial)}$, and many more similar couplings whereby the `level' $p=0,1,2,\cdots$ is indicative of the $\flcb_{\mu\nu}$-power in the corresponding field monomial. Since the background-quantum split symmetry is broken by the IR cutoff, the different levels evolve independently under the RG flow.

A key observation in this context is the following. The partial differential equation which determines self-consistent backgrounds, that is, backgrounds which once prepared by external means are not modified by the intrinsic quantum fluctuations, involve only the couplings of {\it level-(1).}
The (thermodynamical, etc.) properties of the backgrounds they imply are determined by the {\it level-(0)} couplings in addition. In the example considered, in fact  {\it only} those of level-(0) happened to be relevant. 
It is therefore  possible to have a bulk-boundary matching of Newton's constant at level-(1), $\Nk_k^{(1)}=\Nk_k^{(1,\partial)}$, hence a standard variational principle, but nevertheless a mismatch at level zero: $\Nk_k^{(0)}\neq \Nk_k^{(0,\partial)}$. This mismatch can encode important information relevant to the effective field theory description of physics at finite scales $k$; black hole thermodynamics turned out to be a prime example.

In both systems we analyzed, full fledged QEG in the single-metric Einstein-Hilbert truncation, and the scalar-induced bi-metric truncation, we found that {\it along all RG trajectories, at every scale $k$, the surface Newton constant is an increasing function of $k$}. It is small in the IR and becomes large in the UV. This behavior is markedly different from the behavior of the bulk Newton constant which, at least in pure QEG, was found to {\it decrease} in the UV \cite{mr}. The resulting weakening of the gravitational interaction at high momentum scales has been interpreted a kind of gravitational {\it anti-screening} due to the attraction of virtual excitations.
Using $\Nk_k^{(0,\partial)}$ in order to define a scale dependent analogue of the ADM mass, $M_k$, we argued that it is precisely this opposite running of the boundary Newton constant that supports the picture of gravitational anti-screening: Both $M_k\propto 1 \slash \Nk_k^{(0,\partial)}$ and the interaction strength given by the bulk Newton constant decrease in the ultraviolet. Consistent with that we found a non-Gaussian fixed point on the full 17-dimensional theory space suitable for the Asymptotic Safety construction.

In (semi-)classical black hole physics there exists only a single mass parameter, $M$, and this parameter plays various conceptually rather different r\^{o}les. It controls, for instance, the semiclassical partition function of a single black hole but it also describes the strength of the Newtonian force between two black holes at large distances, say. In quantum gravity, in the context of the average action, there does not exist a single running mass  which serves all these purposes at a time. Likewise, there is not a single, but actually quite many different running Newton constants. Each of them takes over one specific r\^{o}le played by the standard Newton constant, or the classical concept of mass, respectively, and depending on this r\^{o}le its $k$-dependence is different in general. We saw that the partition function and the related thermodynamics of an isolated black hole is governed by $M_k\propto 1\slash \Nk_k^{(0,\partial)}$. The interaction of two bodies, scalar $A$-particles, say, is governed by another member of the `$\Nk_k$ zoo', however. Within the effective field theory description, the one graviton exchange between these bodies is governed by the effective Einstein equation \eqref{eqn:05_10}, and so it is the level-(1) bulk coupling $\Nk_k^{(1)}$ that controls the quantum gravity effects.

Based upon the concept of self-consistent backgrounds, solutions of the running tadpole equation, we proposed and tested a natural device counting the number of field modes integrated out by the average action. It is a partition function which is associated to a given RG trajectory and  the background picked, and which describes the statistical mechanics of the metric and matter fluctuations relative to this background. While this mode count is of interest also on spacetimes without boundary, here it led us to an intriguing scale dependent generalization of black hole thermodynamics which represents the physical basis and motivation for the specific definition of $M_k$. We shall come back to a more phenomenological analysis of the corresponding quantum effects in black hole spacetimes elsewhere \cite{daniel-prep}.

\paragraph{Acknowledgment:} We are grateful to Professor Abhay Ashtekar for helpful and inspiring discussions.

\clearpage
\begin{appendix}
\section*{Appendix}
\section{Matrices, operators, and traces}
In particular when dealing with bi-metric truncations it is crucial to distinguish the two metrics $g_{\mu\nu}$ and $\bg_{\mu\nu}$, and to keep track of which one is used to raise and lower indices, or to convert tensors to tensor densities, and vice versa, by means of the corresponding volume element.
Therefore we compile in this appendix various definitions and notations that are particularly relevant in this respect.

We consider the example of a quantized scalar $A(x)$ (or, under the functional integral, $\hat{A}(x)$) which interacts with the dynamical metric $g_{\mu\nu}(x)$, kept classical, but is coarse grained by means of 
\begin{align}
 \Delta_k S[A;\bg]&= \frac{1}{2} \int \md^d x \sqrt{\bg}\, A(x)\, \RegA_k[\bg]\, A(x) \label{eqn:A_01}
\end{align}
which contains the background metric $\bg_{\mu\nu}$. In this respect, the scalar is a prototype of any dynamical (i.e., non-background) field appearing in full-fledged quantum gravity. Correspondingly, its generating functional is given by an abridged version of the QEG functional integral:
\begin{align}
 e^{\CGF_k[J,g,\bg]}&\equiv \int \mathcal{D} \hat{A}\, e^{-S[\hat{A},g]-\Delta_k S[\hat{A};\bg]} \,\exp\left\{\int \md^d x\sqrt{\bg}\,\, J(x)\hat{A}(x)\right\} \label{eqn:A_02}
\end{align}
As a general rule \cite{mr}, all source functions, by definition, transform as tensors (rather than tensor densities) under diffeomorphisms, and the integrand of the source coupling term is made a scalar density by means of the volume element of the {\it background} metric.
As a consequence, the field expectation value $A(x)\equiv \langle \hat{A}(x)\rangle$ and the connected 2-point function
\begin{align}
G(x,y)&\equiv \langle \hat{A}(x)\hat{A}(y)\rangle-\langle \hat{A}(x)\rangle\langle  \hat{A}(y)\rangle \label{eqn:A_03}
\end{align}
are given by functional derivatives involving explicit factors of $\sqrt{\bg}$:
\begin{align}
&A(x)= \frac{1}{\sqrt{\bg(x)}} \, \frac{\var \CGF_k[J,g,\bg]}{\var J(x)} \label{eqn:A_04} \\
&G(x,y)= \frac{1}{\sqrt{\bg(x)}}\frac{1}{\sqrt{\bg(y)}} \, \frac{\var^2 \CGF_k[J,g,\bg]}{\var J(x)\var J(y)} \label{eqn:A_05}
\end{align}

Upon inverting the field-source relationship \eqref{eqn:A_04} we Legendre-transform $\CGF_k$ with respect to $J$, at fixed $g$ and $\bg$, arriving at the functional $\widetilde{\EAA}_k[A,g,\bg]$, with the property
\begin{align}
\frac{1}{\sqrt{\bg(x)}} \, \frac{\var \widetilde{\EAA}_k[A,g,\bg]}{\var A(x)}=J(x) \label{eqn:A_06}
\end{align}
so that the actual effective average action writes simply $\EAA_k[A,g,\bg]=\widetilde{\EAA}_k[A,g,\bg]-\Delta_k S[A,\bg]$.

It is often convenient to employ a symbolic bra-ket formalism in which functions such as $A$ are represented by a ket vector $| A\rangle$ and the 2-point function by a matrix, which is considered the matrix representation (in the position eigenbasis $| x\rangle$) of an abstract operator:
\begin{align}
 &\langle x | A\rangle \equiv A_x \equiv A(x) \label{eqn:A_07}\\
&\langle x | G| y\rangle \equiv G_{xy} \equiv G(x,y) \label{eqn:A_08}
\end{align}
Matrix multiplication is defined by 
\begin{align}
\langle x | V\, W| y\rangle &= \int \md^d z \sqrt{\bg(z)}\, \,\langle x | V| z\rangle \, \langle z | W| y\rangle \label{eqn:A_09}
\end{align}
or, in a more compact notation,
\begin{align}
 \left(V\,W\right)_{xy} &= \int \md^d z\sqrt{\bg(z)}\,\,V_{xz} \,W_{zy} \label{eqn:A_10}
\end{align}
In our conventions, vector components ($A_x,\, \cdots$) and matrix elements ($V_{xy},\, \cdots$) are always genuine tensors under coordinate transformations, but all integrations and functional differentiations come with explicit factors of $\sqrt{\bg}$. Assuming it exists, the trace of an operator writes
\begin{align}
 \Tr\left(V\right)&=\int \md^d x \sqrt{\bg(x)}\,\, \langle x| V| x\rangle \equiv \int\md^d x \sqrt{\bg(x)}\,\, V_{xx} \label{eqn:A_11}
\end{align}
The unit operator $\Id$, satisfying $V\Id=\Id V = V$, has the matrix elements
\begin{align}
 \Id_{xy}\equiv \langle x| \Id | y\rangle \equiv \langle x| y\rangle \equiv \frac{\delta(x-y)}{\sqrt{\bg(x)}}\, \label{eqn:A_12}
\end{align}
while the completeness relation of the basis vectors reads
\begin{align}
 \int \md^d x \sqrt{\bg(x)} \, \, | x\rangle\langle x| = \Id \label{eqn:A_13}
\end{align}

The matrix elements of the operators $G$ and $\widetilde{\EAA}_k^{(2)}$, the latter defined by
\begin{align}
 \langle x | \widetilde{\EAA}_k^{(2)} | y\rangle \equiv \left(\widetilde{\EAA}_k^{(2)}\right)_{xy}\equiv \frac{1}{\sqrt{\bg(x)}\sqrt{\bg(y)}}\,\, \frac{\var^2 \widetilde{\EAA}_k[A,g,\bg]}{\var A(x)\var A(y)}\, , \label{eqn:A_14}
\end{align}
being the second derivatives of functionals related by a Legendre transformation, satisfy
\begin{align}
 \int \md^d y \sqrt{\bg(y)}\,\, \langle x | G| y\rangle \, \langle y | \widetilde{\EAA}_k^{(2)}| z\rangle = \frac{\delta(x-z)}{\sqrt{\bg(z)}} \label{eqn:A_15}
\end{align}
In operator notation this equation has the standard appearance $G \widetilde{\EAA}_k^{(2)} = \Id$.
In fact, the FRGE given in equation \eqref{eqn:03_05} of the main text is written down using these rules. In particular $\EAA_k^{(2)}$ is defined by \eqref{eqn:A_14} with $\widetilde{\EAA}_k$ replaced by $\EAA_k$. 

The rules we employ have the advantage of giving a simple appearance to the FRGE, but it is very important to keep in mind that the underlying calculus `hides' certain dependencies on the background metric.

For more general sets of dynamical fields, a few more rules must be observed if we want the FRGE to keep this simple form.
Consider for instance the case that $A\equiv \left(A_M\right)$ carries an index, $M$, acted upon by some arbitrary spacetime and \slash or internal transformation group. Then the Hessian $\var^2 \EAA_k \slash \var A_M(x) \var A_N(y)$, even with the factors of $\sqrt{\bg}$ added, merely defines a {\it quadratic form}, an integral kernel, {\it but not an operator}. 

To obtain an operator we need an isomorphism which relates, at least formally, vectors `$| x \rangle^M$' to dual vectors `$_M\langle x |$'. In the standard applications of the FRGE this isomorphism is provided by a {\it metric in the space of fields} which can be used to `pull down' one of the two $M$, $N$ indices. Usually this metric is taken ultra-local so that it boils down to a field $G_{MN}(x)$ on spacetime, and the appropriate generalization of \eqref{eqn:A_14} reads
\begin{align}
 _M\langle x | \EAA_k^{(2)} | y\rangle^N &= \frac{G_{MK}(x)}{\sqrt{\bg(x)}\sqrt{\bg(y)}}\,\,\frac{\var^2 \EAA_k[A,g,\bg]}{\var A_K(x) \var A_N(y)} \label{eqn:A_16}
\end{align}
An example of \eqref{eqn:A_16} is equation \eqref{eqn:03_06} in the main text where $M\hat{=}(\mu,\nu)$ is a pair of spacetime indices, and the ultra-local metric in field space is induced by the spacetime metric or appropriate tensor products thereof: $G_{MK}\hat{=}\bg_{\mu\rho}\bg_{\nu\sigma}$. We emphasize that also this r\^{o}le is played by the {\it background} rather than the dynamical metric.

This is indeed the general rule whenever the quantum field $A$ is a spacetime tensor:
The {\it operator} $\EAA^{(2)}_k$ is obtained from the {\it quadratic form} (second functional derivative of $\EAA_k$) by pulling indices up and down with suitable products of $\bg_{\mu\nu}$'s.

If $M$ is an internal index, $G_{MN}$ is unrelated to any spacetime metric, and the usual FRGE can be set up only after a certain tensor $G_{MN}$ has been specified as an additional, externally provided input.
An example is the $\On{n}$ symmetric scalar field theory studied in the present paper where we adopt the $\On{n}$ invariant choice $G_{MN}=\delta_{MN}$.

Sometimes it is convenient to represent an operator $W$ as a {\it differential operator} $W^{\text{diffop}}$ rather than by its matrix elements $\langle x| W| y\rangle$. The action of $W$ on a vector $| A \rangle$ with $\langle x| A\rangle= A(x)$ can be written as
\begin{align}
 \langle x | W | A\rangle = \int \md^d y \sqrt{\bg(y)} \, \, \langle x | W | y\rangle \langle y | A \rangle \equiv \left(W^{\text{diffop}}_x A\right)(x) \label{eqn:A_17}
\end{align}
whereby the second equality defines a (pseudo) differential operator acting on the argument $x$ of the function $A$. 
We are mostly interested in operators of the type
\begin{align}
 \langle x | W | y\rangle = F \left(\bZ_{\mu}^{(y)}\right) \, \frac{\delta(x-y)}{\sqrt{\bg(y)}} \label{eqn:A_18}
\end{align}
where $F$ is any function of the background covariant derivative $\bZ_{\mu}^{(y)}$ acting on $y$. In this case equation \eqref{eqn:A_17} yields, after an integration by parts and taking advantage of the metricity condition $\bZ_{\mu}\bg_{\alpha\beta}=0$,
\begin{align}
W^{\text{diffop}}_x =F\left(\bZ_{\mu}^{(x)}\right) \label{eqn:A_19}
\end{align}
Usually we omit the superscript `diffop' when no confusion can arise.

 \section{Beta-functions for the matter induced \\bi-metric truncation}\label{sec:B}
In this appendix we sketch the  derivation of the beta-functions for the matter induced bi-metric truncation considered in Section  \ref{sec:04}. 
Essentially, the evaluation of the $\RHS$ of the flow equation 
\begin{align}
\partial_t \EAA_k[\flcb,A;\bg] &= \frac{1}{2}\Tr\left[\partial_t R_k[\bg] \left(\EAA_k^{(2)}[\flcb,A;\bg]+R_k[\bg]\right)_{AA}^{-1}\right] \label{eqn:B_01}
\end{align}
proceeds along similar lines as in ref. \cite{MRS1}. Differences arise as a consequence of the subtleties due to the different topology of the spacetime manifold, i.e. a non-vanishing boundary $\partial\MaFs$, and the presence of non-minimal coupling terms $\xi \SRb$. 

The inverse operator on the $\RHS$ of \eqref{eqn:B_01} is obtained by a second  functional derivative with respect to the scalar fields $A$ only, since in the truncation ansatz considered the gravitational fluctuation $\flcb_{\mu\nu}$ contributes just up to linear order to the effective average action, $\EAA_k[\flcb,A;\bg]=\EAA^{\background}_k[A;\bg]+\EAA^{\lin}_k[\flcb,A;\bg]$. We omit the mixed contributions to the Hessian to focus on the induced gravity effects due to the matter sector alone. 

The $\RHS$ of the FRGE, eq. \eqref{eqn:B_01} has to be projected onto the subspace of theory space spanned by the monomials of the truncation ansatz \eqref{eqn:04_02} - \eqref{eqn:04_05}. Equating the coefficients of the basis monomials on the $\LHS$ and $\RHS$ of the flow equation then yields the set of beta-functions describing the running of the, in general, dimensionful couplings.

\subsection{The fluctuation expansion }
On the $\RHS$ of \eqref{eqn:B_01} the inverse $(\cdots)^{-1}$ generates arbitrarily high powers of $\flcb_{\mu\nu}$ which however have no counterparts on the $\LHS$. 
It is sufficient to expand the $\RHS$ of \eqref{eqn:B_01} to first order in the fluctuation field $\flcb_{\mu\nu}$. Hereby we repeatedly exploit that we assume Dirichlet boundary conditions $\left.\flcb_{\mu\nu}\right|_{\partial \MaFs}=0$ to hold.
The formal expansion as a Taylor series in $\flcb_{\mu\nu}$,  
$\RHS= \left. \RHS\right|_{g=\bg} +\int  \left. \fD{\RHS}{g_{\mu\nu}}\right|_{g=\bg} \flcb_{\mu\nu} + \Order{\flcb^2} $, can be explicitly written in terms of the `background plus linear part decomposition' of the effective average action, $\EAA_k[\flcb,A;\bg]=\EAA^{\background}_k[A;\bg]+\EAA^{\lin}_k[\flcb,A;\bg]$, as follows:
\begin{align}
&\partial_t \EAA_k^{\background}[A;\bg]+ \partial_t \EAA_k^{\lin}[\flcb,A;\bg]= \frac{1}{2}\Tr\left[\partial_t R_k[\bg]\,\left(\EAA_k^{\background\,(2)}[A;\bg]+R_k[\bg]\right)^{-1}\right] \label{eqn:B_02} \\
& \qquad \qquad \qquad \quad -\frac{1}{2}\Tr\left[\partial_t R_k[\bg]\,\left(\EAA_k^{\background\,(2)}[A;\bg]+R_k[\bg]\right)^{-2} \, \EAA_k^{\lin\,(2)}[\flcb,A;\bg] \right]  + \Order{\partial^4,\flcb^2} \nonumber
\end{align}
The inverse operator under the traces is completely determined by $\EAA_k^{\background\, (2)}[A;\bg]$ and the cutoff operator. The couplings of  level-$(1)$ enter the $\RHS$ of equation \eqref{eqn:B_02} solely by its second term. 

\subsection{The Hessian operator}
Due to Dirichlet conditions for the field $A$, while not vanishing but kept fixed on the boundary, i.e. $\left. A\right|_{\partial\MaFs}=\AS$, all potential surface contributions vanish in the second functional derivative of $\EAA_k$ and we can easily extract the associated Hessian operator which contains operators of the bulk sector only. 
Considering the level-$(0)$ part of the action, denoted $\EAA^{\background}_k[A;\bg]$, the operator associated to its Hessian  is given by
\begin{align}
\EAA_k^{\background\,(2)}[A;\bg]&=  -\bZ^2  +\xiA \SRb  + \PotAd{(0)\prime\prime}  \label{eqn:B_03}
\end{align}
The remaining ingredient stems from the second functional derivative, with respect to $A$, of the level-$(1)$ sector, i.e.  $\EAA_k^{\lin\,(2)}[\flcb,A;\bg]$.  Again all boundary terms vanish due to Dirichlet conditions for either $\var A$ or $\flcb_{\mu\nu}$ and the associated operator assumes the form:
\begin{align}
\EAA_k^{\lin\,(2)}[\flcb,A;\bg]&=\frac{1}{2}\left(\xiBg \SRb + \PotAd{(1)\prime\prime}\right)\flcb^{\mu}_{\phantom{\mu}\mu}
-\xiBR \Ricb^{\mu\nu}\,\flcb_{\mu\nu}
\nonumber \\
& \quad  - \left(\var\Z^2+\frac{1}{2} \flcb^{\mu}_{\phantom{\mu}\mu}\bZ^2\right) \label{eqn:B_04}
\end{align}
Notice that neither $\var\Z^2$ nor $\tfrac{1}{2} \flcb^{\mu}_{\phantom{\mu}\mu}\bZ^2$ is Hermitian with respect to the scalar product $(\psi_1,\psi_2)=\int_{\MaFs}\md^d x \sqrt{\bg}\, \psi_1 \psi_2$ but the combination in the last line of \eqref{eqn:B_04} is, however.
Though we could make the variation $\var \Z^2$ explicit, we keep this more compact form which will be useful in the evaluation of the trace via heat kernel techniques later on. 

\subsection{Expansion in the number of derivatives}
Going back to equation \eqref{eqn:B_03} we see that the scalar curvature $\SRb$ is part of the Hessian $\EAA^{\background\, (2)}[A;\bg]$. Since it appears in the inverses under the trace it can produce arbitrarily high orders in $\SRb$. The truncated theory space under considerations is spanned by monomials of at most linear order in $\SRb$. Hence all relevant invariants are still covered after expanding the inverse operator in the following truncated Taylor series  in $\SRb$:
\begin{align}
\left(\EAA_k^{\background\,(2)}[A;\bg]+R_k[\bg]\right)^{-1}&  = \left(-\bZ^2 + \PotAd{(0)\,\prime\prime} +R_k[\bg]\right)^{-1} \nonumber\\ 
& \quad  - \left(-\bZ^2 + \PotAd{(0)\,\prime\prime} +R_k[\bg]\right)^{-2} \, \xiA\, \SRb + \Order{\SRb^2} \label{eqn:B_05}
\end{align}
Notice that the first term of the expansion yields contributions starting with zeroth order in $\partial^2$, i.e. no derivatives acting on $\bg_{\mu\nu}$, whereas the second part of equation \eqref{eqn:B_05} is at least of order $\partial^2$, because $\SRb$ contains two derivatives acting on $\bg_{\mu\nu}$. 
We substitute this expansion into the $\RHS$ of equation \eqref{eqn:B_02} and neglect all terms leading to higher derivatives. Thus, we obtain:
\begin{subequations} \label{eqn:B_05B}
\begin{align}
\partial_t \EAA_k[\flcb,A;\bg]&= +\frac{1}{2}\Tr\left[\left. \partial_t R_k[\bg]\left(\EAA_k^{\background\,(2)}[A;\bg]+R_k[\bg]\right)^{-1}\right|_{\SRb=0}\right] \\
&\quad  -\frac{1}{2}\Tr\left[\left.\partial_t R_k[\bg] \left(\EAA_k^{\background\,(2)}[A;\bg]+R_k[\bg]\right)^{-2}\right|_{\SRb=0} \, \xiA \SRb\right] 
\\
& \quad -\frac{1}{2}\Tr\left[\left.\partial_t R_k[\bg] \left(\EAA_k^{\background\,(2)}[A;\bg]+R_k[\bg]\right)^{-2}\right|_{\Ricb^{\mu\nu}=0}\, \EAA_k^{\lin\,(2)}[\flcb,A;\bg] \right] \\
& \quad +\,\Tr\left[\left.\partial_t R_k[\bg] \left(\EAA_k^{\background\,(2)}[A;\bg]+R_k[\bg]\right)^{-3}\, \EAA_k^{\lin\,(2)}[\flcb,A;\bg] \right|_{\Ricb^{\mu\nu}=0} \, \xiA \SRb \right]  \\
& \quad + \Order{\partial^4,h^2}  \nonumber
\end{align}
\end{subequations}
Since the structure of all traces in equation \eqref{eqn:B_05B} is essentially the same, it simplifies matters to absorb the common parts into a new function:
\begin{align}
\TOp{p}{-\bZ^2} &\equiv \left. \partial_t R_k[\bg]\left(\EAA_k^{\background\,(2)}[A;\bg]+R_k[\bg]\right)^{-p}\right|_{\SRb=0}  \label{eqn:B_07}
\end{align}
Next, we specify the cutoff operator by $R_k[\bg]=k^2 \,\RegX{-\frac{\bZ^2}{k^2}}$ with some shape function  $\RegX{z\slash k^2}$, where $z\equiv -\bZ^2$,  and insert the Hessian of the background effective action:
\begin{align}
\TOp{p}{z}&=2\, k^{2\,(1-p)}\left[\RegX{z} - z \, \RegXd{z}\right]\,\left(z  +\RegX{z}+ k^{-2}\PotAd{(0)\,\prime\prime}\right)^{-p} \label{eqn:B_07b}
\end{align}
By virtue of these definitions  the projection of the  $\RHS$ of equation \eqref{eqn:B_01} assumes the following form now:
\begin{align}
\RHS&= +\frac{1}{2}  \Tr \left[  W_1(-\bZ^2;A) \right] - \frac{1}{2}  \Tr \left[  W_2(-\bZ^2;A)\cdot \xiA\, \SRb \right] \nonumber\\
&\quad - \frac{1}{4} \Tr \left[ \xiBg\,\SRb\,\flcb^{\mu}_{\phantom{\mu}\mu}  \,\,\, W_2(-\bZ^2;A)\right] \nonumber \\
&\quad  - \frac{1}{4} \Tr \left[\flcb^{\rho}_{\phantom{\rho}\rho}\, V_k^{(1)\,\prime\prime}(A) 	\,\, W_2(-\bZ^2;A)\right]  + \frac{1}{2} \Tr \left[\flcb^{\rho}_{\phantom{\rho}\rho}\, V_k^{(1)\,\prime\prime}(A) 	\,\, W_3(-\bZ^2;A)\cdot \xiA\, \SRb\right]
\nonumber\\
&\quad + \frac{1}{2} \Tr \left[ \var\Z^2\,	\,W_2(-\bZ^2;A)\right] -  \Tr \left[ \var\Z^2\,\,W_3(-\bZ^2;A)\cdot\xiA\,\SRb\right] \nonumber \
\\
&\quad + \frac{1}{4} \Tr \left[\flcb^{\rho}_{\phantom{\rho}\rho}\bZ^2	\,\,\, W_2(-\bZ^2;A)\right] - \frac{1}{2} \Tr \left[\flcb^{\rho}_{\phantom{\rho}\rho}\bZ^2	\,\,\, W_3(-\bZ^2;A) \cdot \xiA\, \SRb\right] \nonumber \\
&\quad - \frac{1}{2} \Tr \left[\xiBR \,\var\SR	\,\,\, W_2(-\bZ^2;A)\right]
 + \Order{\partial^4,\flcb^2} \, \label{eqn:B_08}
\end{align}
The two traces in the first line of equation \eqref{eqn:B_08} will be the source of all beta-functions in the background sector since they involve no fluctuation field $\flcb_{\mu\nu}$. The remaining terms encode the scale dependence of the couplings corresponding to invariants linear in $\flcb_{\mu\nu}$.

\subsection{The asymptotic heat kernel series}
All traces to be computed now refer to operators which are functions of the background Laplacian $\bZ^2$. The heat kernel techniques are an appropriate tool to project out the relevant basis invariants considered in our truncation since they provide a systematic expansion in terms of diffeomorphism invariants. 

The corresponding representation of a trace over an operator function $W_p(-\bZ^2;A)$ multiplied by some scalar $f$, i.e. $\Tr[f\,W_p(-\bZ^2;A)] = \sum_{j\geq 0} a_j(f;-\bZ^2) \, Q_{\frac{d-j}{2}}[W(\,\cdot\,;A)]$, consists of two ingredients: First, $a_j(f;-\bZ^2)$ is the heat kernel expansion  coefficient that contains all invariants of $j$th order in $\bZ^2$ along with some topology specific prefactors, and second,  $Q_{n}[W_p(\,\cdot\,;A)]$, defined in equation \eqref{eqn:03_12}, is the Mellin transform of $W(z;A)$ in which information about the function $W_p$ is absorbed. 

Explicitly, for an Euclidean signature manifold with non-vanishing boundary and Dirichlet boundary conditions, $\left.f\right|_{\partial\MaFs}=0$, the  terms in this expansion relevant for our truncation ansatz are given by \cite{vassi-review}:
\begin{align}
\Tr\left[f\, W_p\left(-\bZ^2;A\right)\right]&=(4\pi)^{-d\slash 2}\tr(\Id) \left\{ \int_{\MaFs}\md^d x \, \sqrt{\bg}\, f \,\cdot\,Q_{d\slash 2}[W_p] \right. \nonumber\\
 &\phantom{=(4\pi)^{-d\slash 2}\tr(\Id)}\quad- \frac{1}{2}\sqrt{\pi}\,\int_{\partial\MaFs} \md^{d-1}x\,\sqrt{\biM} \, f\,\cdot\,Q_{(d-1)\slash 2}[W_p]\label{eqn:B_09}\\
&\phantom{=(4\pi)^{-d\slash 2}}\,
+\frac{1}{6}\, \int_{\MaFs}\md^d x\, \sqrt{\bg}\, \SRb \,\cdot\, Q_{d\slash 2 -1}[W_p]\nonumber \\
& \left.\phantom{=(4\pi)^{-d\slash 2}}
+\frac{1}{6}\,
\int_{\partial\MaFs}\md^{d-1}x \,\sqrt{\biM}\left( 2\bEC\, f +3 n^{\lambda}\bZ_{\lambda}f \right)\,\cdot\, Q_{d\slash 2 -1}[W_p]+\cdots\right\}\nonumber 
\end{align}
Here, the trace $\tr(\Id)$ is over the unit matrix in field space and is thus equal to the number of scalar fields: $\tr(\Id)=\ns$. In the sequel, all invariants that are not part of the truncation ansatz will be omitted.

Further, notice that we did not allow for a renormalization of the kinetic term of the matter fields here, i.e. $A \bZ^2 A$ has no running prefactor. Thus, 
we can project out all relevant contributions by setting the scalar fields constant in the spacetime variables: $A(x)=A=\const.$ 

The first 5 traces as well as the last one in eq. \eqref{eqn:B_08} can be  evaluated by means of equation \eqref{eqn:B_09} and the following identity:
\begin{align}
Q_n\left[\TOp{p}{\,\cdot\,}\right] &\equiv 2\, k^{2(n+1-p)}\ThrfA{p}{n}{{ \mk^{(0)\, 2} +\tfrac{1}{2}\,\omgA^{(0)}A^2}} \, . \label{eqn:B_10}
\end{align}
Here $\ThrfA{p}{n}{w}$ is the standard threshold function introduced in \cite{mr}. For the 8$^{\text{th}}$ and 9$^{\text{th}}$ trace of \eqref{eqn:B_08} we use in addition the identity $Q_n[z F(z)]=  n \,Q_{n+1}[F(z)]$. 

Finally, we have to take care of the not yet tackled  traces of equation \eqref{eqn:B_08}, which are more involved since they contain $\var\bZ^2$.  However, we can exploit the fact that $\var \Tr[F(\Omega)]=\Tr[F^{\prime}(\Omega)\,\var \Omega]$ for any function $F$ of  a Hermitian operator $\Omega$ and
solve this problem by exchanging the order of variation and trace expansion (see \cite{MRS1} for more details):
\begin{align}
\Tr\left[F^{\,\prime}(-\Z^2)\var \Z^2\right]&=-\var \Tr\left[F(-\Z^2)\right] =
-\sum_{j\geq 0}  Q_{\frac{d-j}{2}}[F]  \, \var  a_j(\idm;-\bZ^2) 		 \label{eqn:B_11}
\end{align}
If we further use $Q_{n}[F]=-Q_{n+1}[F^{\,\prime}]$ we finally obtain
\begin{align}
\Tr\left[\TOp{p}{\,\cdot\,}\var \Z^2\right]&=\sum_{j\geq 0}  Q_{\frac{d-j+2}{2}}[\TOp{p}{\,\cdot\,}]\, \var a_j(\idm;-\bZ^2) \label{eqn:B_12}
\end{align}
 Since the variation operator `$\var$' acts on all invariants generated by the heat kernel coefficients $a_j(\idm;-\bZ^2)$, this trace affects only the level-(1) sector. In particular, the boundary terms cancel upon variation due to the relative coefficient of $+2$ between the Einstein-Hilbert and the Gibbons-Hawking term in the heat kernel expansion. The final form of the required trace, $ \Tr\left[\TOp{p}{\,\cdot\,}\var \Z^2\right]$, is therefore given by
 \begin{align}
 \Tr\left[\TOp{p}{\,\cdot\,}\var \Z^2\right]
&=(4\pi)^{-\frac{d}{2}}\ns \int_{\MaFs} d^d x \sqrt{\bg}\,  \left\{- \frac{1}{2}\, Q_{\frac{(d+2)}{2}}[\TOp{p}{\,\cdot\,}] \, \flcb^{\rho}_{\phantom{\rho}\rho} \right. 
\nonumber \\
& \left.\phantom{=(4\pi)^{-\frac{d}{2}}\ns \int_{\MaFs} d^d x \sqrt{\bg}\, \quad}
+\frac{1}{6}\, Q_{\frac{d}{2}}[\TOp{p}{\,\cdot\,}] \, {\bar G}^{\mu\nu}\,\flcb_{\mu\nu} \right\}+\cdots	\label{eqn:B_12B}
\end{align}
which can again be evaluated using \eqref{eqn:B_10}.

\subsection{The expansion in powers of  $A^2$}
So far we have projected the $\RHS$ of the flow equation \eqref{eqn:B_01} onto a subspace of diffeomorphism invariant functions. 
Still there is an infinite number of superfluous  terms present  due to the scalar potential in the denominator of the threshold-functions. In the truncation ansatz we have at most bilinear couplings between matter and gravity and the potential consists of a mass term and a four vertex only. In fact, the effective average action has an $\On{\ns}$ symmetry that is preserved under the RG evolution, so that we can expand the $\RHS$ directly in terms of $A^2$. The scalar fields enter via the argument of the threshold function $ \ThrfA{p}{n}{f(A^2)}$ that can be expanded in powers of $A^2$ as follows:
\begin{align}
\ThrfA{p}{n}{k^{-2}\bmk^{(0)\, 2} + \frac{1}{2}k^{-2} \bomg^{(0)} A^2}&
=+ \ThrfA{p}{n}{\mkB } 
-  \frac{p}{2\,k^2} \cdot \ThrfA{p+1}{n}{\mkB}
 \cdot  \bomg^{(0)} A^2 \nonumber \\
&\quad 
+ \frac{p\,(p+1)}{8\, k^4} \cdot \ThrfA{p+2}{n}{\mkB}\, \cdot  \bomg^{(0)\, 2} A^4
 + \Order{A^2}^{3}  \label{eqn:B_13}
\end{align}
Here we made use of the relation $\md\ThrfA{p}{n}{f}\slash\md f= (-p)\cdot \ThrfA{p+1}{n}{f}$ between the threshold-functions and their derivatives.

Finally, we can read off the beta-functions for the dimensionful couplings by comparing the coefficients for equal basis invariants. This leads to the equations \eqref{eqn:04_06} - \eqref{eqn:04_08} in Section \ref{sec:04} of the main text where also  the conversion to dimensionless couplings, as described there, has been performed.

\end{appendix}

\end{document}